\documentclass[12pt,draftcls, onecolumn,journal]{IEEEtran}
\usepackage{graphicx}
\usepackage{amsmath}
\usepackage{amsfonts}
\usepackage{amssymb}
\usepackage{cite}
\usepackage{color}
\usepackage{multirow}
\usepackage{epstopdf}
\usepackage{tabularx}

\usepackage{xcolor}
\usepackage{amssymb,bm,upgreek}
\usepackage{algorithm}
\usepackage{algorithmic}
\usepackage{multirow}
\usepackage{transparent}
\newcolumntype{L}[1]{>{\raggedright\arraybackslash}p{#1}}
\newcolumntype{C}[1]{>{\centering\arraybackslash}p{#1}}
\newcolumntype{R}[1]{>{\raggedleft\arraybackslash}p{#1}}

\usepackage{subfig}
\newcommand{\Prob}[1]{{\mathbb{P}}\left({#1}\right)}

\newcommand{\FB}{\mathsf{FB}}

\newcommand{\paren}[1]{\left(#1\right)}
\newcommand{\sqparen}[1]{\left[#1\right]}
\newcommand{\brparen}[1]{\left\{#1\right\}}

\newcommand{\norm}[1]{\left\| #1\right\|}
\newcommand{\field}[1]{\ensuremath{\mathbb{#1}}}
\newcommand{\N}{\ensuremath{\field{N}}} % natural numbers
\newcommand{\R}{\ensuremath{\field{R}}} % real numbers
\newcommand{\I}[1]{\ensuremath{\mathsf{1}_{\left\{#1\right\}}}} % indicator function
\newcommand{\ra}{\ensuremath{\rightarrow}} % abbreviation for right arrow
\newcommand{\PR}[1]{\ensuremath{\mathsf{Pr}\left\{#1\right\}}} % probability with braces
\newcommand{\PRP}[1]{\ensuremath{\mathsf{Pr}\left(#1\right)}} %\probability with parentheses
\newcommand{\EW}{\ensuremath{\mathsf{E}}} % Plain expectation without any parentheses and braces

\newcommand{\Arg}[1]{\ensuremath{\mathsf{Arg}\left(#1\right)}} %\probability with parentheses

\newcommand{\snr}{\ensuremath{{\sf SNR}}}
\newcommand{\snrPow}{\ensuremath{{\sf SNR}^{\rm pow}}}

\newcommand{\Floor}[1]{\ensuremath{\left\lfloor #1\right\rfloor}}
\renewcommand{\vec}[1]{\ensuremath{\boldsymbol{#1}}} %Re-define \vec command to generate vectors in bold
\newcommand{\pol}{\ensuremath{\mathcal{P}}}

\newcommand{\polOptPow}{\ensuremath{\mathcal{P}_{\rm pro}^{\ast}}}
\newcommand{\polOptExp}{\ensuremath{\mathcal{P}_{\rm sum}^{\ast}}}
\newcommand{\polOptPowFB}{\ensuremath{\mathcal{P}_{\rm pro}^{\FB}}}
\newcommand{\polOptExpFB}{\ensuremath{\mathcal{P}_{\rm sum}^{\FB}}}
\newcommand{\NPowFB}{\ensuremath{N_{\rm pro}^{\FB}}}
\newcommand{\NExpFB}{\ensuremath{N_{\rm sum}^{\FB}}}
\newcommand{\XiPow}{\ensuremath{\Xi_{\rm pro}}}
\newcommand{\XiExp}{\ensuremath{\Xi_{\rm sum}}}
\newcommand{\Tpow}{\ensuremath{T_{\rm pro}}}
\newcommand{\Texp}{\ensuremath{T_{\rm sum}}}
\newcommand{\ESZ}[1]{\ensuremath{\mathsf{E}_{Z}\left[#1 \right]}} %Expectation with square parentheses

\newcommand{\TX}{\ensuremath{\vec{x}_{\rm s}}}
\newcommand{\RX}{\ensuremath{\vec{x}_{\rm d}}}

\newcommand{\Rave}{\ensuremath{{\sf R}_{\rm ave}}}
\newcommand{\RavePow}{\ensuremath{{\sf R}_{\rm ave}^{\rm pow}}}
\newcommand{\RaveExp}{\ensuremath{{\sf R}_{\rm ave}^{\rm exp}}}

\newcommand{\XoptPow}{\ensuremath{\vec{X}_{\times}^{\ast}}}
\newcommand{\XoptExp}{\ensuremath{\vec{X}_{+}^{\ast}}}

\newcommand{\Gpow}{\ensuremath{G^{\rm pow}}}
\newcommand{\Gexp}{\ensuremath{G^{\rm exp}}}

\newcommand{\xs}{\ensuremath{\vec{x}_{\rm s}}}
\newcommand{\xd}{\ensuremath{\vec{x}_{\rm d}}}

\newcommand{\Xr}{\ensuremath{\vec{X}_{\pol}}}

\newcommand{\Pout}{\ensuremath{{\sf P}_{\rm out}}}
\newcommand{\PoutPow}{\ensuremath{{\sf P}_{\rm out}^{\rm pow}}}
\newcommand{\PoutExp}{\ensuremath{{\sf P}_{\rm out}^{\rm exp}}}

\newcommand{\RISselect}{\ensuremath{\widehat{s}_{}}}
\newcommand{\RISselectPow}{\ensuremath{\widehat{s}_{\rm pro}}}
\newcommand{\RISselectExp}{\ensuremath{\widehat{s}_{\rm sum}}}

\newcommand{\expo}{\rm sum}
\newcommand{\pow}{\rm pro}

\newcommand{\m}{\textnormal{m}}

\DeclareMathOperator{\arctang}{arctan}

\usepackage{amsmath}

\DeclareMathOperator*{\argmin}{arg\,min}

\newtheorem{definition}{Definition}

\newtheorem{selection}{Selection Policy}

\newtheorem{theorem}{Theorem}

\newtheorem{lemma}{Lemma}
\newtheorem{remark}{Remark}

\newlength{\figwidth}
\newcommand{\Fr}[1]{{\color{black}{#1}}}

\begin{document}
\title{Optimum Reconfigurable Intelligent Surface Selection for Wireless Networks}

\vspace{-5mm}
\author{\IEEEauthorblockN{Yuting Fang, {\it  Member, IEEE}, Saman Atapattu, {\it Senior Member, IEEE},\\
Hazer Inaltekin, {\it  Member, IEEE}, and Jamie Evans, {\it Senior Member, IEEE}\\}
\vspace{-5mm}
%\thanks{Yuting Fang, S. Atapattu, and J.~Evans are with the Department of Electrical and Electronic Engineering, University of Melbourne, Parkville, VIC 3010, Australia.
%Email:\{yuting.fang, saman.atapattu, jse\}@unimelb.edu.au.}
%\thanks{H.~Inaltekin is with the School of Engineering, Macquarie University, North Ryde, NSW 2109, Australia.
%Email: hazer.inaltekin@mq.edu.au.}
}

\maketitle
\vspace{-15mm}
\begin{abstract}
The reconfigurable intelligent surface (RIS) is a promising technology that is anticipated to enable high spectrum and energy efficiencies in future wireless communication networks. This paper investigates optimum location-based RIS selection policies in RIS-aided wireless networks to maximize the end-to-end signal-to-noise ratio for product-scaling and sum-scaling path-loss models where the received power scales with the \emph{product} and \emph{sum} of the transmitter-to-RIS and RIS-to-receiver distances, respectively. These scaling laws cover the important cases of end-to-end path-loss models in RIS-aided wireless systems. %and an exponential path-loss model in indoor communications. %The main communications scenario of interest is indoor communications with exponentially decaying path-loss function.
The random locations of all available RISs are modeled as a Poisson point process. To quantify the network performance, the outage probabilities and average rates attained by the proposed RIS selection policies are evaluated by deriving the distance distribution of the chosen RIS node as per the selection policies for both product-scaling and sum-scaling path-loss models. %Since feeding the information of all RIS locations back to the source for an optimum selection can
%To make the network operation distributed, we further assume the extra signal processing capability at RISs for feedback %Feedback could incur heavy signaling overhead. To reduce the overhead, we
%and
We also propose a limited-feedback RIS selection framework to achieve distributed network operation. %policies by limiting the average number of RISs that feed back their location information to the source. %we also propose limited-feedback RIS selection policies to reduce the overhead. In this case, only a part of RISs feedback their locations to the source, and then the best RIS is selected from those feeding back RISs.
The outage probabilities and average rates obtained by the limited-feedback RIS selection policies are derived for both path-loss models as well. The numerical results show notable performance gains obtained by the proposed RIS selection policies. %and demonstrate that the conventional relay selection policies are not suitable for RIS-aided wireless networks.
\end{abstract}
\vspace{-5mm}

%\newpage
\section{Introduction}

\subsection{Background and Motivation}
Globally, mobile and machine-to-machine data traffic is expected to grow at a rate of about 55\% per year from 2020 to 2030, reaching 5,000\, Exabytes per month in 2030~\cite{ITU-R}. While supporting 1\,Terabyte per second speeds, the sixth-generation (6G) wireless networks are expected to facilitate sensing, localization, and computing in real-time by using a smart wireless environment. One of the key enablers to realizing a smart environment is a reconfigurable intelligent surface (RIS), which includes many nearly passive elements having ultra-low power consumption. Each element can electronically control the phase of the reflected radio waves to concentrate energy in the desired spatial directions~\cite{renzo2019eurasip}. As such, an RIS dynamically adapts to changing wireless channel conditions to create a favorable propagation environment and to increase the energy efficiency of wireless networks~\cite{9133157}. Moreover, RIS greatly decreases hardware costs
and power consumption. This is because the spatial feeding method of the RIS avoids the immoderate power loss due to the massive feeding networks of phased arrays \cite{9020088}. A critical milestone to realize the full scale of these advantages in a network setting is to have adaptive algorithms optimizing the selection and activation of RISs to enhance smart wireless connectivity, which will be an essential feature of future wireless systems.

Due to the possible irregular terrain for deployment of RISs, it is naturally expected that multiple RISs will be deployed according to a potentially random-looking topology to provide connectivity in future wireless networks. In these cases, the best RISs will need to be selected for achieving high quality connection between given source and destination nodes, in contrast to popular single and fixed RIS deployments. Similar to relay networks, utilizing multiple RISs for single-user communication increases the overall system complexity and signaling overhead. %, which are usually much worse than the case of RIS networks due to passive RISs.
Thus, a well-designed adaptive RIS-selection policy is of particular importance to achieve the benefits of the multiple RIS deployment topologies. Further, RISs are proposed to be nearly passive network elements with limited computing power to support signal processing and edge computing. %RISs may have limited computing power to support signal processing or edge computing. %the task of channel estimation (as well as timely feedback of this information to a central entity for RIS selection) is an onerous operation in RIS-aided networks.
Therefore, a selection policy utilizing only location information of available RISs is more practical
%and desirable
in this context, which can be assumed to be time-invariant. {Motivated by these facts, this paper aims to focus on the location-based optimum RIS selection problem in RIS-aided future wireless networks that consist of multiple randomly distributed RISs and derives performance metrics under the optimum selection policy.}
\vspace{-3mm}
\subsection{Related Work}
%\subsubsection{Fixed RIS Networks}
Most previous work on performance analysis of RIS-aided wireless communications considers different wireless network scenarios with single and fixed RIS setup, e.g., \cite{Huang2019twc,Jung2020twc}. %\cite{Huang2019twc,Jung2020twc,Wang2020spl,Badiu2020wcl,atapattu2020reconfigurable,Zhao2020wcl,9066923,Dharmawansa2020wcl}. and references therein.
%\com {More specifically, \cite{Huang2019twc} focuses on the joint optimization of the transmit power allocation and the RIS phase shifts.
%\cite{Yang2020tcom} studies channel estimation and performance analysis for an RIS-based orthogonal frequency division multiplexing (OFDM) network. \cite{Zheng2020wcl} also focuses on OFDM system with channel estimation and the RIS phase shifts optimization. \cite{Jung2020twc} presents an asymptotic analysis of the uplink data rate of an RIS-based multiple-input multiple-output (MIMO) system. \cite{Wang2020spl} conducts channel estimation for RIS-aided millimeter wave (mmWave) systems. \cite{Badiu2020wcl} studies the transmission through an RIS with phase errors. \cite{atapattu2020reconfigurable} conducts the performance analysis and optimization of the two-way communication assisted by an RIS. \cite{9066923} optimizes the performance under a power constraint for any given matrix of RIS phases.} %focus on channel estimation and performance analysis. power allocation at the transmitter (active beamforming) phase optimization at the RIS (passive beamforming), hybrid beamforming (active and passive beamforming),
%
%\new {include following papers appropriately, specially \cite{yang2020outage}}
\cite{9129476,xiu2020secure,yang2020energyefficient,zheng2020doubleirs,9060923,9427474,9366805,9410457} considers a given set of locations of multiple RISs without RIS selection. Specifically, the signal-to-noise ratio ($\snr$), achievable sum-rate, secrecy rate, and energy efficiency of RIS-assisted networks are maximized %by jointly optimizing
%the transmit power vector at the transmitter and the phase shift matrix with passive beamforming at RISs
in \cite{9129476,xiu2020secure,yang2020energyefficient,zheng2020doubleirs}, respectively. {\cite{9060923} proposes and analyzes a double-RIS aided system. \cite{9427474} investigates the capacity region of a two-user  network with one access point aided by multiple RIS elements. \cite{9366805} proposes a channel estimation framework for a RIS-aided multi-user system. \cite{9410457} designs a novel hybrid beamforming scheme for a RIS-aided multi-hop network.} {Very recently, the RIS-user associations with and without the BS power control are optimized in a multi-RIS aided network in \cite{9448236}.} The RIS with the highest instantaneous end-to-end $\snr$ is selected among multiple fixed RISs to aid the communication in \cite{yang2020outage}, where the outage probability and average sum-rate are investigated. This paper only focuses on deterministic RIS deployment without modeling potential randomness in RIS locations.

%\subsubsection{Random RIS Networks}

Spatial point processes are regarded as tractable analytical tools to model the locations of the network elements (e.g., base stations, users, or relays), with a good statistical fit to physical wireless network deployments~\cite{Andrews2011tcom}. However, only a few papers have so far focused on spatial network models for the deployment of RISs or the distribution of users in RIS-assisted networks; see~\cite{Renzo2019jwcn,hou2019arxiv,kishk2020arxiv,9110835,nemati2020risassisted,shafique2021stochastic}. In \cite{Renzo2019jwcn}, environmental objects are assumed to be coated as RISs where the deployment is modeled as a modified line process with random locations and orientations. %The authors in \cite{Renzo2019jwcn}  show that the probability a typical random object acts as a reflector is independent of the length of the object itself.
In \cite{hou2019arxiv}, they propose a joint design for the detection weight at randomly distributed users and passive beamforming weight at RISs. \cite{kishk2020arxiv} considers a cellular network where the midpoints of the blockages are distributed according to a Poisson point process (PPP) and the blockages are equipped with RISs. %They show that how equipping a subset of the blockages with RISs enhances the performance of the cellular network.
\cite{9110835} provides performance analysis of a large-scale
mmWave cellular network where the locations of base stations and RISs are modeled using independent PPPs. %\cite{9129476} maximizes the sum-rates of an RIS-empowered network that consists of multiple source-RIS-destination pairs where RISs are located at fixed locations.
Recently, \cite{nemati2020risassisted} studies the coverage of an RIS-aided large-scale mmWave cellular network where the buildings with RISs are distributed according to a PPP. \cite{shafique2021stochastic} analyzes the coverage probability, ergodic capacity, and energy efficiency for indirect RIS-aided
network with where the locations of the RISs follow a binomial point process. {Very recently, \cite{9500724} analyzes the outage probability of an RIS-aided network where multiple IRSs are randomly distributed with different association policies activating a random RIS, the closest RIS to the transmitter, or all available RISs. However, none of these papers considered a selection strategy choosing the best RIS from a collection of randomly distributed RISs to optimize connectivity between source and destination nodes. Performance characterization of RIS-aided random wireless networks with such optimum RIS selection is an open problem in the literature, which we tackle in this paper.}

%However, it has not been investigated how the optimal subset of blockages can be obtained through its analysis.

%While \cite{Renzo2019jwcn,hou2019arxiv,kishk2020arxiv,9110835, 9129476, nemati2020risassisted} stand on their own merits,  \cite{Renzo2019jwcn,9110835,nemati2020risassisted} utilize all randomly distributed RISs without selecting an RIS, \cite{hou2019arxiv,9129476} consider that RISs are located at fixed locations, \cite{kishk2020arxiv} considers a random criterion for selecting the blockages equipped with RISs. Therefore, the study of an RIS-aided wireless network with the optimum RIS selection from multiple randomly distributed RISs and characterization of the resulting network performance, has not been studied in the literature.

%An important contribution of this paper is the stochastic geometry based modelling of an RIS-aided wireless network and characterization of the resulting network performance when RISs are optimally selected for data relaying.
\vspace{-3mm}
\subsection{Our Approach and Contributions}
%To solve this open research problem on the topic of RIS-aided wireless networks,
In this paper, we consider an RIS-aided wireless network where multiple RISs are randomly distributed and an RIS is optimally selected to assist data transmission from a transmitter (TX) to a receiver (RX) based on the relative locations of RISs with respect to TX and RX nodes. As analyzed in \cite{9119122,9206044,9154326}, the scaling law of received power through the reflection of an RIS is a function of the TX-RIS and RIS-RX distances. We note that the product-scaling law where the path-loss scales according to the \emph{product} of the TX-RIS and RIS-RX distances and the sum-scaling law where the path-loss scales according to the \emph{sum} of these distances have been established as the fundamental path-loss models to model the end-to-end signal power attenuation when the TX and RX nodes are connected by an intermediate RIS. They will be the path-loss models that we follow in this paper. %are worth of analysis since these two scaling laws cover the most cases of RIS-aided end-to-end path-loss models. %To accommodate different types of signal propagation, this paper considers the classical power-law propagation model for outdoor communications and exponential-law (which we refer to as exp-law for brevity for the rest of the paper) propagation model for indoor communications as well.
%To make the network operation distributed, we further assume the extra signal processing capability at RISs for feedback and propose limited-feedback RIS selection policies.
% We note that selecting an optimum RIS from all RISs requires feeding the information of all RIS locations back to the TX, which could incur heavy signaling overhead. To reduce the overhead,
%we propose limited-feedback RIS selection policies
%that a part of superior RISs feed back their location information to the source and then the source selects an RIS among these RISs feeding back.
Using the tools from stochastic geometry, we develop a tractable theoretical framework to obtain the outage probability and average rate for the RIS-aided wireless network under the optimum RIS selection policies for both path-loss scaling laws. We also analyze the limited feedback case to achieve distributed network operation under the assumption of availability of RIS feedback capability for selection. We emphasize that the derivations of our theoretical results are challenging since we need to tackle the \emph{random} instantaneous $\snr$ values at the random RIS locations coupled with the \emph{randomness} over the fading process.
%\com{Let's write contribution here...}
%{\it Notation:} The main notation being used throughout the paper is as follows, with remaining some of which to be introduced later in the paper when needed.
Despite these challenges, we make the following novel contributions:
\begin{itemize}
\item Based on the nature of the $\snr$ in RIS-aided wireless networks, we propose location-based optimum RIS selection policies that maximize the end-to-end $\snr$ of the link connecting a TX and a RX via an RIS for product-scaling and sum-scaling path-loss models. These selection policies make RIS selection decisions based on the insight that the optimum RIS given the \emph{product-scaling} path-loss law is the RIS that has the \emph{minimal product} of the distances of the TX-RIS and RIS-RX links among all randomly distributed RISs. On the other hand, the optimum RIS given the \emph{sum-scaling} path-loss law is the RIS that has the \emph{minimal sum} of these distances.

\item We derive the distance distribution of the optimum RIS node for both path-loss models. These distributions are of critical importance to obtain the outage probability of RIS-aided wireless networks. The derived distributions have broader applicability where a node that has the minimal product or sum of the distances is selected. Using the derived distance distribution and the gamma approximation for fading channels, we derive theoretical expressions for the outage probability and average rate of the optimum RIS selection policies for product-scaling and sum-scaling path-loss models.

\item To characterize the system performance given limited-feedback RIS selection policy, we derive the average number of RISs feeding back and confirm the number of RISs feeding back is a Poisson random variable (RV) for both path-loss models. Using this result, we obtain theoretical expressions for the outage probability and average rate under the limited-feedback RIS selection policies for both path-loss models.
\end{itemize}

We verify the derived analytical results by means of extensive numerical analysis and simulations. The potential performance improvement obtained by the optimum RIS selection policy is demonstrated by comparing the performance gains obtained by the optimum policy with those of heuristic sub-optimum policies via simulations. {Our results reveal that a selection policy performing optimally for decode-and-forward (DF) relay networks (i.e., the min-max policy) can perform very poorly for RIS networks due to fundamentally different signal propagation characteristics.}
From a system design point of view, this is an important result. Numerical results also demonstrate that limited-feedback RIS selection policies achieve almost the same outage and data rate performance as the optimum  centralized RIS selection policy while significantly reducing the feedback and signaling load. Through these theoretical and numerical results, this paper provides important guidelines for selecting the optimal RIS towards a reliable yet practical RIS-aided wireless network.

%We acknowledge that the idea of location-based optimum selection criteria and its limited-feedback one are inspired by \cite{atapattu2020locationbased}, but the results in \cite{atapattu2020locationbased} cannot be applied to this paper.
The results in \cite{atapattu2020locationbased} focus on the node distance distribution given the min-max optimum selection criteria, and cannot be directly applied to this paper.
%the idea of location-based optimum selection criteria and its limited-feedback one are inspired by , but
%This is due to the fact that the expressions of $\snr$ via a given RIS is different from the one in \cite{atapattu2020locationbased} that considers relay-aided communications. This leads to solving the distance distributions of the optimum RIS node and the corresponding performance metrics for power-law and exp-law path-loss models are fundamentally new challenging problems. Also, the fading channel distribution is different from the one considered in \cite{atapattu2020locationbased}.
Our key technical challenges are to solve the node distance distribution given the min-product and min-sum optimum selection criteria. These are fundamentally different problems with their own particular technical challenges requiring new solution approaches when compared to those investigated in
\cite{atapattu2020locationbased}. Moreover, there are also fundamental differences between RIS-aided networks and relay-aided networks in terms of their end-to-end path-loss scaling laws as a function of the TX-to-RIS and RIS-to-RX distances; see \cite{9119122,9095301}. Thus, this paper provides novel theoretical results and numerical insights that have not been studied before.

\vspace{-5mm}

\subsection{Notation}
We use boldface letters to represent vector quantities. $\N$ denotes the set of natural numbers and $\R^2$ denotes the two-dimensional Euclidean space. $| x |$ and $\|\vec{x}\|$ denote
the absolute value of a scalar quantity $x$ (real or complex) and the Euclidean norm of a vector quantity $\vec{x}$, respectively.
%$f_{A}(x)$ is the density probability function (PDF) of RV $A$.
$\Prob{\cdot}$ denotes probability. $\ESZ{\cdot}$ is the expectation over RV $Z$. $\EW_\Phi\sqparen{\cdot}$ is the expectation over the point process $\Phi$. %spatial

%We organize the rest of the paper as follows: Section \ref{Sec:System} describes the system model and performance metrics. Section \ref{sec:power}, which focuses on the product-scaling and sum-scaling path-loss models, covers the performance analysis of the optimum RIS selection. Section \ref{sec:limited} focuses on performance analysis of RIS selection policy with limited-feedback. Section \ref{sec: numerical} presents detailed numerical study, justifying the derived analytical results and revealing the impact of the key system parameters on system performance. Finally, we conclude the paper in Section \ref{sec:con}.
\vspace{-3mm}
\section{System and Analytical Models}\label{Sec:System}

\subsection{System Model}
\begin{figure}[!t]
\centering
\includegraphics[width=0.4\textwidth]{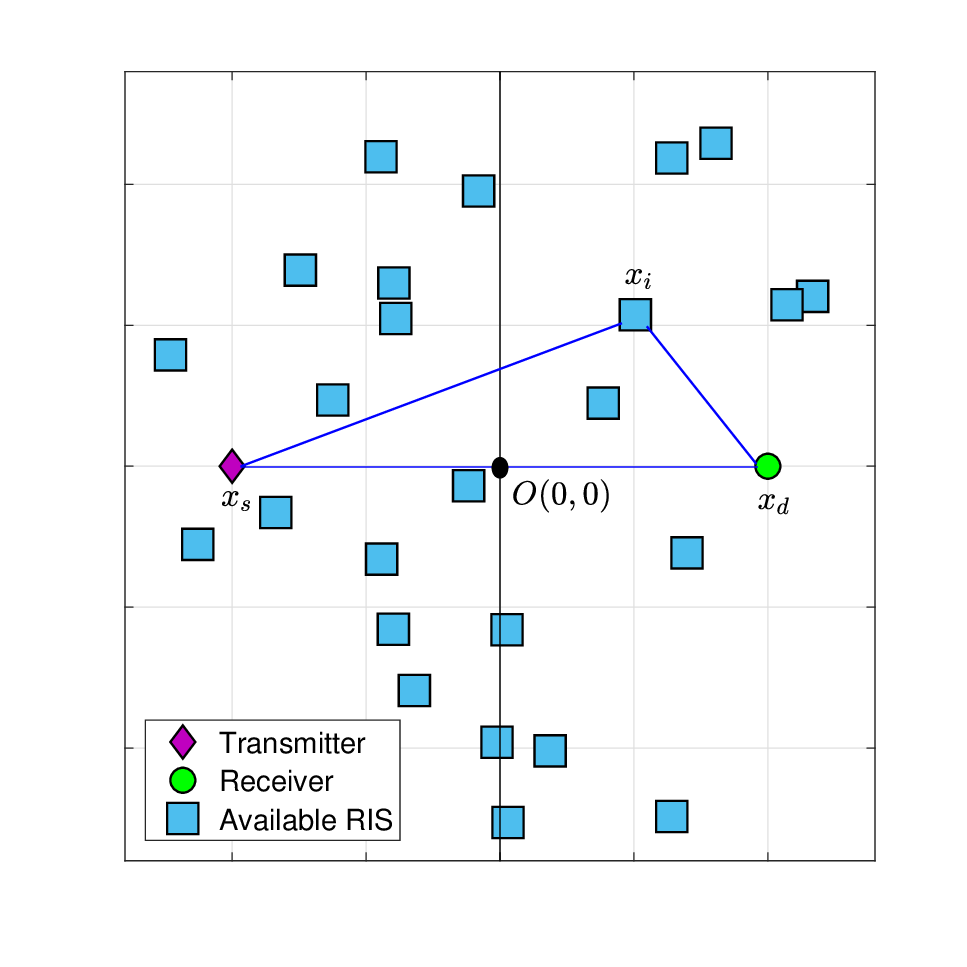}\vspace{-8mm}
\caption{An example illustration for an RIS-aided wireless network.}
\label{fig:model}
\vspace{-10mm}
\end{figure}

We consider an RIS-aided wireless system in $\R^2$, as illustrated in Fig. \ref{fig:model}, where the TX and RX are located at arbitrary locations, denoted by $\TX \in \R^2$  and $\RX \in \R^2$, respectively. Potential RISs are randomly distributed according to a spatial homogeneous Poisson point process (PPP) $\Phi$ with density $\lambda>0$. The locations of available RISs are denoted by $\phi = \brparen{\vec{x}_1, \vec{x}_2, \ldots}$, where $\vec{x}_i \in \R^2$ is the $i$th RIS location for $i \in \N$ and $\phi$ is a particular realization of $\Phi=\brparen{\vec{X}_1, \vec{X}_2, \ldots}$. We consider that the locations of available RISs are known by a central network controller to make a decision for selecting an RIS to aid the communication between the TX and RX\footnote{{Our system setup can be applied in the scenarios where RISs are irregularly deployed and one RIS is selected to enhance the communication from the source to destination. For example, in a RIS-enhanced cellular network, potential RISs are employed on irregularly-located trees or buildings and the optimal RIS is selected to strengthen the received signal power of a cell-edge user.}}. %there exists a central network controller knows the locations of RISs and makes the selection based on this knowledge of locations of RISs
{As in \cite{9448236,9500724}, we assume when one RIS is selected (activated), the remaining RISs remain switched off.} {We assume the signal power received over the longer TX-RX link is much smaller than the one received over the shorter RIS-RX link, i.e., there is no direct link between TX and RX\footnote{{This assumption is reasonable when the direct link is severely shadowed by an object in the environment or the decay of path-loss with distance is very sharp.}}.}

We assume nearly passive RISs as introduced in \cite{9119122} where each RIS is implemented with $N$ number of reflecting elements which can be adjusted individually for adapting to fading conditions. We denote $\mathcal{W}=\textrm{diag}\paren{\exp\paren{j\varpi_{i,1}}, \ldots, \exp\paren{j\varpi_{i,N}}}$ as phase shifts of the $i$th RIS. {Further,
we denote $h_{i,n}=\alpha_{i,n}\exp(-j\psi_{i,n})$ and $g_{i,n}=\beta_{i,n}\exp(-j\varphi_{i,n})$ as fading channels between the TX and the $n$th reflecting element of the $i$th RIS and that between the $n$th reflecting element of the $i$th RIS and the RX, respectively. As in \cite{9095301,8796365}, we assume $h_{i,n}$ and $g_{i,n}$ are independent and identically distributed (i.i.d.) complex Gaussian RVs with zero mean and unit variance i.e., $h_{i,n}, g_{i,n}\sim \mathcal{CN}(0,1)$.}
%Since fading channels are independent and identically distributed (i.i.d.) complex Gaussian RVs with zero mean and unit variance, i.e., $h_{i,n}, g_{i,n}\sim \mathcal{CN}(0,1)$,
Hence, magnitudes $\alpha_{i,n}$ and $\beta_{i,n}$ follow the Rayleigh distribution.
%\com{We assume that random coefficients $\alpha_{i,n}$ and $\beta_{i,n}$ change at a much higher speed than the RIS node locations.}
{Then, the instantaneous received signal\footnote{{In practice, the obliquity factor (i.e., the incidence and reflected angles of signals at the RIS) mentioned in \cite{9433568} and blockages may affect the power received at the RX via the RIS. For tractability, as in \cite{hou2019arxiv,yang2020outage,shafique2021stochastic}, we ignore the obliquity factor and blockages when modeling the received power via the RIS.}
} at time $t$ via the $i$th RIS located at $\vec{X}_i$ is given by \cite[eq.~(11)]{8796365}
\begin{eqnarray}\label{eq: RxSig1}
y_i(t) = \frac{\sqrt{P}\sum_{n=1}^{N}\alpha_{i,n}\beta_{i,n}\exp(j(\varpi_{i,n}-\psi_{i,n}-\varphi_{i,n}))}
{\sqrt{G\paren{\norm{\xs -\vec{X}_i},\norm{\vec{X}_i -\xd}}}}s(t)+w(t),
\end{eqnarray}
where $P$ denotes the transmit power, $s(t)$ is a unit energy signal, $w(t)$ is additive white Gaussian noise (AWGN) having complex Gaussian distribution with mean zero and variance $N_0$. $G\paren{\norm{\xs -\vec{X}_i},\norm{\vec{X}_i -\xd}}$ is a path-loss function and we will discuss its dependence on the TX-RIS distance $\norm{\xs -\vec{X}_i}$ and the RIS-RX distance $\norm{\vec{X}_i -\xd}$ in Sec. \ref{subsec:RIS-selection}.
A careful inspection of the structure of \eqref{eq: RxSig1} reveals that the optimal choice of $\varpi_{i,n}$ that maximizes the instantaneous received signal $y_i(t)$ is $\varpi_{i,n}=\psi_{i,n}+\varphi_{i,n}, i \in \N$.
This is feasible at RISs since they can obtain the knowledge of the channel phases by various methods mentioned in \cite{8796365,renzo2019eurasip,8879620}, e.g., by embedding low-power sensors throughout the RISs.
%A careful inspection of the structure of \eqref{eq: RxSig1} reveals that the optimal choice of $\varpi_{i,n}$ that maximizes the instantaneous received signal $y_i(t)$ is $\varpi_{i,n}=\psi_{i,n}+\varphi_{i,n}, i \in \N$. This is feasible at RISs since the channel phases can be known to RISs by embedding low-power sensors throughout the RISs, possibly powered by energy harvested modules, which are in charge of sensing the channels . Another promising method to estimate the optimal phases without assuming the knowledge of the individual channels between RISs and the TX and the RX was recently reported in \cite{} by considering only the combined (or product) channel between them. %Or the channel state information is provided for the RISs via a communication software.
Thus, the assumption of optimal phase shifting is widely adopted in existing RIS studies, e.g., \cite{9095301,8796365,atapattu2020reconfigurable,9094171}. Using this assumption, \eqref{eq: RxSig1} can be rewritten as
\begin{eqnarray}\label{eq: RxSig}
y_i(t) = \frac{\sqrt{P}Z_i}
{\sqrt{G\paren{\norm{\xs -\vec{X}_i},\norm{\vec{X}_i -\xd}}}}s(t)+w(t),
\end{eqnarray}
where $Z_i = \sum_{n=1}^{N}\alpha_{i,n}\beta_{i,n}$.} We note that $Z_i$ are i.i.d. RVs for different RISs. Thus, for compactness, we remove subscript ``$i$'' from $Z_i$, i.e., $Z_i=Z$, $\forall i$, in the rest of the paper, and it will be clear from the context that $Z$ is the power gain associated with the RIS selected to connect TX and RX.

%Furthermore, we assume the $n$th reflecting element of the $i$th RIS always operates at, i.e., Using this assumption,

\subsection{RIS Selection Policies}\label{subsec:RIS-selection}

%The aim of this paper is to investigate the communication performance of an RIS-assisted wireless network where one RIS having the maximal instantaneous $\snr$ is selected for the communication between the TX and RX.
%In this subsection, we define present the expression of instantaneous $\snr$ and the definitions of outage probability and data rate which facilitates our derivation of network performance of power-law and exp-law path-loss in Sections \ref{sec:power} and \ref{sec:limited}.

In this subsection, we first give the instantaneous $\snr$ of an RIS-aided wireless network, then discuss the path-loss scaling laws, % as functions of the TX-RIS and RIS-RX distances,
and finally formulate the location-based optimum RIS selection policies that maximize the instantaneous $\snr$ for different path-loss scaling laws.

%power-law and exp-law path-loss models.

%\subsubsection{Instantaneous $\snr$}\label{subsubsec:Ins-SNR}
We consider the case where a single RIS is selected for aiding the communication between the TX and RX according to a given RIS selection policy which is defined as follows:
\begin{definition} \label{Def: Selection Policy}
an RIS selection policy $\pol: \Sigma \mapsto \R^2$ is a mapping from the set of all countable locally finite subsets of $\R^2$, denoted by $\Sigma$, to $\R^2$, that satisfies the condition $\pol\paren{\phi} \in \phi$ for all $\phi \in \Sigma$.
\end{definition}

%\com{we can directly write \eqref{SNR} saying to max instantaneous $\snr$ by setting $\varpi_{i,n}=\psi_{i,n}+\varphi_{i,n}$. Index $i$ is confusing as it does not have any relation to  $\pol$.}

With a slight abuse of notation, we will denote the RIS selected by $\pol$ as $\Xr$. %We denote $\varpi_{\pol,n}$, $h_{\pol,n}=\alpha_{\pol,n}\exp(-j\psi_{\pol,n})$, and $g_{\pol,n}=\beta_{\pol,n}\exp(-j\varphi_{\pol,n})$ as the reflection coefficients,
Using \eqref{eq: RxSig}, we write the instantaneous $\snr$ associated with RIS $\Xr$ according to
%\begin{eqnarray}\label{SNR-1}
%\snr_{\rm inst}\paren{\pol,\Phi} = \frac{\bar\gamma|\sum_{n=1}^{N}\alpha_{i,n}\beta_{i,n}\exp\paren{j\mathcal{Z}_{i,n}}|^2}
%{G\paren{\norm{\xs -\Xr}}G\paren{\norm{\Xr -\xd}}},
%\end{eqnarray}
%Since the corresponding instantaneous $\snr$ associated with RIS $\xr$ is given by
\begin{eqnarray}\label{SNR}
\snr_{\rm inst}\paren{\pol,\Phi,Z} = \frac{\bar\gamma Z^2}{G(\norm{\xs - \Xr},\norm{\Xr-\xd})},
\end{eqnarray}
where $\bar\gamma=\frac{P}{N_0}$ is the average $\snr$.
%\com{Definitions for $G_1(\cdot)$ and $G(\cdot)$? Better use same $G$ if you really do not consider non-identical cases. RV $Z$ can even be Incorporated to \eqref{SNR}. $Z$ should have relation with $i$, so it may be $Z_i$.}

We next discuss the dependence of a path-loss function $G(\norm{\xs - \Xr},\norm{\Xr-\xd})$ on the TX-RIS distance $\norm{\xs - \Xr}$ and the RIS-RX distance $\norm{\Xr-\xd}$. %the relationship between the path-loss function $G$ of RIS-assisted wireless communications and TX-RIS and RIS-RX distances.
Based on \cite{9119122,9154326,9206044,9433568}, the scaling law of the end-to-end received power through the reflection of an RIS as a function of the TX-RIS and RIS-RX distances, depends on the relation between the geometric size of the RIS, the wavelength of the radio wave, and the relative TX-RIS and RIS-RX distances. Notably, two path-loss scaling laws of the TX-RIS and RIS-RX distances are worth of analysis.

\emph{1) Product-Scaling path-loss models:} If the size of the RIS is not large enough as compared with the wavelength and the transmission distances $\norm{\xs - \Xr}$ and $\norm{\Xr-\xd}$, the end-to-end received power at the receiver scales, as $4L_{1}^2(
\norm{\xs - \Xr}\norm{\Xr-\xd})$ \cite{9154326}, where $2L_{1}$ is the length of one-dimensional RIS.
\cite{9206044} also gives similar scaling laws\footnote{{For example, in \cite{9206044}, a large RIS with the size of $1\,\m\times1.2\,\m$ and the carrier frequency of 10.5 GHz at the TX-RIS distance of 100\,m and the RIS-RX distance of 100\,m and a small RIS with the size of $0.384\,\m\times0.096\,\m$ and the carrier frequency of 4.25 GHz at the TX-RIS distance of 3.5\,m and the RIS-RX distance of 10\,m are shown to operate as the product-scaling path-loss model.}} with different power of product distances in the far-field case\footnote{When the distance between the TX(RX) and the center of the RIS is less than $\xi=\frac{2 D^2}{\lambda_s}$, the RIS is considered to be in the near-field of the TX(RX). Otherwise, the RIS is said to be in the far-field of the TX(RX).}, %$\xi$ the boundary of the far-field and the near-field of the RIS. $D^2=4L_{x}L_{y}$ is the largest dimension of the antenna array, where an RIS has length $2L_{x}$ and $2L_{y}$ parallel to the x-axis and y-axis, respectively. $\lambda_s$ is the wavelength of the signal.}
where the path loss of through the reflection of an RIS is proportional to $(\norm{\xs - \Xr}\norm{\Xr-\xd})^{2}$. The path-loss given in \cite{9206044} also relates to the antenna gains, the RIS element gain, and other system parameters. {Moreover, \cite{9433568} shows for focusing lenses of RISs, a single scaling law is observed, i.e., the product path-loss model is \emph{sufficiently accurate for short and long distances}}. On the other hand, for a classical power-law path-loss model, signal power decays as $\Gpow(d)=d^{\eta}$, where $d$ is the link distance and $\eta>2$ is the path-loss exponent. The end-to-end path-loss for the power-law model is $\Gpow(\norm{\xs - \Xr},\norm{\Xr-\xd})=\left(\norm{\xs - \Xr}\norm{\Xr - \xd}\right)^{\eta}$. Overall, the end-to-end path-loss for all aforementioned cases is proportional to the \emph{product of the distances between TX-RIS and RIS-RX}, although being with different exact path-loss expressions.  Our key analytical results below will hold for all path-loss functions resulting in product-scaling for the end-to-end $\snr$ achieved via an intermediate RIS. For performance evaluation, %except for the evaluation of performance analysis where
we will consider a specific product-scaling path-loss model.

%in different cases are different.Please refer to \cite{9154326,9206044} for more details.

\emph{2) Sum-Scaling path-loss models:} If the geometric size of the RIS is large enough as compared with the wavelength $\lambda_s$ and the transmission distances $\norm{\xs - \Xr}$ and $\norm{\Xr-\xd}$, the end-to-end received power scales according to  $(\mu k\norm{\xs - \Xr} + \nu k\norm{\Xr-\xd})$, where $k=\frac{2\pi}{\lambda_s}$ and the coefficients $\mu$ and $\nu$ depend on the angles of incidence and reflection of the radio waves \cite{9119122,9154326}. Further, \cite{9206044} also empirically validates that the path-loss function $G$, {when the TX and RX both or only one of them are in the near-field of RIS}\footnote{{For example, a RIS prototype in \cite{9206044}, whose size is $0.34\,\m\times0.5\,\m$ and whose carrier frequency is 10.5 GHz, at the TX-RIS distance of 0.5\,m and the RIS-RX distance of 1\,m is shown to operate as the sum-scaling path-loss model.}}, is approximately proportional to $(\norm{\xs - \Xr}+\norm{\Xr-\xd})^{2}$. $G$ in this particular case depends on antenna gains, wavelength, and the amplitude value of RIS elements as well. Also, for the exponential-law path-loss model, which we refer to as exp-law for brevity for the rest of the paper, signal power decays over a link as $\Gexp(d)=\exp(\alpha d^\beta)$, where $\alpha>0$ and $\beta>0$ are tunable parameters \cite{8122033}. We note that the exp-law model is suitable for modeling short-range communication, e.g., indoor communication\footnote{{For indoor environments, mmWave transmissions would be more appropriate due to having shorter transmission distances. Hence, the RIS size can be considered as large when compared to the transmission wavelength for indoor environments.}}, which is one of the important scenarios that RISs can be deployed to assist communications. For $\beta=1$\footnote{$\beta=1$ suits for indoor communications when the number of obstacles scales linearly with the distance of the link between $\xs$ and $\Xr$ and the one between $\Xr$ and $\xd$.}, the exp-law path-loss model can be expressed as $\Gexp(\norm{\xs - \Xr},\norm{\Xr-\xd})=\exp\paren{\alpha\paren{\norm{\xs - \Xr}+\norm{\Xr - \xd}}}$. In the cases mentioned above, we note that the path-loss function $G$ scales with \emph{the sum of the distance between TX-RIS and RIS-RX} and our results below hold correct for such cases.

\Fr{Due to the random locations of RISs in a PPP, the TX-RIS and RIS-RX distances are random.
Thus, RISs in a PPP may be in the regime where the product-scaling law holds or the
regime where the sum-scaling law holds. We note that considering mixture path-loss models (e.g., product-scaling, sum-scaling, or other models which do not directly depend on distances) for potential RISs and analytically deriving the performance metrics of corresponding optimum selection policies is important future work. For tractable performance analysis, we assume all RIS-aided links in a PPP have the product-scaling law or all RIS-aided links in a PPP have the sum-scaling law for solving the optimum RIS selection problems. We note that this assumptions is reasonable because in some cases, a single scaling law is accurate. For example, based on \cite{9433568}, for focusing lenses of RISs, a single product-scaling law is observed, and for anomalous reflectors, the sum-scaling law may be accurate up a few tens of meters. We also have conducted simulations with mixture path-loss models and compare these simulations with our analysis with a single path-loss model for all potential RISs, to show the feasibility of the assumption. The reasonableness is shown by the facts that i) the results derived by assuming all RISs have the sum-scaling law can be good approximations of the results that assumes the mixture product and sum scaling path-loss models for large RISs, and ii) the results assuming all RISs have the product-scaling path-loss model can be good approximations of the results assuming the mixture product and sum scaling path-loss models for small RISs. Our simulation results and detailed discussions are presented in Appendix \ref{Appendix: simu}. %show that the product-scaling laws generally hold for all RIS-aided links in a PPP when the RISs are small and frequencies are low and the sum-scaling laws generally hold for all RIS-aided links in a PPP when the RISs are large and frequencies are high.
Moreover, this assumption is widely used in the papers that consider random locations of RISs or users, e.g., \cite{kishk2020arxiv,9500724,hou2019arxiv,shafique2021stochastic}.}

\Fr{We can maximize the instantaneous $\snr$ by selecting the RIS that has the \emph{minimal product} of the distances of the TX-RIS and RIS-RX links over the set of RIS locations in $\Phi$, when all available RIS-aided links follow product-scaling path-loss models (i.e., only the product-based path-loss scaling law is observed). Similarly, the optimum RIS that maximizes the instantaneous $\snr$ for the sum-scaling path-loss model is the one that has the \emph{minimal sum} of these distances, when all available RIS-aided links follow sum-scaling path-loss models (i.e., only the sum-based path-loss scaling law is observed).} %the RIS selection function $\RISselectPow\paren{\vec{X}}$ over the set of RIS locations in $\Phi$,
Thus, we can formulate the optimum RIS selection problem for the product-scaling path-loss model as follows:
\begin{eqnarray}
\begin{array}{ll}
\underset{\vec{X} \in \R^2}{\mbox{minimize}} & \RISselectPow\paren{\vec{X}} \\
\mbox{subject to} & \vec{X} \in \Phi
\end{array}, \label{Eqn: RIS Selection Problem power}
\end{eqnarray}
where $\RISselectPow\paren{\vec{X}}$ is given by
%\begin{eqnarray}
$\RISselectPow\paren{\vec{X}} = \norm{\xs - \vec{X}}\times\norm{\vec{X} - \xd}$.  %\label{Eqn: RIS Selection Function}
%\end{eqnarray}
With the aim of maximizing the instantaneous $\snr$ for the sum-scaling path-loss model, the optimum RIS selection problem can be formulated as:
\begin{eqnarray}
\begin{array}{ll}
\underset{\vec{X} \in \R^2}{\mbox{minimize}} & \RISselectExp\paren{\vec{X}} \\
\mbox{subject to} & \vec{X} \in \Phi
\end{array}, \label{Eqn: RIS Selection Problem}
\end{eqnarray}
where $\RISselectExp\paren{\vec{X}}$ is given by
%\begin{eqnarray}
$\RISselectExp\paren{\vec{X}} = \norm{\xs - \vec{X}}+\norm{\vec{X} - \xd}$. %, \label{Eqn: RIS Selection-exp}
%\end{eqnarray}
%We start with the goal of maximizing $\snrExp\paren{\pol}$ in \eqref{SNR-exp} by optimizing $\pol$.
%In this paper, we assume that random fading coefficients $\alpha_{i,k}$ and $\beta_{i,k}$ change at a much faster time-scale than the RIS node locations. Hence, our selection criterion will be based only on RIS locations for selecting the RIS node, which is also embodied in Definition \ref{Def: Selection Policy}.
%Since we have $\snrExp\paren{\pol}\sim{\exp\paren{\alpha\paren{\norm{\xs - \Xr}+\norm{\Xr - \xd}}}}^{-1}$, where $\vec{x} \in \phi$, we minimize the RIS selection function $\RISselectExp\paren{\vec{x}}$ given by
%\begin{eqnarray}
%\RISselectExp\paren{\vec{x}} = \norm{\xs - \vec{x}}+\norm{\vec{x} - \xd}, \label{Eqn: RIS Selection-exp}
%\end{eqnarray}
%over the set of RIS locations in $\phi$ to maximize $\snrExp\paren{\pol}$.
The corresponding optimum RIS selection policies for the product-scaling and sum-scaling path-loss models are formally defined as follows:
%The optimum RIS selection policy can then be formally defined as follows:
\begin{selection}\label{opt-RIS-power}
The optimum RIS selection policy for the product-scaling path-loss law, denoted by $\polOptPow$, is the one solving \eqref{Eqn: RIS Selection Problem power} for all realizations of $\Phi$ in $\Sigma$. The optimum RIS location maximizing the instantaneous $\snr$ for the product-scaling path-loss law, $\XoptPow$, is
%\begin{eqnarray}
$\XoptPow= \argmin_{\vec{X} \in \Phi} \RISselectPow\paren{\vec{X}}$, which is unique with probability one.
%\end{eqnarray}
\end{selection}

\begin{selection}\label{opt-RIS-exp}
The optimum RIS selection policy under the sum-scaling path-loss law, denoted by $\polOptExp$, is the one solving \eqref{Eqn: RIS Selection Problem} for all realizations of $\Phi$ in $\Sigma$. The optimum RIS location to maximize the instantaneous $\snr$ for the sum-scaling path-loss law, $\XoptExp$, is
%\begin{eqnarray}
$\XoptExp= \argmin_{\vec{X} \in \Phi} \RISselectExp\paren{\vec{X}}$, which is also unique with probability one.
%\end{eqnarray}
\end{selection}

\subsection{Performance Metrics}

In this paper, we aim to characterize the performance metrics associated with $\polOptPow$ and $\polOptExp$. To this end, we first define the performance metrics of a given RIS selection policy $\pol$ in this subsection. We will use the averaged $\snr$ to determine outage probability and data rate over the fading process to characterize the data performance of an RIS selection policy
$\pol$\footnote{These are relevant metrics when the permissible decoding delay is large enough to average over the fading process.}.  We use $\ESZ{\snr_{\rm inst}\paren{\pol,\Phi,Z}}$ and $\ESZ{\log_2\paren{1 + \snr_{\rm inst}\paren{\pol,\Phi,Z}}}$ as the $\snr$ and data rate averaged over the fading process, respectively. %, where $\ESZ{\cdot}$ is the expectation over RV $Z$.
For compactness, we define $\ESZ{\snr_{\rm inst}\paren{\pol,\Phi,Z}}\triangleq\snr\paren{\pol,\Phi}$ and $\ESZ{\log_2\paren{1 + \snr_{\rm inst}\paren{\pol,\Phi,Z}}}\triangleq
{\sf R}\paren{\pol,\Phi}$ in the rest of the paper. We note that $\snr\paren{\pol,\Phi}$ and ${\sf R}\paren{\pol,\Phi}$ are still random quantities since they depend on random RIS locations. Using $\snr\paren{\pol,\Phi}$, we define the outage probability %for delay-sensitive traffic
as follows:
\begin{definition} \label{Def: Outage Probability}
For a target $\snr$ $\rho$, the $\snr$-outage probability $\Pout\paren{\pol}$ achieved by an RIS selection policy $\pol$ is given by
\begin{eqnarray}
\Pout\paren{\pol} = \PR{\snr\paren{\pol,\Phi} \leq \rho}.  \label{Eqn: Outage Probability}
\end{eqnarray}
\end{definition}

Using ${\sf R}\paren{\pol,\Phi}$, we define the average rate %for delay-sensitive traffic
as follows:
\begin{definition} \label{Def: rate}
The average rate achieved by an RIS selection policy $\pol$ is given by
\begin{equation}\label{Eqn: rate}
\begin{split}
\Rave\paren{\pol} & = \EW_\Phi\sqparen{{\sf R}\paren{\pol,\Phi}}.%\nonumber \\
%& = \int_{}^{}\ESZ{\log_2\paren{1 + \bar\gamma y Z^2}} f_Y(y) dy
\end{split}
\end{equation}
%where we emphasize that $\EW_\Phi\sqparen{\cdot}$ is the expectation over the spatial point process $\Phi$.
\end{definition}

In the next sections, we will evaluate $\Pout$ and $\Rave$ for the optimum RIS selection policies $\polOptPow$ and $\polOptExp$. This is a challenging problem since we need to derive the distribution of optimum RIS distance functions $\RISselectPow\paren{\XoptPow}$ and $\RISselectExp\paren{\XoptExp}$ over the random spatial point process $\Phi$ and the averaged performance metrics over the random fading process.
%the channel quality indicators are random variables obtained by minimising a function over a Poisson field.

%We derive $\Pout\paren{\pol}$ and $\Rave\paren{\pol}$ with the different path-loss models in the following sections.

\vspace{-3mm}
\section{Optimum RIS Distance Distribution and Performance Analysis}\label{sec:power}

In this section, we first obtain the optimum RIS distance distributions under the product-scaling and sum-scaling path-loss models. %The obtained distribution can be generally applied in any context where a node that has the minimal product or sum of the distances of the TX-RIS and RIS-RX links among all randomly distributed nodes is selected.
We second derive the averaged performance metrics over the fading process for general path-loss functions $G(\norm{\xs - \Xr},\norm{\Xr-\xd})$.

Using the optimum RIS distance distribution and the averaged performance metrics, we will evaluate the outage probability and average rate for the given optimum RIS selection policy under specific path-loss models. We note that there are multiple potential models for product-scaling and sum-scaling path-loss models, as discussed in Sec. \ref{subsec:RIS-selection}. For performance analysis, we will consider classic power-law path-loss and exp-law path-loss models. %The approach used in evaluating the performance metrics under power-law path-loss and exp-law path-loss models can also be used to evaluate the performance metrics under other specific path-loss models.
It is important to note that the analytical results we obtain for the distributions of $\RISselectPow\paren{\XoptPow}$ and  $\RISselectExp\paren{\XoptExp}$ can be used to obtain the performance metrics for general path-loss models obeying the product-scaling and sum-scaling property.

\vspace{-3mm}
\subsection{Distance Distribution for Optimum RIS Selection}\label{subsec:dis}
\begin{figure}
	\centering
	\subfloat{
		\label{fig:cdf-product}
		\includegraphics[width=0.4\textwidth]{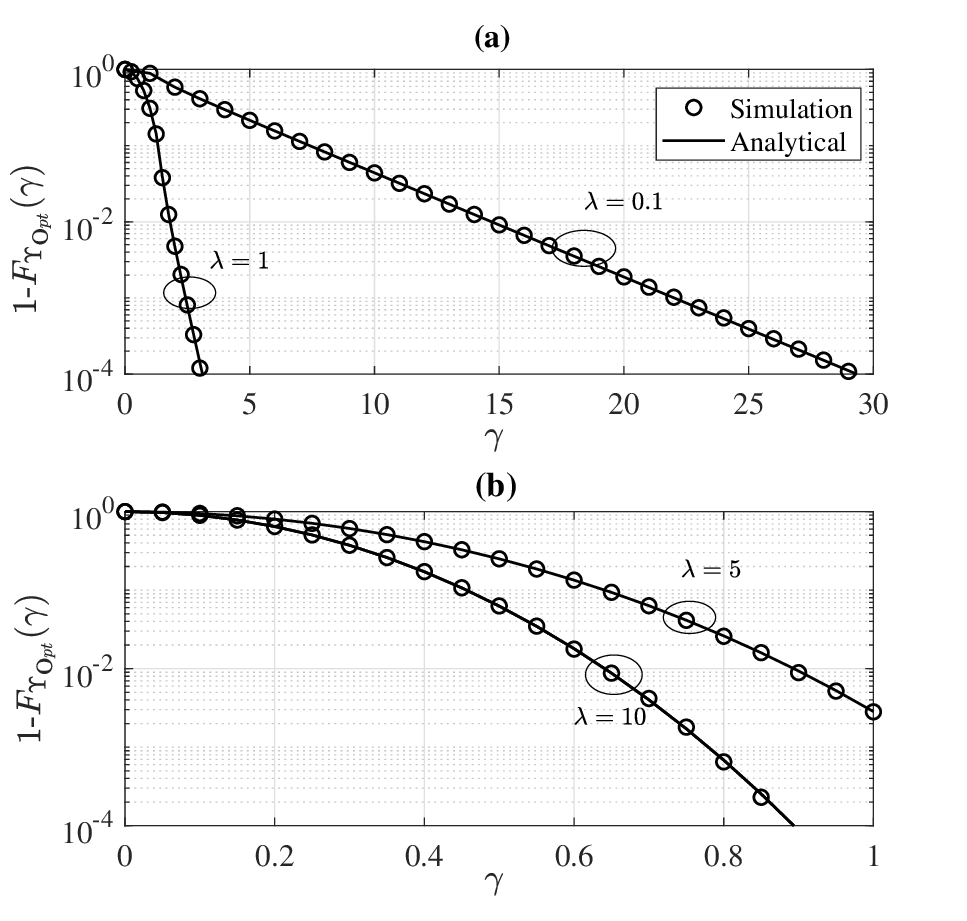}}
        	\subfloat{
		\label{fig:cdf-sum}
		\includegraphics[width=0.4\textwidth]{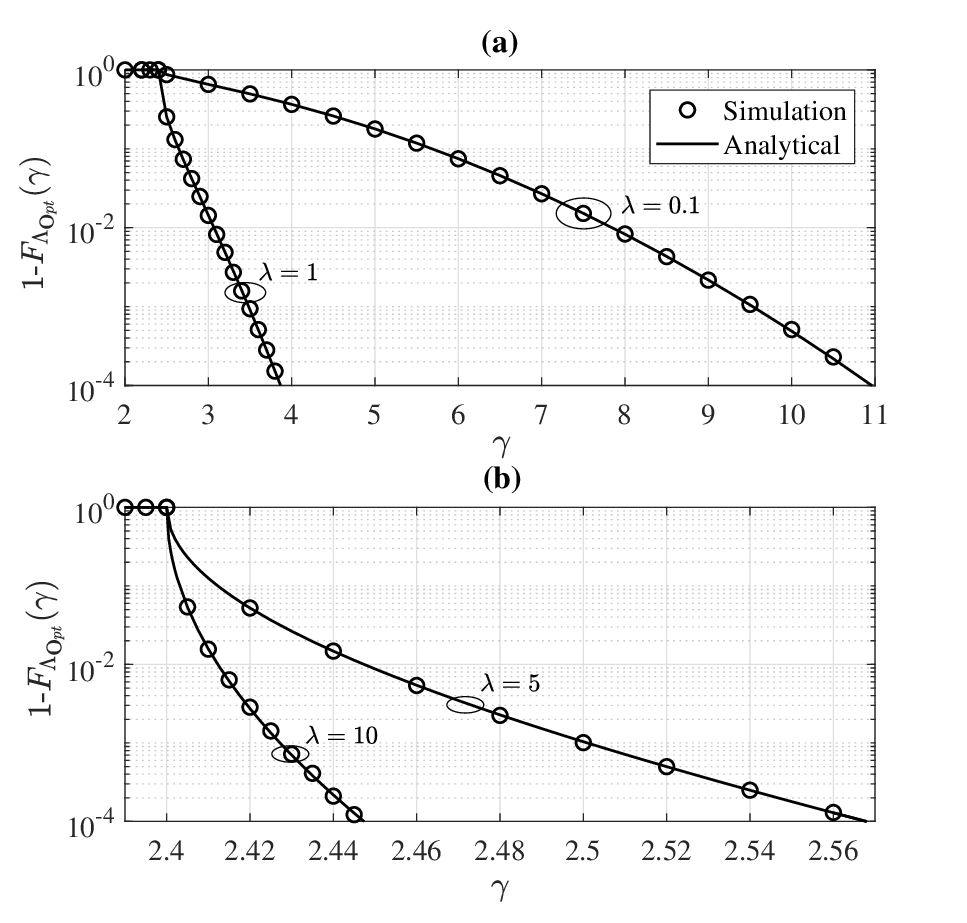}}
\vspace{-3mm}
		\caption{{The complementary CDF of $\Upsilon_{\rm opt}$ and $\Lambda_{\rm opt}$ for different RIS densities $\lambda=0.1, 1, 5, 10$ when $d=1.2$, for validating \eqref{cdf-product} and \eqref{cdf-d-minSum}, respectively.}}
	\label{Fig-cdf}\vspace{-8mm}
\end{figure}

For the sake of simplicity, we define
%\begin{eqnarray}\label{def-Upsilon}
$\Upsilon_{\rm opt} \triangleq \RISselectPow\paren{\XoptPow}$
%\end{eqnarray}
and
%To derive $\PoutExp\paren{\polOptExp}$ and $\RaveExp\paren{\polOptExp}$ for $\polOptExp$ under the exp-law path-loss model, we first define
%\begin{eqnarray}\label{def-Lambda}
$\Lambda_{\rm opt} \triangleq \RISselectExp\paren{\XoptExp}$.
%\end{eqnarray}
We now derive the distribution functions for $\Upsilon_{\rm opt}$ and $\Lambda_{\rm opt}$ which are key to characterize the performance of the RIS selection policies $\polOptPow$ and $\polOptExp$, respectively. We will consider $\xs = \paren{-d, 0}^\top$ and $\xd = \paren{d, 0}^\top$ without loss of generality due to the stationary nature of HPPPs \cite{Kingman93}. %Using $\Upsilon_{\rm opt}$, the instantaneous $\snr$ for the optimum RIS selection policy $\polOptPow$ is given by
%\begin{eqnarray}\label{SNR-power-opt}
%\snrPow\paren{\polOptPow} = \frac{\Bar\gamma Z^2}{\Upsilon_{\rm opt}^{\eta}}.
%\end{eqnarray}
%derive the analytical expressions for $\PoutPow\paren{\polOptPow}$ and $\RavePow\paren{\polOptPow}$ since $\PoutPow\paren{\polOptPow}$ and $\RavePow\paren{\polOptPow}$ rely on $\snrPow\paren{\polOptPow}$ based on \eqref{Eqn: Outage Probability} and \eqref{Eqn: rate}.
In the following theorems, we provide the distribution of $\Upsilon_{\rm opt}$ and $\Lambda_{\rm opt}$. We also numerically verify these distribution in Fig. \ref{Fig-cdf}.
\begin{theorem} \label{Theorem: Optimal dis distribution}
The CDF of $\Upsilon_{\rm opt}$ is given by
\begin{equation}\label{cdf-product}
F_{\Upsilon_{\rm opt}}\paren{\gamma}
=\left\{
\begin{array}{ll}
1-\exp\paren{\frac{-2\lambda}{d^2}\left(d^4 E(\frac{\gamma^2}{d^4})+(\gamma^2-d^4)K(\frac{\gamma^2}{d^4})\right)}& \mbox{ if }\gamma < d^2 \\
1-\exp\paren{-2\lambda\gamma E(\frac{d^4}{\gamma^2})
}  & \mbox{ if } \gamma \geq d^2
\end{array},\right.
\end{equation}
where %$\lambda$ is the density of HPPP distributed RISs, we assume $\xs=(-d,0)$ and $\xd=(d,0)$,
$E(\cdot)$ is the complete elliptic integral of the second kind and $K(\cdot)$ is the complete elliptic integral of the first kind \cite{Abramowitz1974}. The PDF of $\Upsilon_{\rm opt}$ is given by
\begin{equation}\label{pdf-product}
f_{\Upsilon_{\rm opt}}\paren{\gamma}
=\left\{
\begin{array}{ll}
\frac{2}{d^2\gamma\lambda} \exp\paren{-\frac{2\lambda}{d^2}\paren{d^4 E\paren{\frac{\gamma^2}{d^4}}+\paren{\gamma^2-d^4} K\paren{\frac{\gamma^2}{d^4}}  }} K\paren{\frac{\gamma^2}{d^4}} & \mbox{ if }\gamma < d^2 \\
2\lambda\exp\paren{-2\gamma \lambda E\paren{\frac{d^4}{\gamma^2}}}K\paren{\frac{d^4}{\gamma^2}}  & \mbox{ if } \gamma \geq d^2
\end{array}.\right.
\end{equation}
\end{theorem}
\begin{IEEEproof}
%The proof is omitted due to space limitations.
See Appendix \ref{Appendix: optimal dis distribution Proof}.
\end{IEEEproof}

%\begin{figure}[!htb]
%\centering
% \includegraphics[width=0.7\textwidth]{CCDF-product}
% \caption{The complementary CDF of $\Upsilon_{\rm opt}$ for different RIS densities $\lambda=0.1, 1, 5, 10$ when $d=1.2$, for validating \eqref{cdf-product}. \com{we may remove this to save space.}}\label{fig:cdf-product}
% \end{figure}

%\subsection{Distance Distribution for Optimum RIS Selection}
%Due to the stationarity property of HPPPs \cite{Kingman93}, we will assume that $\xs = \paren{-d, 0}^\top$ and $\xd = \paren{d, 0}^\top$ without loss of generality.
%For the sake of simplicity,

%In the following theorem, we present the analytical expressions for the CDF and PDF of $\Lambda_{\rm opt}$.
\begin{theorem} \label{theorem: minSum dis distribution}
The CDF of $\Lambda_{\rm opt}$ is given by
\begin{equation}\label{cdf-d-minSum}
\begin{split}
\hspace{-2mm}F_{\Lambda_{\rm opt}}\paren{\gamma} = \left\{
\begin{array}{ll}
\hspace{-2mm}0 & %\mbox{ if }
\gamma < 2d\\
\hspace{-2mm}
1-\exp\paren{-\frac{\lambda \pi\gamma\sqrt{\gamma^2-4d^2}}{4}} & %\mbox{ if }
\gamma \geq 2d
\end{array}\right..
\end{split}
\end{equation}
The PDF of $\Lambda_{\rm opt}$ is given by
\begin{equation}\label{pdf-d-minSum}
\begin{split}
\hspace{-2mm}f_{\Lambda_{\rm opt}}\paren{\gamma} = \left\{
\begin{array}{ll}
\hspace{-2mm}0 & %\mbox{ if }
\gamma < 2d\\
\hspace{-2mm}
 \frac{\pi  \lambda  \left(\gamma^2-2 d^2\right)}{2 \sqrt{\gamma^2-4 d^2}} \exp\paren{-\frac{\lambda\pi    \gamma \sqrt{\gamma^2-4 d^2}}{4} }& %\mbox{ if }
\gamma \geq 2d
\end{array}\right..
\end{split}
\end{equation}
% where
% \begin{eqnarray}\label{z-eq2}
% \mathcal{Z}(d,\gamma)=\sqrt{-4d^2+\gamma^2}.
% \end{eqnarray}
\end{theorem}
\begin{IEEEproof}
%The proof is omitted due to space limitations.
See Appendix \ref{Appendix: minSum dis distribution Proof}.
\end{IEEEproof}

% \begin{figure}[!htb]
% \centering
% \includegraphics[width=0.7\textwidth]{CCDF-sum}
% \caption{The complementary CDF of $\Lambda_{\rm opt}$ for different densities $\lambda=0.1, 1, 5, 10$ when $d=1.2$, for validating \eqref{cdf-d-minSum}. \com{may remove to save space.}}\label{fig:cdf-sum}
% \end{figure}
\vspace{-3mm}
\subsection{Averaged Performance Metrics over Fading Channel}\label{subsec:aver}
Based on \eqref{Eqn: Outage Probability} and \eqref{Eqn: rate}, we recall that $\Pout\paren{\pol}$ and $\Rave\paren{\pol}$ relate to $\snr\paren{\pol,\Phi}$ and ${\sf R}\paren{\pol,\Phi}$, respectively. To facilitate the derivation of $\Pout$ and $\Rave$ for the optimum RIS selection policies $\polOptPow$ and $\polOptExp$, we derive $\snr\paren{\pol,\Phi}$ and ${\sf R}\paren{\pol,\Phi}$ in this subsection. We derive $\snr\paren{\pol,\Phi}$ as
\begin{eqnarray}\label{snr-fading}
\snr\paren{\pol,\Phi}&=&\ESZ{\frac{\bar\gamma Z^2}{G(\norm{\xs - \Xr},\norm{\Xr-\xd})}}=\frac{\bar\gamma  \ESZ{Z^2} }{G(\norm{\xs - \Xr},\norm{\Xr-\xd})},
\end{eqnarray}
where $\ESZ{Z^2}$ is given by
\begin{eqnarray}\label{expectation-z}
\ESZ{Z^2} &=& N+N(N-1)\frac{\pi^2}{16}.
\end{eqnarray}
%\begin{eqnarray}\label{expectation-z}
%\ESZ{Z^2} &=& \sum_{n=1}^{N}\mathsf{E}\left[\alpha_{i,n}^2 \right]\mathsf{E}\left[\beta_{i,n}^2 \right]+ \sum_{n=1}^{N}\sum_{m=1,m\neq n}^{N}\mathsf{E}\left[\alpha_{i,n} \right]\mathsf{E}\left[\alpha_{i,m} \right]
%\mathsf{E}\left[\beta_{i,n} \right]\mathsf{E}\left[\beta_{i,m} \right]\nonumber \\
%&\overset{(a)}{=}& N\mathsf{E}\left[\alpha_{i,1}^2 \right]^2+N(N-1)\mathsf{E}\left[\alpha_{i,1}\right]^4
%% \overset{(b)}{=} N+N(N-1)\mathsf{E}\left[\alpha_{i,1}\right]^4\nonumber \\
%% &\overset{(c)}{=}& N+N(N-1)\frac{\pi^2}{16},
%\overset{(b)}{=} N+N(N-1)\frac{\pi^2}{16},
%\end{eqnarray}
%where equality (a) exploits the fact that $\alpha_{i,n}$ and $\beta_{i,n}$ are i.i.d. RVs. Equality (b) is due to the fact $h_{i,1}, \sim \mathcal{CN}(0,1)$ and $\mathsf{E}\left[\alpha_{i,1}^2 \right]=1$, and Rayleigh distributed with the scale parameter $\sigma=\frac{1}{\sqrt{2}}$ and $\mathsf{E}\left[\alpha_{i,1}\right]=\sigma\sqrt{\frac{\pi}{2}}=\sqrt{\frac{\pi}{4}}$.
Applying \eqref{expectation-z} to \eqref{snr-fading}, we have
\begin{eqnarray}\label{snr-fading1}
\snr\paren{\pol,\Phi}
&=&\frac{\bar\gamma N(16+(N-1)\pi^2)}{16 G(\norm{\xs - \Xr},\norm{\Xr-\xd})}.
\end{eqnarray}

%To this end, we study the distribution of $Z$. Since RV $Z\geq 0$ is a $Z\geq 0$ is a sum of the product of two i.i.d. Rayleigh RVs with parameter $\sigma=1$, its exact distribution is difficult to determine for $N>1$. We thus approximate it by a gamma RV with the shape parameter given by $k = \frac{N\pi^2}{16-\pi^2}$ and the scale parameter given by $\theta=\frac{16-\pi^2}{4\pi}$~\cite{atapattu2020reconfigurable}. It is important to mention that, even at large $N$, this gamma approximation appears to be more accurate than the central-limit theorem (CLT) approach. Using the PDF of RV $Z$ which $Z$ which is accurately approximated with a gamma RV, %given by
%%\begin{eqnarray}\label{expectation}
%%f_{Z}(x) = \frac{x^{k-1}\exp\paren{-\frac{x}{\theta}}}{\theta^{k}\Gamma\paren{k}},
%%\end{eqnarray}
%we obtain the $\snr$ averaged over the fading process, $\snr\paren{\pol,\Phi}$, as
%\begin{eqnarray}\label{snr-fading}
%\snr\paren{\pol,\Phi}&=&\ESZ{\frac{\bar\gamma Z^2}{G(\norm{\xs - \Xr})G(\norm{\Xr-\xd})}}\nonumber \\
%&=& \int_{-\infty}^{\infty}\frac{\Bar\gamma x^2}{G(\norm{\xs - \Xr})G(\norm{\Xr-\xd}) } f_{Z}(x)dx\nonumber \\
%&\doteq& \int_{-\infty}^{\infty}\frac{\Bar\gamma x^2}{G(\norm{\xs - \Xr})G(\norm{\Xr-\xd}) } \frac{x^{k-1}\exp\paren{-\frac{x}{\theta}}}{\theta^{k}\Gamma\paren{k}}dx\nonumber \\
%&\doteq& \frac{\Bar\gamma \theta^2 \Gamma(2+k)}{G(\norm{\xs - \Xr})G(\norm{\Xr-\xd})  \Gamma(k)},
%\end{eqnarray}
%where we emphasize that in this paper we use $\doteq$ to mean that the equality is valid when the gamma approximation is applied.

We next derive ${\sf R}\paren{\pol,\Phi}$ as
\begin{align}\label{rate-fading}
\hspace{-5mm}{\sf R}\paren{\pol,\Phi}={\sf R}\paren{\pol,Y}=
\ESZ{\log_2\paren{1 + \bar\gamma Y Z^2}}
= \int_{0}^{\infty} \log_2\paren{1 + \bar\gamma Y z^2}f_Z(z) dz,
\end{align}
where $Y=\frac{1}{G(\norm{\xs - \Xr},\norm{\Xr-\xd})}$ and $Y$ is a statistic summarising the overall effect of the point process $\Phi$ on the data rate. Based on \eqref{rate-fading}, we study the distribution of $Z$. Since $Z$ is a sum of the product of two i.i.d. Rayleigh RVs, %with parameter $\sigma=1$
its exact distribution is difficult to determine for $N>1$. However, we can still approximate its distribution by using a gamma distribution with the shape parameter given by $k = \frac{N\pi^2}{16-\pi^2}$ and the scale parameter given by $\theta=\frac{16-\pi^2}{4\pi}$ very tightly, as established in~\cite{9095301,atapattu2020reconfigurable}. %It is important to mention that, even at large $N$, this gamma approximation appears to be more accurate than the central-limit theorem (CLT) approach.
Using suggested gamma distribution approximation and the result in \cite[Eq.~(20)]{atapattu2020reconfigurable}, we can express %${\sf R}\paren{\pol,\Phi}$ in
\eqref{rate-fading} as
%using the gamma approximation, which is given by %For simplicity, we first rewrite $\snr\paren{\pol,\Phi,Z}$ in \eqref{SNR} as $\snr\paren{\pol,\Phi,Z}=\bar\gamma y Z^2$, where $y$ is the realization of RV $Y$ and $Y=\frac{1}{G(\norm{\xs - \Xr})G(\norm{\Xr-\xd})}$. %We then have %the data rate averaged over the fading process given by
% To derive , $\ESZ{\log_2\paren{1 + \snr\paren{\pol,\Phi,Z}}}$, we first write
% \begin{align}\label{rate1}
%$\ESZ{\log_2\paren{1 + \snr\paren{\pol,\Phi,Z}}} = \ESZ{\log_2\paren{1 + \bar\gamma y Z^2}}$.
% \end{align}
%We then derive ${\sf R}_{Y}(y)=\ESZ{\log_2\paren{1 + \bar\gamma y Z^2}}$ as
%Using this relation, we have
\begin{align}\label{e_tp_L2}
{\sf R}\paren{\pol,Y}
= \,&\int_{0}^{\infty} \log_2\paren{1 + \bar\gamma Y z^2}f_Z(z) dz
\doteq \,\int_{0}^{\infty} \log_2\paren{1 + \bar\gamma Y z^2}\frac{z^{k-1}\exp\paren{-\frac{z}{\theta}}}{\theta^{k}\Gamma\paren{k}} dz \nonumber \\
\doteq \,& \frac{1}{\log(2)} \Biggl[ 2 \log (\theta )+\log (\bar\gamma\,Y )+2 \psi ^{(0)}(k) + \frac{\, _2{\sf F}_3\left(1,1;2,\frac{3}{2}-\frac{k}{2},2-\frac{k}{2};-\frac{1}{4 \theta ^2 \bar\gamma\,Y }\right)}{\theta ^2 \bar\gamma\,Y  \left(k^2-3 k+2\right)} \nonumber \\
&
+ \frac{\pi   (\bar\gamma\,Y) ^{-\frac{k}{2}}}{\theta ^{k}\Gamma (k+1)}
\Biggl(
\frac{ _1{\sf F}_2\left(\frac{k}{2};\frac{1}{2},\frac{k}{2}+1;-\frac{1}{4 \theta ^2 \bar\gamma\,Y }\right)}{ \left(\csc \left(\frac{\pi  k}{2}\right)\right)^{-1}}
-\frac{ _1{\sf F}_2\Biggl(\frac{k}{2}+\frac{1}{2};\frac{3}{2},\frac{k}{2}+\frac{3}{2};-\frac{1}{4 \theta ^2 \bar\gamma\,Y }\Biggr)}{\sqrt{\bar\gamma\,Y } \theta(1 +  k) \left(k \sec \left(\frac{\pi  k}{2}\right)\right)^{-1}}
\Biggr)
\Biggr]
\end{align}
where we use $\doteq$ to indicate that the equality is valid when the gamma approximation is applied, %In \eqref{e_tp_L2}, $Y=\frac{1}{G(\norm{\xs - \Xr})G(\norm{\Xr-\xd})}$ and $Y$ is a single statistic summarising the overall effect of point process $\Phi$ on the data rate.
%The third equality in \eqref{e_tp_L2} comes from the results of \cite[eq.~(22)]{atapattu2020reconfigurable},
$\log(\cdot)$ denotes natural logarithm, $_p{\sf F}_q\left(\cdot;\cdot;\cdot\right)$ denotes the generalized hypergeometric functions~\cite{gradshteyn2007}, and $\psi^{(n)}(z)$ denotes the $n$th derivative of the digamma function \cite{gradshteyn2007}. Numerical results in Section \ref{sec: numerical} will confirm the accuracy of this approximation used in \eqref{e_tp_L2}.

We note that the calculation of \eqref{e_tp_L2} requires high computational complexity. To ease the complexity, using Jensen's inequality, we can also derive an upper bound on ${\sf R}\paren{\pol,\Phi}$ as
\begin{align}\label{rate-fading,up}
{\sf R}\paren{\pol,\Phi}=\ESZ{\log_2\paren{1 + \snr_{\rm inst}\paren{\pol,\Phi,Z}}}\leq \log_2\paren{1 + \ESZ{\snr_{\rm inst}\paren{\pol,\Phi,Z}}}.
\end{align}

By applying \eqref{snr-fading1} to \eqref{rate-fading,up}, we rewrite \eqref{rate-fading,up} as
\begin{align}\label{rate-fading,up1}
{\sf R}\paren{\pol,Y}\leq \log_2\paren{1 + \frac{\bar\gamma N Y(16+(N-1)\pi^2)}{16}},
\end{align}
where we write ${\sf \tilde{R}}\paren{\pol,Y}=\log_2\paren{1 + \frac{\bar\gamma N Y(16+(N-1)\pi^2)}{16}}$ in the rest of the paper. We will use ${\sf \tilde{R}}\paren{\pol,Y}$ to calculate the upper bounds on the average rate in Section \ref{subsub:averRate}.

\vspace{-3mm}
\subsection{Performance Analysis}
Based on subsections \ref{subsec:dis} and \ref{subsec:aver}, we evaluate the outage probability and average rate achieved by the optimum RIS selection policies $\polOptPow$ and $\polOptExp$. To illustrate specific applications of our results, we will consider classical power-law and exp-law path-loss models for the performance evaluation in this subsection.

\subsubsection{Outage Probability}
%We first derive the $\snr$ averaged over the fading process $\ESZ{\snrPow\paren{\polOptPow}}$ to derive outage probability. Combining \eqref{path-loss-power}, \eqref{Eqn: RIS Selection Function}, and \eqref{def-Upsilon}, we have
%\begin{eqnarray}\label{g1g2-power}
%\Upsilon_{\rm opt}^\eta = \Gpow_1(\norm{\xs - \xoptPow})\Gpow_2(\norm{\xoptPow-\xd})
%\end{eqnarray}
%
%Substituting \eqref{g1g2-power} into \eqref{snr-fading}, we derive $\ESZ{\snrPow\paren{\polOptPow}}$ as
%\begin{eqnarray}\label{expect-SNR-power}
%\ESZ{\snrPow\paren{\polOptPow}} &\doteq& \frac{\Bar\gamma\theta^2 \Gamma(2+k)}{\Upsilon_{\rm opt}^\eta \Gamma(k)}.
%\end{eqnarray}
%%\com{integral limit may be from $0$?}
%%Due to the fact that $\ESZ{\snrPow}$ is a decreasing function with respect to $\Upsilon_{\pol}$,
%%We substitute \eqref{expectation} into \eqref{outage},
%With the aid of \eqref{Eqn: Outage Probability} and \eqref{expect-SNR-power},
We denote $\PoutPow\paren{\polOptPow}$ and $\PoutExp(\polOptExp)$ as the outage probabilities achieved under the power-law and exp-law path-loss models by the optimum RIS selection policies, respectively. Using \eqref{Eqn: Outage Probability} and applying\footnote{{We assume that the path-loss exponents of the TX-RIS and RIS-RX channels are the same. This assumption is reasonable when the TX-RIS and RIS-RX channels have similar propagation environments.}
} $\Gpow(\norm{\xs - \Xr},\norm{\Xr-\xd})=\left(\norm{\xs - \Xr}\norm{\Xr - \xd}\right)^{\eta}$ to \eqref{snr-fading1}, we derive $\PoutPow\paren{\polOptPow}$ as
\begin{eqnarray}\label{outage2}
\PoutPow\paren{\polOptPow} &=& \PR{\snr\paren{\polOptPow,\Phi}\leq \rho}
= \PR{\frac{\bar\gamma N(16+(N-1)\pi^2)}{16
\Upsilon_{\rm opt}^\eta}\leq \rho}\nonumber \\
%&=& \PR{\Upsilon_{\pol}>\paren{{\frac{P_t \theta^2 \Gamma(2+k)}{\rho N_0 \Gamma(k)}}}^\frac{1}{\eta}}\nonumber \\
&=& 1-F_{\Upsilon_{\rm opt}}\paren{\paren{{\frac{\Bar\gamma N(16+(N-1)\pi^2)}{16\rho }}}^\frac{1}{\eta}},
\end{eqnarray}
%\com{we can save a lot of space by writing these set of expressions in same line.}
where $F_{\Upsilon_{\rm opt}}\paren{\gamma}$ is the CDF of RV $\Upsilon_{\rm opt}$ given in \eqref{cdf-product}. Applying
$\Gexp(\norm{\xs - \Xr},\norm{\Xr-\xd})=\exp\paren{\alpha\paren{\norm{\xs - \Xr}+\norm{\Xr - \xd}}}$ to \eqref{snr-fading1}, %Since $\snr\paren{\polOptExp,\Phi}$ is a decreasing function with respect to $\Lambda_{\rm opt}$ and $\Lambda_{\rm opt}$ is a continuous RV,
we derive $\PoutExp(\polOptExp)$ as
%We substitute \eqref{expectation} into \eqref{outage}, we have
\begin{eqnarray}\label{outage2opt}
\PoutExp(\polOptExp) &=& \PR{\snr\paren{\polOptExp,\Phi}\leq \rho}
= \PR{\frac{\bar\gamma N(16+(N-1)\pi^2)}{16\exp\paren{\alpha\Lambda_{\rm opt}}}\leq \rho}\nonumber \\
&=& \PR{\Lambda_{\rm opt}>\log\paren{{\frac{\bar\gamma N(16+(N-1)\pi^2)}{16 \rho }}}\alpha^{-1}}\nonumber \\
&=& 1-F_{\Lambda_{\rm opt}}\paren{\frac{1}{\alpha}\log\paren{{\frac{\bar\gamma N(16+(N-1)\pi^2)}{16 \rho}}}}.
\end{eqnarray}
where $F_{\Lambda_{\rm opt}}\paren{\gamma}$ is the CDF of RV $\Lambda_{\rm opt}$ given in \eqref{cdf-d-minSum}.

\subsubsection{Average Rate}\label{subsub:averRate}
We denote $\RavePow\paren{\polOptPow}$ and $\RaveExp\paren{\polOptExp}$ as the data rate obtained under the power-law and exp-law path-loss models by the optimum RIS selection policies, respectively. Based on the definition of $\Rave\paren{\pol}$ in \eqref{Eqn: rate}, we evaluate $\RavePow\paren{\polOptPow}$ as
\begin{eqnarray}\label{Eqn: rate-power}
\RavePow\paren{\polOptPow} &=& \EW_\Phi\sqparen{{\sf R}\paren{\pol,Y_{\rm pow}}}
%=\sum {\sf R}\paren{\polOptPow,\phi} \Prob{\phi}\nonumber \\
= \int_{}^{} {\sf R}\paren{\pol,y} f_{Y_{\rm pow}}(y) d \,y,
%= \int_{}^{} {\sf R}_{Y_{\circ}}(y_{\circ})f_{Y_{\circ}}(y_{\circ}) d \,y_{\circ},
\end{eqnarray}
where $y$ is a realization of RV ${Y_{\rm pow}}$ and ${Y_{\rm pow}}=\frac{1}{\Gpow(\norm{\xs - \XoptPow},\norm{\XoptPow-\xd})}={\Upsilon_{\rm opt} }^{-\eta}$. %based on \eqref{g1g2-power}.
We note that ${\sf R}\paren{\pol,Y_{\rm pow}}$ is given by \eqref{e_tp_L2}. Using %$f_{Y_{\circ}}(y_{\circ})=\frac{\partial 1-F_{\Upsilon_{\rm opt}}\paren{y_{\circ}^{-\frac{1}{\eta}}} }{\partial y_{\circ}}$,
the distribution of $\Upsilon_{\rm opt}$,
we derive the PDF of ${Y_{\rm pow}}={\Upsilon_{\rm opt}}^{-\eta}$ as
%\begin{figure*}
%{\small
%\begin{equation}\label{e_pdfmaxminD}
%f_{{Y_{\circ}}}\paren{y_{\circ}}
%=\left\{
%\begin{array}{ll}
%{2\exp\paren{-2y_{\circ}^{-\frac{1}{\eta}}\lambda E(d^4y_{\circ}^{\frac{2}{\eta}})}y_{\circ}^{-1-\frac{1}{\eta}}\lambda K(d^4 y_{\circ}^{\frac{2}{\eta}})\eta^{-1}}
%& \mbox{ if } 0 < y_{\circ} \leq d^{-2\eta} \\
%{2\exp\paren{-{2\lambda d^{-2}\paren{d^4 E\paren{\frac{y_{\circ}^{\frac{-2}{\eta}}}{d^4}}+\paren{-d^4+y_{\circ}^{\frac{-2}{\eta}}}K\paren{\frac{y_{\circ}^{\frac{-2}{\eta}}}{d^4}}
%}}}y_{\circ}^\frac{-2+\eta}{\eta}\lambda K\paren{\frac{y_{\circ}^{\frac{-2}{\eta}}}{d^4}}
%d^{-2} \eta^{-1}}
%& \mbox{ if }  y_{\circ} \geq d^{-2\eta}
%\end{array}.\right.
%\end{equation}
%\hrule
%\end{figure*}
\begin{equation}\label{e_pdfmaxminD}
f_{Y_{\rm pow}}\paren{y}
=\left\{
\begin{array}{ll}
S_1\paren{y}=\frac{2\exp\paren{-2y^{-\frac{1}{\eta}}\lambda E(d^4y^{\frac{2}{\eta}})}y^{-1-\frac{1}{\eta}}\lambda K(d^4 y^{\frac{2}{\eta}})}{\eta}
& \mbox{ if } 0 < y \leq d^{-2\eta} \\
S_2\paren{y}=\frac{2\exp\paren{-\frac{2\lambda\paren{d^4 E\paren{\frac{y^{\frac{-2}{\eta}}}{d^4}}+\paren{-d^4+y^{\frac{-2}{\eta}}}K\paren{\frac{y^{\frac{-2}{\eta}}}{d^4}}
}}{d^2}}y^\frac{-2+\eta}{\eta}\lambda K\paren{\frac{y^{\frac{-2}{\eta}}}{d^4}}
}{d^2 \eta}
& \mbox{ if }  y \geq d^{-2\eta}
\end{array}.\right.
\end{equation}

There is no closed form expression for $\RavePow\paren{\polOptPow}$ but it can be evaluated numerically with the aid of \eqref{e_tp_L2}, \eqref{Eqn: rate-power}, and \eqref{e_pdfmaxminD} by calculating the integrals given below:
\begin{eqnarray}\label{Eqn: rate-power1}
\RavePow\paren{\polOptPow} &=& \int_{0}^{d^{-2\eta}} {\sf R}\paren{\pol,y} S_1\paren{y} d\,y+\int_{d^{-2\eta}}^{\infty} {\sf R}\paren{\pol,y} S_2\paren{y} d\,y.
\end{eqnarray}

%calculated as in \eqref{Eqn: rate-power}

%\subsection{Performance Analysis}
%Now we use Theorem \ref{theorem: minSum dis distribution} to evaluate the outage probability and the average rate %in \eqref{outage2} with the aid of \eqref{cdf-d-minSum} for
%of the optimum RIS selection policy $\polOptExp$.
%
%
%\subsubsection{Outage Probability}
%Employing \eqref{path-loss-exp},
%%$\Gexp_1(\norm{\xs - \Xr})=\exp\paren{\alpha\norm{\xs -\Xr}^{\beta}}$, $\Gexp_2(\norm{\Xr-\xd})=\exp\paren{\alpha\norm{\Xr -\xd}^{\beta}}$,
%\eqref{Eqn: RIS Selection-exp}, and \eqref{def-Lambda}, we have
%\begin{eqnarray}\label{g1g2-exp}
%\exp\paren{\alpha\Lambda_{\rm opt}}=\Gexp(\norm{\xs - \XoptExp})\Gexp(\norm{\XoptExp-\xd}).
%\end{eqnarray}
%
%We then apply \eqref{g1g2-exp} to \eqref{snr-fading}, $\snr\paren{\polOptExp,\Phi}$ is given by
%\begin{align}\label{expectation}
%\snr\paren{\polOptExp,\Phi} %=&\; \int_{0}^{\infty}\frac{\bar\gamma x^2}{\exp\paren{\alpha\Lambda_{\rm opt}}} f_{Z}(x)dx\nonumber \\
%%=&\; \int_{0}^{\infty}\frac{\bar\gamma x^2}{\exp\paren{\alpha\Lambda_{\rm opt}}}\frac{x^{k-1}\exp\paren{-\frac{x}{\theta}}}{\theta^{k}\Gamma\paren{k}}dx\nonumber \\
%\doteq&\; \frac{\bar\gamma \theta^2 \Gamma(2+k)}{\exp\paren{\alpha\Lambda_{\rm opt}}\Gamma(k)},
%\end{align}
%
%
%
%
%
%\subsubsection{Average Rate}
%Here, we evaluate the average rate $\RaveExp\paren{\polOptExp}$ obtained by the optimum RIS selection policy $\polOptExp$.

As \eqref{Eqn: rate-power}, we write $\RaveExp\paren{\polOptExp}$ as
\begin{eqnarray}\label{throughput given phi}
\RaveExp\paren{\polOptExp} %=\int_{}^{}\ESZ{\log_2\paren{1 + \bar\gamma y_{\diamond} Z^2}} f_{Y_{\diamond}}(y_{\diamond}) d\,y_{\diamond}
%&=& \EW_\Phi\sqparen{{\sf R}\paren{\polOptExp,\Phi}}
%=\sum {\sf R}\paren{\polOptExp,\phi} \Prob{\phi}\nonumber \\
&=& \int_{}^{} {\sf R}\paren{\pol,y} f_{Y_{\rm exp}}(y) d \,y,
\end{eqnarray}
where $y$ is a realization of $Y_{\rm exp}$ and $Y_{\rm exp}=\exp\paren{-\alpha\Lambda_{\rm opt}}$.  ${\sf R}\paren{\pol,Y_{\rm exp}}$ is given by \eqref{e_tp_L2}. We then derive the PDF of $Y_{\rm exp}$. %=\exp\paren{-\alpha\Lambda_{\rm opt}}$.
By using variable transformation and $f_{\Lambda_{\rm opt}}\paren{\gamma}$ in \eqref{pdf-d-minSum}, the PDF of $Y_{\rm exp}$ can be derived %for $y\leq e^{-2\alpha d}$
as
\begin{align}\label{e_Y_pdf}
    f_{Y_{\rm exp}}(y)=\frac{ \pi  y^{\frac{\pi  \lambda  \sqrt{\frac{\log ^2\left(\frac{1}{y}\right)}{\alpha^2}-4 d^2}}{4 \alpha}-1} \lambda  \left(\frac{\log ^2\left(\frac{1}{y}\right)}{\alpha^2}-2 d^2\right)}{2 \alpha \sqrt{\frac{\log ^2\left(\frac{1}{y}\right)}{\alpha^2}-4 d^2}},
\end{align}
for $y\leq e^{-2\alpha d}$. $f_{Y_{\rm exp}}(y)$ is zero for $y> e^{-2\alpha d}$. %For $y> e^{-2\alpha d}$, the PDF is zero.
With the aid of \eqref{e_tp_L2}, \eqref{throughput given phi} and \eqref{e_Y_pdf}, the average rate can be numerically calculated by evaluating the integral
\begin{equation}\label{throughput given phi 2}
\begin{split}
\RaveExp\paren{\polOptExp} & = \int_{0}^{e^{-2\alpha d}}{\sf R}\paren{\pol,y} f_{Y_{\rm exp}}(y) d\,y.
\end{split}
\end{equation}

\begin{remark}\label{remark:upper}
The upper bounds on $\RavePow\paren{\polOptPow}$ and $\RaveExp\paren{\polOptExp}$ via Jensen's inequality can be obtained by replacing ${\sf R}\paren{\pol,y}$ by ${\sf \tilde{R}}\paren{\pol,y}$ in \eqref{Eqn: rate-power1} and \eqref{throughput given phi 2}, where ${\sf \tilde{R}}\paren{\pol,Y}$ is given in \eqref{rate-fading,up1}.
\end{remark}

\section{Limited-Feedback RIS Selection}\label{sec:limited}
In previous sections, we consider that there exists a central entity  (i.e., a network controller) knowing the locations of all RISs to perform the selection based on this knowledge by optimising either product or sum distances without feedback. Different from previous sections, in this section, we assume the existence of extra feedback capability at RISs for distributed operation. % and the feedback capability makes the network operation distributed.
We assume that the RISs can feed back their channel quality indicators\footnote{{An RIS can know its channel quality by employing some active sensors among the passive reflective elements at the RISs as proposed in \cite{9370097} or deploying anchor nodes near the RISs as proposed in \cite{9603291}.}}, {\em whilst} still functioning as the nearly passive elements for communications after the feedback phase.

In this section, we propose a limited-feedback RIS selection policy that selects the best RIS from a limited number of RISs feeding back. We will first derive the distribution of the number of RISs feeding back under the product-scaling and sum-scaling path-loss models. We will then evaluate the average rate and outage probability attained by the proposed limited-feedback RIS selection policy under the specific path-loss models. To illustrate specific applications of our results, we will consider classic power-law and exp-law path-loss models for the performance evaluation in this section. We note that the derivation method used to obtain the performance metrics under power-law and exp-law path-loss models can be also used for other path-loss models.

\subsection{Limited-Feedback Strategy}\label{subsec: feedback}
%In our discussion in the previous sections, we observe that a central entity knowing location information from all RISs picks the best one. In this section, we assume that the RISs can feed back their channel quality indicators and the feedback capability makes the network operation distributed. %We note that the feedback task here may not be practical. %which change at a much slower time-scale than fading.

\begin{figure}[!t]
\centering
\includegraphics[width=0.45 \textwidth]{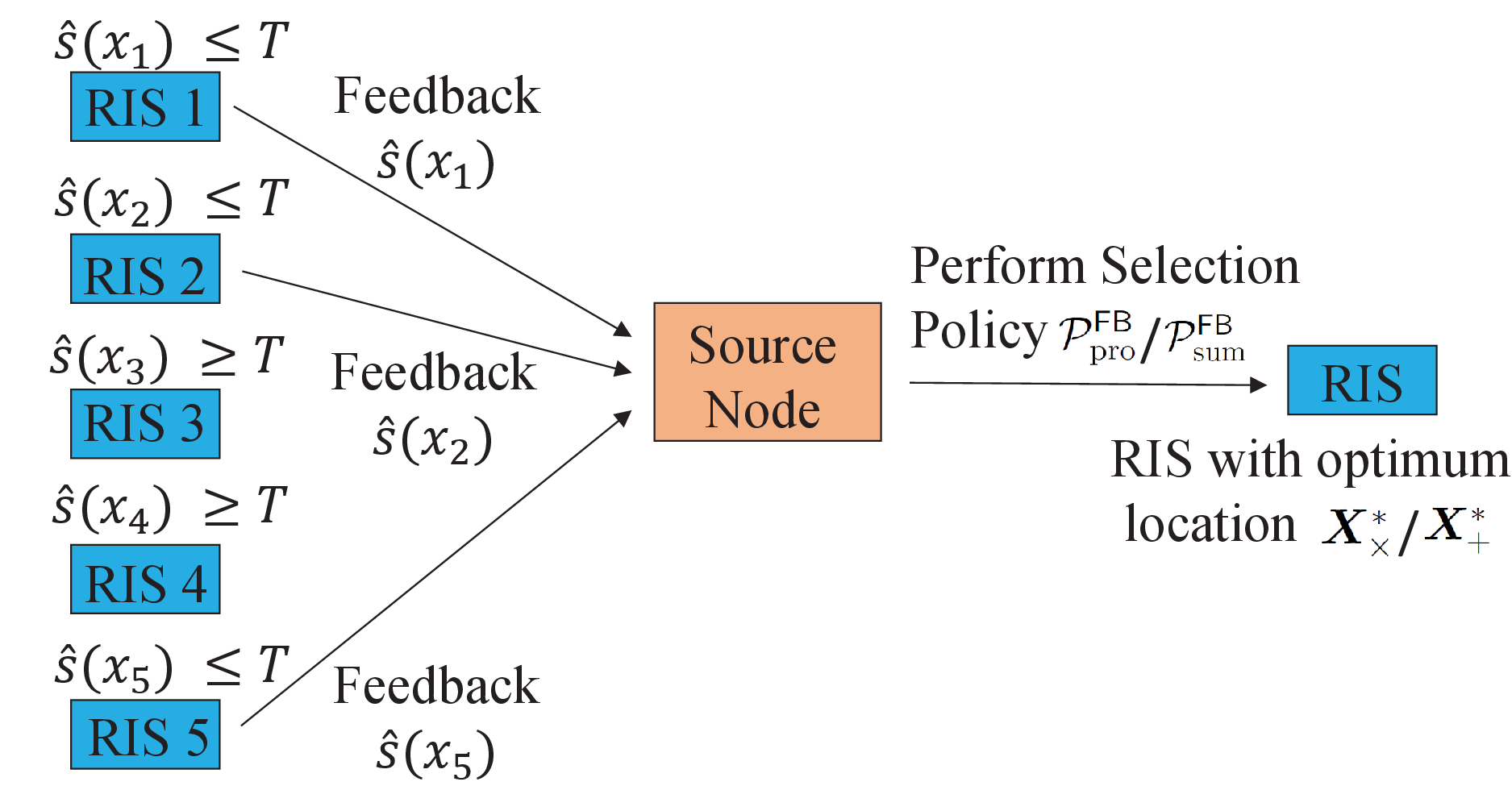}
\caption{{An example illustration for the limited-feedback RIS selection with 5 potential RISs. RIS1, RIS2, and RIS5, whose the channel quality indicators $\RISselect\paren{\vec{x}_1}$, $\RISselect\paren{\vec{x}_2}$, $\RISselect\paren{\vec{x}_5}$ are less than the given threshold value $T$, send their channel quality indicators to the source node. The source node then performs selection policy $\polOptPowFB$ or $\polOptExpFB$ to select the RIS with optimum location $\XoptPow$ or $\XoptExp$.}}
\label{fig:spatial}\vspace{-8mm}
\end{figure}

To alleviate the feedback overhead, we will consider an effective yet simple limited-feedback strategy (illustrated in Fig. \ref{fig:spatial}), which is formally put forward as follows.
%to control the number of RIS nodes feeding their channel states back to the source node. %To describe the feedback strategy, we also name the RIS selection functions and $\RISselectExp\paren{\vec{x}}$ as channel quality indicator of each RIS node $\vec{x}$ since value of the RIS selection functions indicates the associated channel quality. %We use $\RISselect\paren{\vec{x}}$ to refer the channel quality of each RIS node $\vec{x}$ for a general path-loss model. Using the notation $\RISselect\paren{\vec{x}}$,
%The limited-feedback strategy

{\it Limited-Feedback Strategy}: {An RIS located at $\vec{X}\in\Phi$ will send its channel quality indicator $\RISselect\paren{\vec{X}}$ back to the source node when $\RISselect\paren{\vec{X}}\leq T$, where $T>0$ is a given threshold value. For the product-scaling path-loss law, $\RISselect\paren{\vec{X}}$ is $\RISselect\paren{\vec{X}}=\RISselectPow\paren{\vec{X}}$, where $\RISselectPow\paren{\vec{X}}$ is given by $\RISselectPow\paren{\vec{X}} = \norm{\xs - \vec{X}}\times\norm{\vec{X} - \xd}$. For the sum-scaling path-loss law, $\RISselect\paren{\vec{X}}$ is $\RISselect\paren{\vec{X}}=\RISselectExp\paren{\vec{X}}$, where $\RISselectExp\paren{\vec{X}}$ is $\RISselectExp\paren{\vec{X}} = \norm{\xs - \vec{X}}+\norm{\vec{X} - \xd}$.} If no RIS feeds back its channel quality indicator, no data is transmitted by the source node. %An example illustration for the limited-feedback strategy is shown in Fig. \ref{fig:spatial}.

We will characterize the number of RISs feeding back given the aforementioned feedback strategy and evaluate the average rate and outage probability attained by limited-feedback RIS selection policies in the following subsections.

\subsection{Distribution of the Feedback Load with Limited-Feedback}\label{subsec:Power-Law path-loss}
Given the feedback strategy proposed in subsection \ref{subsec: feedback}, we denote the total number of RISs feeding back under the
product-scaling and the sum-scaling RIS selection function by $\NPowFB$ and $\NExpFB$, respectively. That is,
$\NPowFB=\sum_{\vec{X}\in\Phi}\I{\RISselectPow\paren{\vec{X}}\leq \Tpow}$ and $\NExpFB=\sum_{\vec{X}\in\Phi}\I{\RISselectExp\paren{\vec{X}}\leq \Texp}$, where $\I{\cdot}$ is the indicator function. $\Tpow$ and $\Texp$ are the thresholds used in the product-scaling and sum-scaling scenarios, respectively. The average number of RISs feeding back is given by $\XiPow = \EW_\Phi\sqparen{\NPowFB}$ and $\XiExp = \EW_\Phi\sqparen{\NExpFB}$. The sets of the RISs feeding back %$\Phi^{\FB}_{\pow}$ and $\Phi^{\FB}_{\expo}$
are given by $\Phi^{\FB}_{\pow} = \{X \in \Phi: \RISselectPow\paren{\vec{X}}\leq \Tpow\}$ and $\Phi^{\FB}_{\expo} = \{X \in \Phi: \RISselectExp\paren{\vec{X}}\leq \Texp\}$. %in a particular realization $\Phi$ is denoted by $\phi^{\FB}_{\pow}$.

We now formulate the RIS selection policy with limited-feedback for the power-law and exp-law path-loss models as follows:
\begin{selection}\label{opt-RIS-power-FB}
The RIS selection policy with limited-feedback strategy under the product-scaling path-loss law, denoted by $\polOptPowFB$, is the one that solves the following optimization problem
\begin{eqnarray}
\begin{array}{ll}
\underset{\vec{X} \in \R^2}{\mbox{minimize}} & \RISselectPow\paren{\vec{X}} \\
\mbox{subject to} & \vec{X} \in \Phi^{\FB}_{\pow}
\end{array}. \label{Eqn: RIS Selection Problem power-FB}
\end{eqnarray}
\end{selection}
\begin{selection}\label{opt-RIS-exp-FB}
For the sum-scaling path-loss law, the RIS selection policy with limited-feedback strategy, denoted by $\polOptExpFB$, is the one that solves the following optimization problem
\begin{eqnarray}
\begin{array}{ll}
\underset{\vec{X} \in \R^2}{\mbox{minimize}} & \RISselectExp\paren{\vec{X}} \\
\mbox{subject to} & \vec{X} \in \Phi^{\FB}_{\expo}
\end{array}. \label{Eqn: RIS Selection Problem exp-FB}
\end{eqnarray}
\end{selection}

%To characterize the outage performance and data rate performance given As shown in Subsection \ref{subsec:Power-Law path-loss},
Derivation of the distributions of $\NPowFB$ and $\NExpFB$ is a key step to quantify the performance attained by $\polOptPowFB$ and $\polOptExpFB$. We present the distributions of $\NPowFB$ and $\NExpFB$ in the following theorems. %we first characterize the distribution of, which is given as follows:
\begin{theorem}\label{theorem:feedback-power}
$\NPowFB$ is a Poisson RV with the mean $\XiPow$ given by
\begin{equation}\label{muT}
\XiPow
=\left\{
\begin{array}{ll}
2\lambda\paren{\frac{1}{d^2}\left(d^4 E(\frac{\Tpow^2}{d^4})+(\Tpow^2-d^4)K(\frac{\Tpow^2}{d^4})\right)}& \mbox{ if }\Tpow \leq d^2 \\
2\lambda\paren{\Tpow E(\frac{d^4}{\Tpow^2})
}  & \mbox{if }\Tpow > d^2
\end{array}.\right.
\end{equation}
\end{theorem}
\begin{IEEEproof}
See Appendix \ref{Appendix: feedback-power}.
\end{IEEEproof}
\begin{theorem}\label{theorem:feedback-exp}
$\NExpFB$ is a Poisson RV with the mean $\XiExp$ given by
\begin{equation}\label{muT-exp}
\XiExp
=\left\{
\begin{array}{ll}
0& \mbox{ if }\Texp \leq 2d \\
\frac{\lambda\pi\Texp}{4}\sqrt{-4d^2+\Texp^2}
& \mbox{ if }\Texp>2d
\end{array}.\right.
\end{equation}
\end{theorem}
\begin{IEEEproof}
The proof is similar to the one given for Theorem \ref{theorem:feedback-power}. Hence, it is omitted to avoid repetitions.
\end{IEEEproof}

%We validate the Theorems \ref{theorem:feedback-power} and \ref{theorem:feedback-exp} in, respectively, by plotting the mean $\XiPow$ versus the feedback threshold $\Tpow$ for different node density $\lambda$ in Fig. \ref{fig:FB-power1} and comparing the simulated distribution of $\NPowFB$ with the Poisson distribution with mean $\XiPow$ in Fig. \ref{fig:FB-power2}.As suggested by Theorem \ref{theorem:feedback-power}, simulated and theoretical distributions are in perfect agreement.

\begin{figure}
	\centering
	\subfloat[The mean number of $\NPowFB$]{
		\label{fig:FB-power1}
		\includegraphics[width=0.25\textwidth]{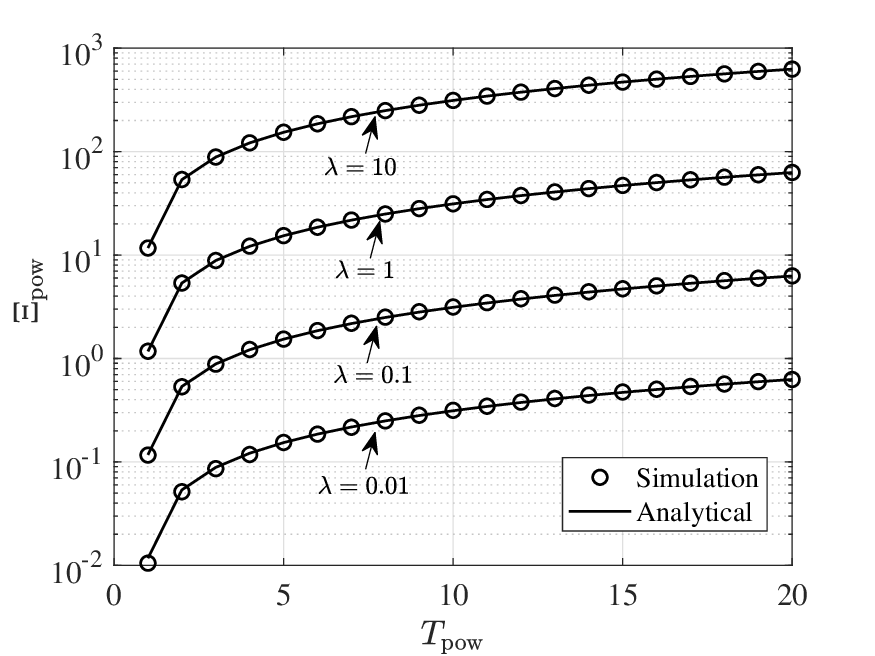}}
        	\subfloat[Distribution of $\NPowFB$ when $\lambda=0.5$ and $T=20$]{
		\label{fig:FB-power2}
		\includegraphics[width=0.25\textwidth]{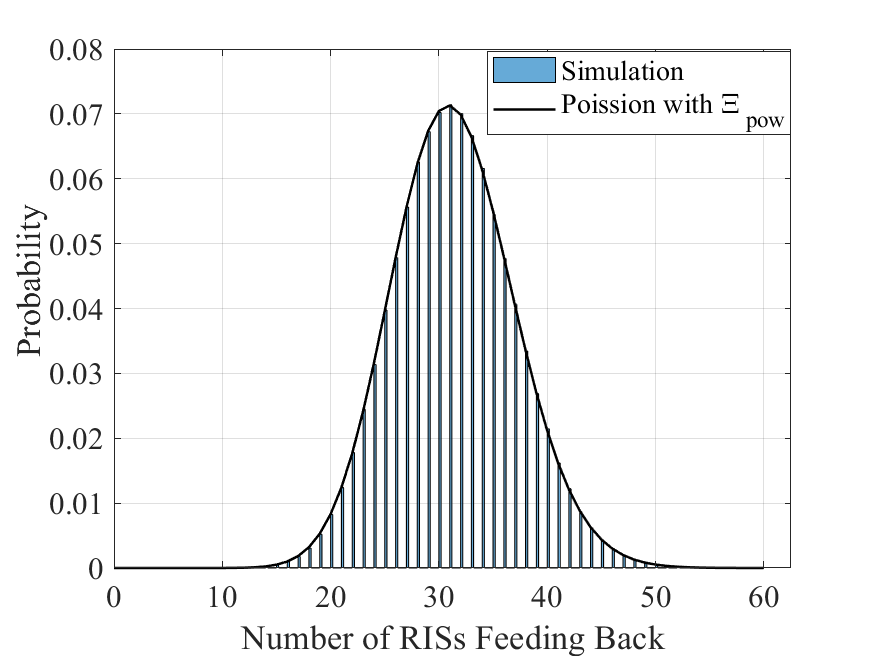}}
        \subfloat[The mean number of $\NExpFB$]{
		\label{fig:FB-exp1}
		\includegraphics[width=0.25\textwidth]{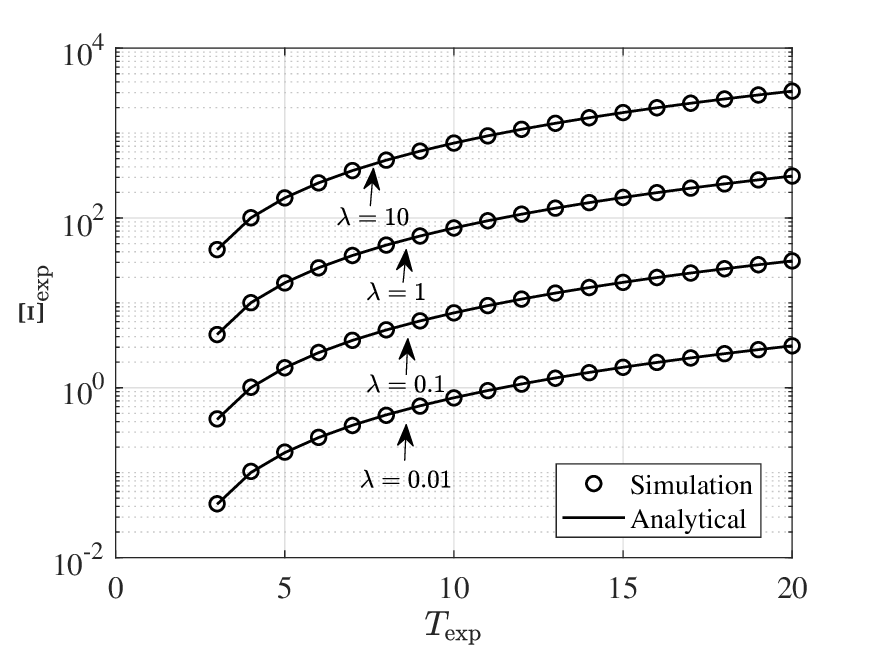}}
        \subfloat[Distribution of $\NExpFB$ when $\lambda=0.5$ and $T=20$]{
		\label{fig:FB-exp2}
		\includegraphics[width=0.25\textwidth]{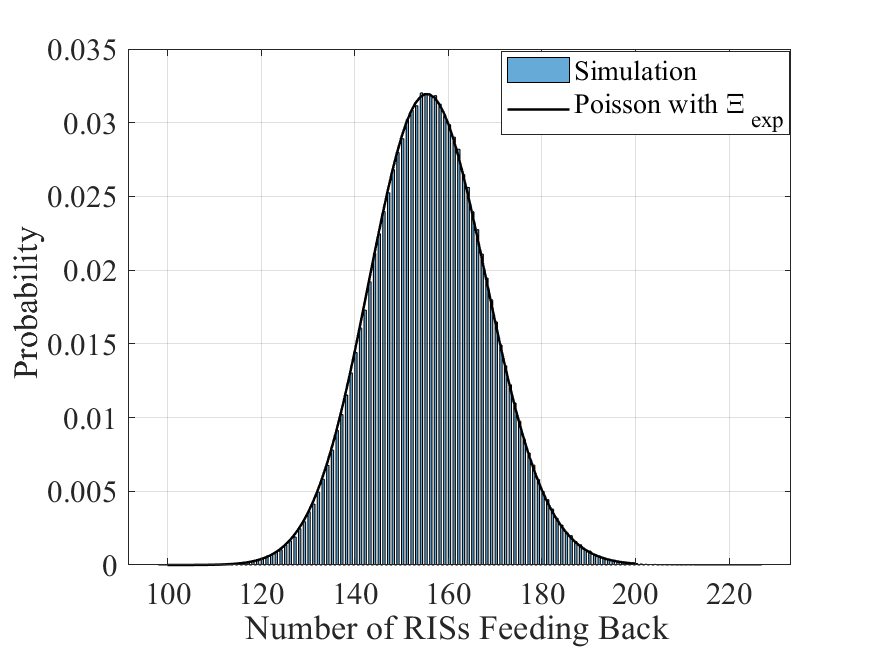}}
\vspace{-2mm}
		\caption{Average number of RISs feeding back and probability distribution of the number of RISs feeding back for $d = 1.2$ for product-scaling path-loss law in Figs. 2(a) and 2(b) and for sum-scaling path-loss law in Figs. 2(c) and 2(d).}\label{fig:FB-power}\vspace{-8mm}
\end{figure}

In Fig. \ref{fig:FB-power}, we plot the expected numbers of RISs feeding back and compare the simulated distribution of the total number of RISs feeding back with the Poisson distribution with analytical means in \eqref{muT} and \eqref{muT-exp}. We observe that simulated distributions match the theoretical ones perfectly, which validates the Theorems \ref{theorem:feedback-power} and \ref{theorem:feedback-exp}. As discussed in \cite{atapattu2020locationbased}, the threshold value $\XiPow\geq5$ enables the limited-feedback RIS selection strategy to achieve very similar outage and data rate performance as the centralized RIS selection strategy, while providing a massive reduction in the feedback overhead. This is because the RIS with optimum location, $\XoptPow$, always feeds its channel quality indicators back to the source node if $\NPowFB\geq1$ and $\PR{\NPowFB\geq1}\geq0.99$ when $\XiPow>5$ due to the exponentially decaying tail of Poisson distribution. The same arguments also apply to the sum-scaling path loss models.

\vspace{-6mm}
\subsection{Performance Analysis}\label{subsec:limited-feedback-performance}
We next evaluate the outage probability and average rate attained by the limited-feedback RIS selection policies, parameterized by the threshold values $\Tpow$ and $\Texp$, for the power-law and exp-law path-loss models, respectively.

The outage probability with limited feedback for the power-law and exp-law path-loss models are given in the following theorems:
\begin{theorem}\label{theorem:power-FB-outage}
The outage probability $\PoutPow\paren{\polOptPowFB}$ for a limited-feedback RIS selection policy $\polOptPowFB$ with threshold $\Tpow$ is equal to
\begin{equation}\label{power-feedback}
\PoutPow\paren{\polOptPowFB}
=\left\{
\begin{array}{ll}
\exp\paren{-\XiPow}& \mbox{ if }\rho \leq \frac{\Bar\gamma N(16+(N-1)\pi^2)}{\Tpow^\eta 16} \\
 1-F_{\Upsilon_{\rm opt}}\paren{\paren{{\frac{\Bar\gamma N(16+(N-1)\pi^2)}{\rho 16}}}^\frac{1}{\eta}}  & \mbox{ otherwise}
\end{array},\right.
\end{equation}
\end{theorem}
where $\XiPow$ is the average feedback load at $\Tpow$ and given in \eqref{muT}.
\begin{IEEEproof}
To prove Theorem \ref{theorem:power-FB-outage}, we first write the outage event as follows:
\begin{align}
\brparen{\ESZ{\snrPow\paren{\polOptPow}}  \leq \rho}
= \brparen{\Upsilon_{\rm opt} > \Tpow} \bigcup \paren{\brparen{\Upsilon_{\rm opt} \leq \Tpow} \bigcap \brparen{\frac{\Bar\gamma N(16+(N-1)\pi^2)}{\Upsilon_{\rm opt}^\eta16} \leq \rho}}. \hspace{-1.0cm} \nonumber
\end{align}
Hence, $\Pout\paren{\polOptPowFB}$ is given by
\begin{align}\label{eq:FB-outage-derive}
\Pout\paren{\polOptPowFB} =& \PR{\Upsilon_{\rm opt} > \Tpow} + \PRP{\brparen{\Upsilon_{\rm opt} \leq \Tpow} \bigcap \brparen{\frac{\Bar\gamma N(16+(N-1)\pi^2)}{\Upsilon_{\rm opt}^\eta16}\leq \rho}} \nonumber \\
=& \exp\paren{-\XiPow} \!+ \!\PR{\frac{\Bar\gamma N(16\!+\!(N-1)\pi^2)}{\Tpow^\eta16}\leq\frac{\Bar\gamma N(16+(N-1)\pi^2)}{\Upsilon_{\rm opt}^\eta16}\leq \rho}.
\end{align}

If $\rho<\frac{\Bar\gamma N(16+(N-1)\pi^2)}{\Tpow^\eta16}$, then the second summand in \eqref{eq:FB-outage-derive} becomes zero, and we have \\$\Pout\paren{\pol_{\rm FB}, \Tpow} = \exp\paren{-\XiPow}$, which is the first condition in \eqref{power-feedback}.  If $\rho>\frac{\Bar\gamma N(16+(N-1)\pi^2)}{\Tpow^\eta16}$, we have $\Pout\paren{\pol_{\rm FB}, T}=\PR{\frac{\Bar\gamma N(16+(N-1)\pi^2)}{\Upsilon_{\rm opt}^\eta16}\leq \rho}=1-F_{\Upsilon_{\rm opt}}\paren{\paren{{\frac{\Bar\gamma N(16+(N-1)\pi^2)}{\rho 16}}}^\frac{1}{\eta}}$, which is the second condition in \eqref{power-feedback}. This completes the proof of Theorem \ref{theorem:power-FB-outage}.
\end{IEEEproof}
%Using Theorem \ref{theorem:feedback-exp}, we evaluate the average rate and outage probability with limited feedback for the exp-law path loss model. We first derive the outage probability expressions.
\begin{theorem}\label{theorem:exp-FB-outage}
The outage probability $\PoutExp\paren{\polOptExpFB}$ for a limited-feedback RIS selection policy $\polOptPowFB$ with threshold $\Texp$ is given by
\begin{equation}\label{exp-feedback}
\PoutExp\paren{\polOptExpFB}
=\left\{
\begin{array}{ll}
\exp\paren{-\XiExp}& \mbox{ if }\rho \leq \frac{\Bar\gamma N(16+(N-1)\pi^2)}{\exp\paren{\alpha \Texp} 16} \\
1-F_{\Lambda_{\rm opt}}\paren{\log\paren{{\frac{\Bar\gamma N(16+(N-1)\pi^2)}{\rho 16}}}\alpha^{-1}} & \mbox{ if }\frac{\Bar\gamma N(16+(N-1)\pi^2)}{\exp\paren{\alpha \Texp} 16}<\rho<\frac{\Bar\gamma N(16+(N-1)\pi^2)}{\exp\paren{\alpha 2d} 16} \\
 1  & \mbox{ if } \rho\geq\frac{\Bar\gamma N(16+(N-1)\pi^2)}{\exp\paren{\alpha 2d} 16}
\end{array}.\right.
\end{equation}
\end{theorem}
\begin{IEEEproof}
Theorem \ref{theorem:exp-FB-outage} can be proven similarly to Theorem \ref{theorem:power-FB-outage}.
\end{IEEEproof}

We evaluate the average rate achieved by $\polOptPowFB$ and $\polOptExpFB$ in the following theorems:
\begin{theorem}\label{theorem:power-FB-rate}
The average rate $\RavePow\paren{\polOptPowFB}$ for a limited-feedback RIS selection policy $\polOptPowFB$ with threshold $\Tpow$ is given by
\begin{eqnarray}\label{throughput given phi,FB}
%\RaveExp\paren{\pol_{\FB},T} & = & \int_{T^{-\eta}}^{\infty}\ESZ{\log_2\paren{1 + \bar\gamma y A^2}} f_Y(y) dy
\RavePow\paren{\polOptPowFB} &=& \int_{\Tpow^{-\eta}}^{d^{-2\eta}} {\sf R}\paren{\pol,y}S_1\paren{y} d\,y+\int_{d^{-2\eta}}^{\infty} {\sf R}\paren{\pol,y}S_2\paren{y} d\,y.
\end{eqnarray}
\end{theorem}
\begin{IEEEproof}
Theorem \ref{theorem:power-FB-rate} can be proven by using the equivalence of events $\brparen{\NPowFB \geq 1}$ and $\brparen{\Upsilon_{\rm opt} > \Tpow}$. % and our assumption that no data will be transmitted if $\Upsilon_{\rm opt} > \Tpow$.
\end{IEEEproof}

%\begin{IEEEproof}
%It can be proven by the equivalence of events $\brparen{\NPowFB \geq 1}$ and $\brparen{\Upsilon_{\rm opt} > \Tpow}$, and our assumption that no data will be transmitted if $\Upsilon_{\rm opt} > \Tpow$.
%\end{IEEEproof}
%\subsection{Exp-Law path-loss}
%We derive the average rate with limited feedback in the following theorem:
\begin{theorem}\label{theorem:exp-FB-rate}
The average rate $\RaveExp\paren{\polOptExpFB}$ for a given limited-feedback RIS selection policy $\polOptExpFB$ with threshold $\Texp$ is equal to
\begin{equation}\label{throughput given phi,FB,exp}
\begin{split}
\RaveExp\paren{\polOptExpFB} %& = \int_{\exp(-\alpha T)}^{\infty}\ESZ{\log_2\paren{1 + \bar\gamma y A^2}} f_Y(y) dy
&= \int_{\exp(-\alpha \Texp)}^{e^{-2\alpha d}}{\sf R}\paren{\pol,y}f_{Y_{\rm exp}}(y) d\,y.
\end{split}
\end{equation}
\end{theorem}
\begin{IEEEproof}
Theorem \ref{theorem:exp-FB-rate} can be proven similarly to Theorem \ref{theorem:power-FB-rate}.
\end{IEEEproof}

We note that the upper bounds on $\RavePow\paren{\polOptPowFB}$ and $\RaveExp\paren{\polOptExpFB}$ using Jensen's inequality can be obtained as in Remark \ref{remark:upper}.

\vspace{-5mm}
\section{Numerical Results}\label{sec: numerical}
In this section, we present simulation and numerical results to verify our derived analytical results, discuss the performance of the proposed centralized and limited-feedback RIS selection policies, and reveal the effect of system parameters on the system performance.

In our simulations, %all distances are normalized to a {\it unit} distance. A circular network coverage area with radius $10$ [units] is considered.
the target $\snr$ and the distance between TX and RX are set to $\rho=5$\,dB and $d=1.2$\,m, respectively. {The unit for $\lambda$ is RISs/$\textrm{m}^2$ and the value of $\lambda$ will be mentioned in each figure.} For the performance evaluation, the path-loss exponent $\eta$ is taken to be $4$ and the tunable parameter $\alpha$ of the exp path-loss model is $1.037$ \cite{8122033}. The simulation results are averaged over many realizations of the random locations of RISs and channel fading.

\vspace{-5mm}

\subsection{Centralized RIS Selection}\label{sec: numerical-nonFB}
%In Figs. \ref{fig:cdf}, \ref{fig:cdf1}, and \ref{fig:cdf2}, we test the accuracy of \eqref{e_cdfmaxminD}, \eqref{Eqn: Distance Distribution 3-mMin-1}, and \eqref{cdf-d-minSum}, respectively, with different density for the fixed $d$. We see that the accuracy of \eqref{e_cdfmaxminD} is good when $\gamma<d^2$. When density is smaller, \eqref{e_cdfmaxminD} has a bit deviation with simulations when $\gamma>d^2$. This deviation is cause by the approximation of $\lim_{\tau \ra \infty} F_{\Upsilon}\paren{\gamma}= 1$ when $\gamma>d^2$.

In this subsection, we focus on the performance of optimum centralized RIS selection policies. %, i.e., the information of all RIS locations is available.
To benchmark the optimum selection policy, we consider three other RIS selection schemes:
\begin{enumerate}
  \item {\it Min-min} scheme selects the closet RIS to the TX and RX set $\{\xs, \xd\}$, i.e.,
\[\underset{\vec{X} \in \R^2}{\mbox{minimize}}\,\,\min\brparen{\norm{\xs - \vec{X}},\norm{\vec{X} - \xd}},\,\,\vec{X} \in \Phi;\]
  \item {\it Min-max} scheme selects an RIS according to
\[\underset{\vec{X} \in \R^2}{\mbox{minimize}}\,\,\max\brparen{\norm{\xs - \vec{X}}, \norm{\vec{X} - \xd}},\,\,\vec{X} \in \Phi;\]
which is the optimum scheme for a decode-and-forward relay network \cite{atapattu2020locationbased};
  \item {\it Mid-point} scheme selects the RIS that has the minimal distance to the mid-point between the TX and the RX.
\end{enumerate}

%\subsubsection{Power-Law path-loss}

\begin{figure}
	\centering
	\subfloat[$\lambda=0.5$]{
		\label{Fig-4-1}
		\includegraphics[width=0.45\textwidth]{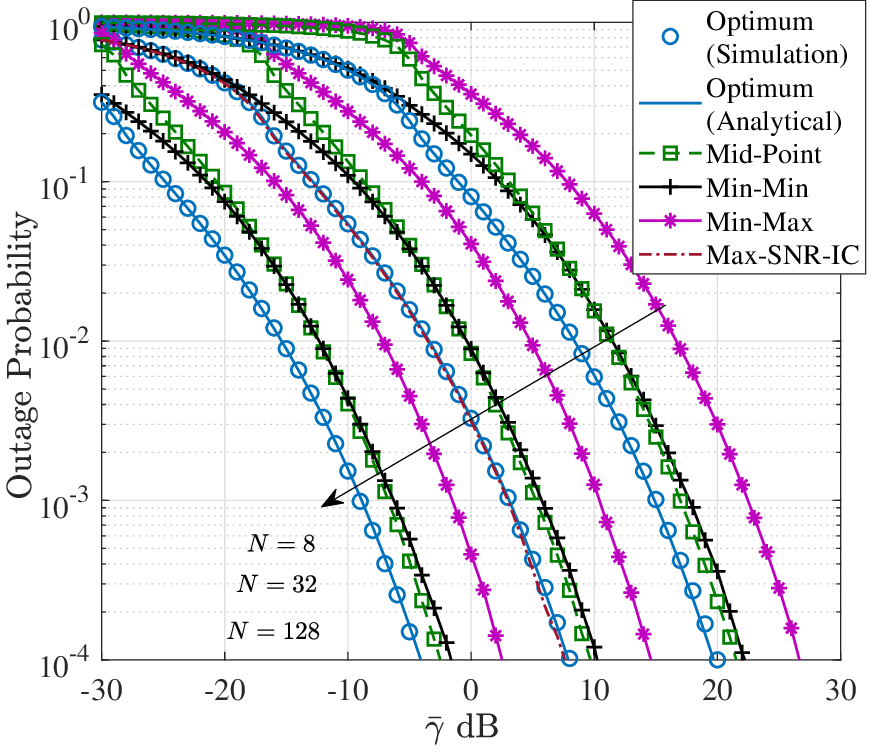}}
        	\subfloat[$N=16$]{
		\label{Fig-4-2}
		\includegraphics[width=0.45\textwidth]{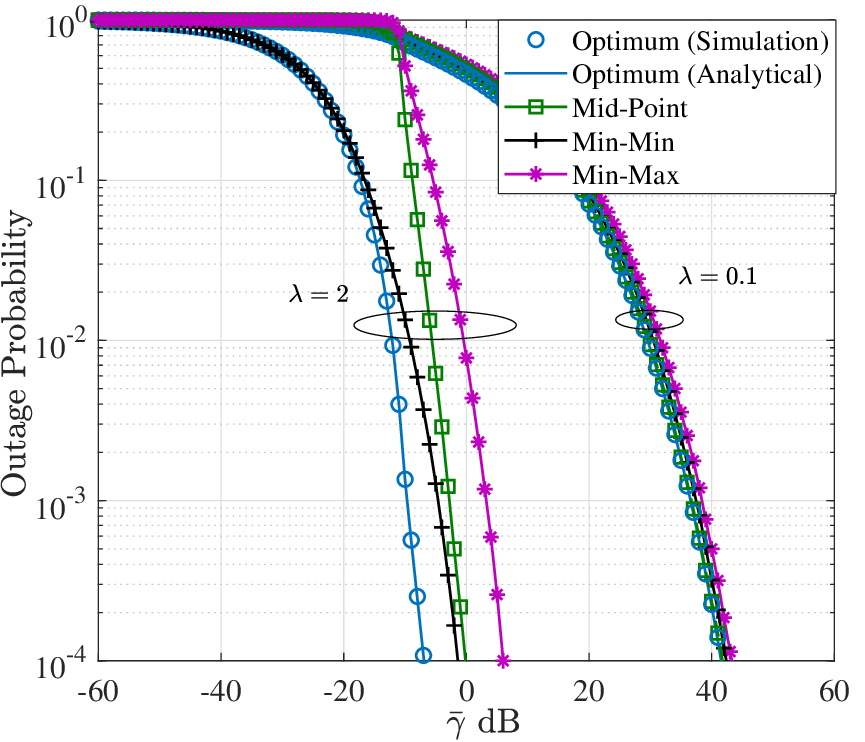}}
         %\subfloat[$\lambda=10$,$N=16$.]{
%		\label{f_tpvslam5}
%		\includegraphics[width=0.33\textwidth]{Pout_pow_snr_lambda10_N32.eps}}
%\subfloat[$\lambda=10$,$N=32$.]{
%		\label{f_tpvslam5}
%		\includegraphics[width=0.33\textwidth]{Pout_pow_snr_lambda10.eps}}
\vspace{-3mm}
		\caption{Outage probability achieved by different RIS selection schemes versus average $\snr$ $\Bar\gamma$, $d=1.2$, $\eta=4$, $\rho=5$ dB, for the power-law path-loss model.}%is used to generate this figure.
	\label{Fig-4}\vspace{-8mm}
\end{figure}
\begin{figure}
	\centering
	\subfloat[$\lambda=0.5$]{
		\label{Fig-5-1}
		\includegraphics[width=0.45\textwidth]{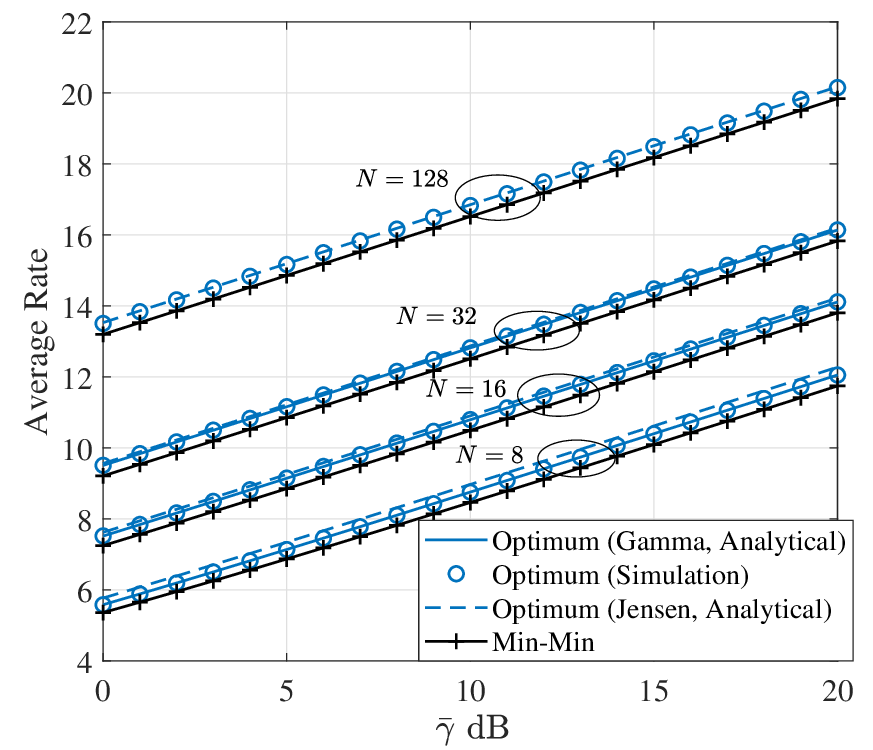}}
        	\subfloat[$\Bar\gamma=0$ dB ]{
		\label{Fig-5-2}
		\includegraphics[width=0.45\textwidth]{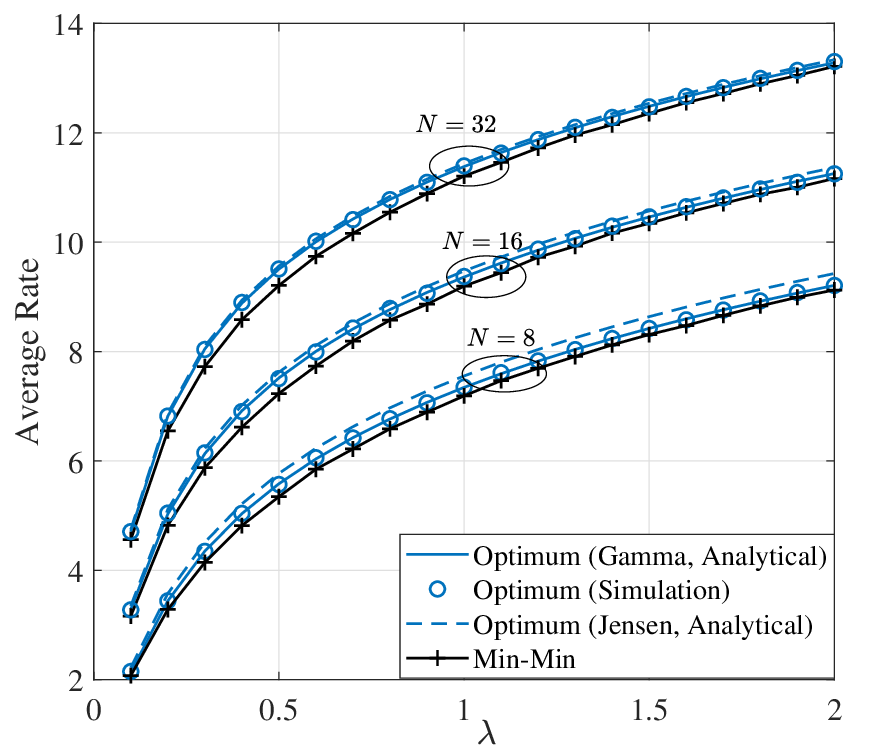}}
        % \subfloat[$\lambda=10$]{
%		\label{f_tpvslam5}
%		\includegraphics[width=0.33\textwidth]{Rate_pow_snr_lambda10.eps}}
%\subfloat[$\lambda=10$,$N=32$.]{
%		\label{f_tpvslam5}
%		\includegraphics[width=0.3\textwidth]{Pout_pow_snr_lambda10.eps}}
\\
\vspace{-3mm}
		\caption{Average rate achieved by different RIS selection schemes for different values of $N$, $d=1.2$, $\eta=4$, for the power-law path-loss model.}%is used to generate this figure.
	\label{Fig-5}\vspace{-8mm}
\end{figure}

In Figs. \ref{Fig-4} and \ref{Fig-5}, we investigate the performance for power-law communications scenarios under the optimum and other different RIS selection schemes as described above. In Fig \ref{Fig-4}, we plot the outage probability curves versus the average $\snr$ for $N=8$, $N=32$ and $N=128$ (i.e., see Fig. \ref{Fig-4-1}) and for $\lambda=0.1\text{ and } 2 $ (i.e., see Fig. \ref{Fig-4-2}). In Fig. \ref{Fig-4}, analytical curves are obtained by using \eqref{outage2}. The perfect agreement between analytical and simulation curves verifies the outage probability expression for the optimum RIS selection scheme in \eqref{outage2}. {In Fig. \ref{Fig-4-1}, we also plot max-SNR-instantaneous-channel (max-SNR-IC) scheme that selects the RIS yielding the largest instantaneous SNR based on the instantaneous channels.} In Fig. \ref{Fig-4-1}, we first see the outage probability decreases significantly when the number of elements increases from $8$ to $32$, where an RIS with $N=8$ requires around $15$ times more power than an RIS with $N=32$ to achieve the outage probability level of $10^{-3}$. For $N=8, 32\text{ and }128$, the optimum scheme has the best performance among all schemes. {We see that max-SNR-IC slightly outperforms the optimum policy. We note that the max-SNR-IC scheme is actually the optimal RIS selection policy, but this scheme is more difficult to be implemented than the optimum scheme proposed in the paper since max-SNR-IC requires the knowledge of instantaneous fading channels of all available RIS-aided links.} Among the other three schemes, min-min is the best suboptimal one when the average $\snr$ $\Bar\gamma$ is less than $0$ dB; otherwise, mid-point is the best suboptimal one.
For example, the optimum scheme only requires around 63\% transmit power that mid-point scheme requires to achieve the outage probability level of $10^{-3}$. Further, min-max always has the worst performance among these four schemes. In Fig. \ref{Fig-4-2}, we observe a significant decrease in outage probability when the density increases from $\lambda=0.1$ to $\lambda=2$. When $\lambda$ is very small as $0.1$, all selection schemes have very similar performance. This observation indicates there is a high probability to select the same RIS by different schemes at a very low RIS density and the superiority of the optimum scheme is not obvious.  However, the performance gap clearly increases with $\lambda$. For example, when $\lambda$ is $2$, the optimum scheme only requires around 32\% transmit power that min-min scheme requires to achieve the outage probability level of $10^{-3}$.

\begin{figure}
	\centering
	\subfloat[$\lambda=0.5$]{
		\label{Fig-7-1}
		\includegraphics[width=0.45\textwidth]{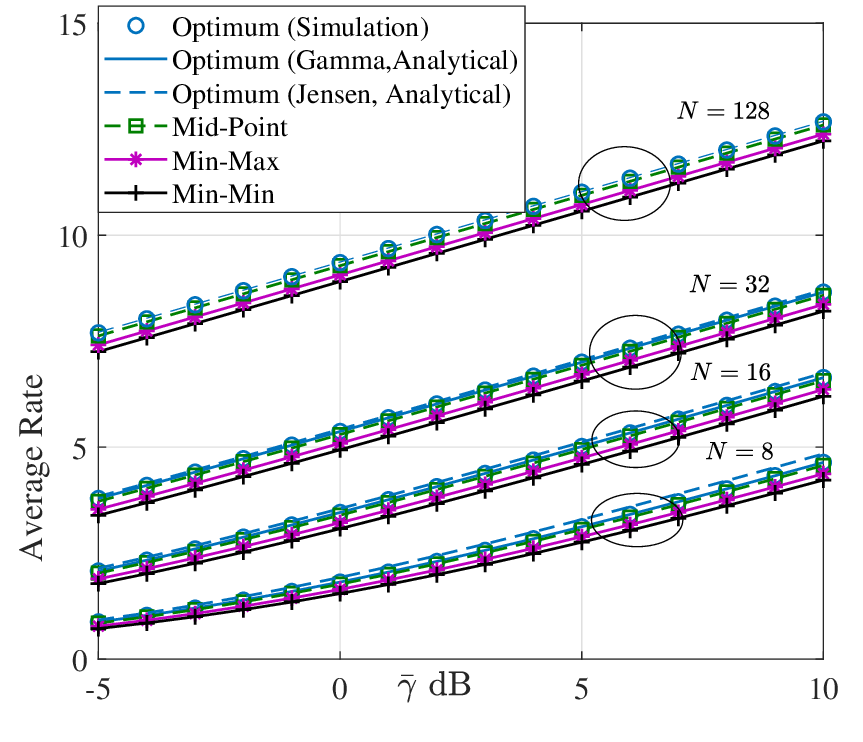}}
\subfloat[$\Bar\gamma=0$ dB]{
		\label{Fig-7-2}
		\includegraphics[width=0.45\textwidth]{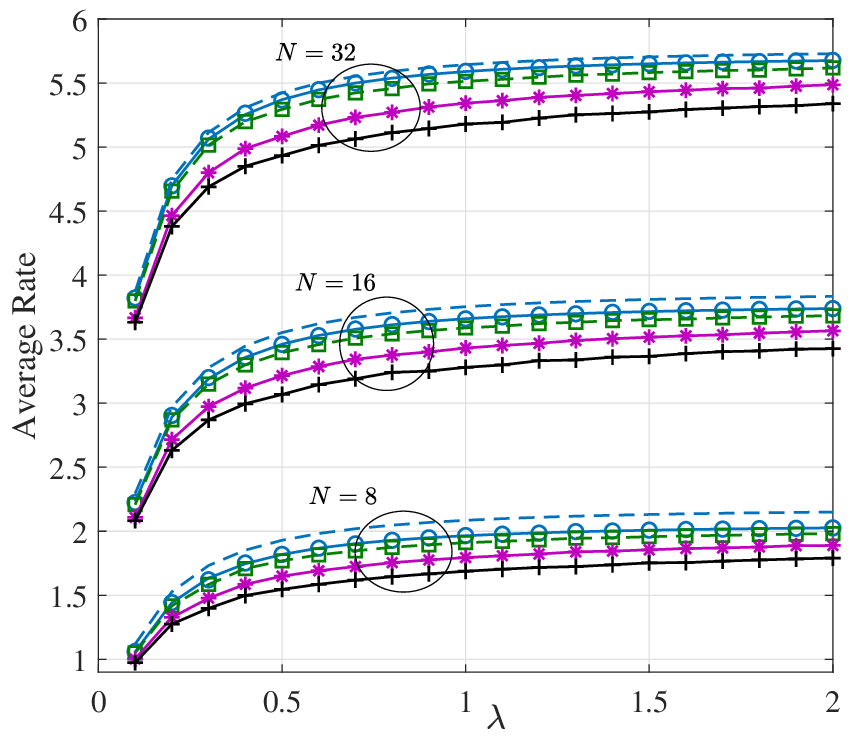}}
\\
		\caption{Average rate achieved by different RIS selection schemes for different values of $N$, $d=1.2$, $\alpha=1.037$, for the exp-law path-loss model.}%is used to generate this figure.
	\label{Fig-7}\vspace{-8mm}
\end{figure}

We plot the average rate versus $\Bar\gamma$ in Fig. \ref{Fig-5-1} and the average rate versus the density in Fig. \ref{Fig-5-2}. We only consider the optimum scheme and min-min scheme since these two schemes generally outperform the other two schemes for different sets of parameters based on Fig \ref{Fig-4}. Analytical values calculated from \eqref{Eqn: rate-power1} has a perfect agreement with the simulation, which verifies the accuracy of our analysis. In Fig.~\ref{Fig-5}, we see that the optimum and min-min selection schemes have small rate differences for different values of $N$ at $\lambda=0.5$. For example, the rate difference is around 0.3\,[bits/sec/Hz] for $N=16$ at $\snr=5$\,dB. Interestingly, the average rate increases almost linearly with $\snr$ and the gap between the optimum scheme and min-min scheme keeps unchanged as the average $\snr$ increases. In Fig. \ref{Fig-5-2}, we see the average rate for both schemes increases as the RIS density increases, but the increase in average rate with the density is not linear as that with the average $\snr$. In addition, we observe that the upper bounds obtained via Jensen's inequality is tighter as $N$ increases. Similar observations will be illustrated in Fig. \ref{Fig-7} as well.

%\subsubsection{Exp-Law path-loss}

\begin{figure}
	\centering
	\subfloat[$\lambda=0.5$]{
		\label{Fig-6-1}
		\includegraphics[width=0.45\textwidth]{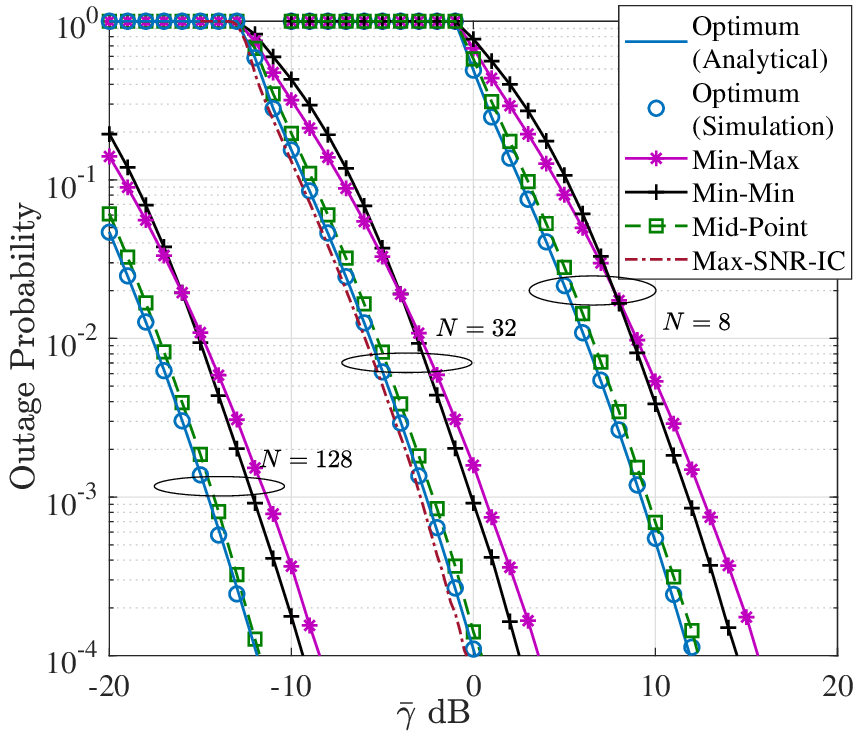}}
        	\subfloat[$N=16$]{
		\label{Fig-6-2}
		\includegraphics[width=0.45\textwidth]{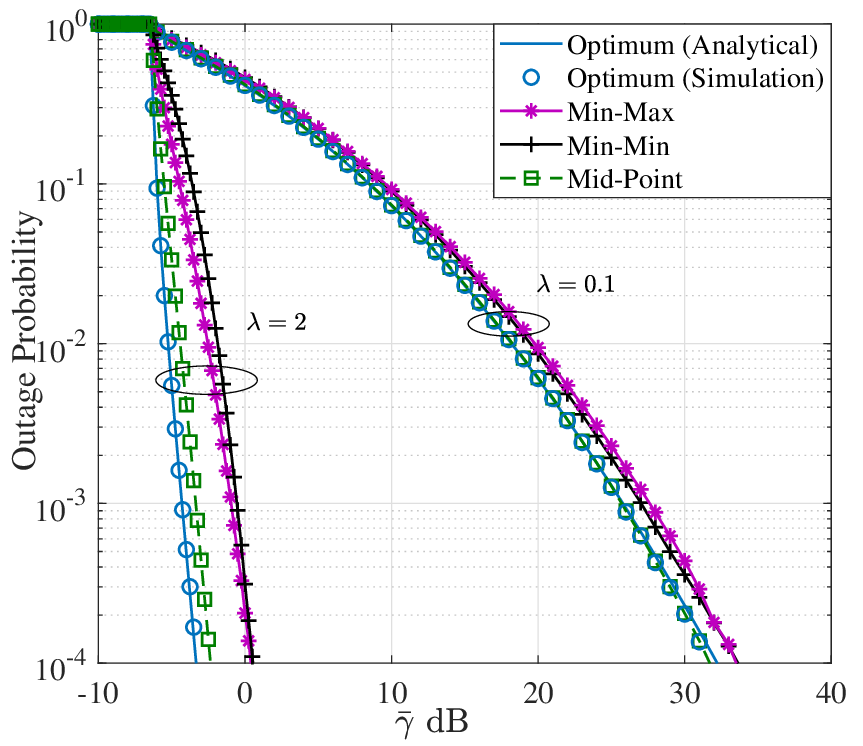}}
        % \subfloat[$\Bar\gamma=-3$ dB.]{
%		\label{f_tpvslam5}
%		\includegraphics[width=0.33\textwidth]{Pout_exp_lambda_snr-3.eps}}
\\
		\caption{Outage probability achieved by different RIS selection schemes versus average $\snr$ $\Bar\gamma$, $d=1.2$, $\alpha=1.037$, $\rho=5$ dB, for the exp-law path-loss model.}%is used to generate this figure.
	\label{Fig-6}\vspace{-8mm}
\end{figure}

In Figs. \ref{Fig-6} and \ref{Fig-7}, we focus on the performance of the exp-law path-loss model. In Fig.~\ref{Fig-6}, we plot the outage probability curves versus $\snr$ for $N=8$, $N=32$ and $N=128$ (i.e., see Fig. \ref{Fig-6-1}) and for $\lambda=0.1\text{ and } 2$ (i.e., see Fig. \ref{Fig-6-2}). Analytical curves obtained by \eqref{outage2opt} match with the simulations, thus validating the accuracy of \eqref{outage2opt}.
The optimum and {max-SNR-IC} schemes outperform the other schemes, which is similar to the observations in Fig. \ref{Fig-4}. However, the other schemes perform differently than the behavior observed in Fig. \ref{Fig-4} for the power-law path-loss model. Specifically, the mid-point scheme always has very close performance with the optimum scheme. For example, the optimum scheme can save around 10\% transmit power with respect to mid-point scheme to achieve the outage probability level of $10^{-3}$. Min-min and min-max have the worst performance among these four schemes, where min-min starts to outperform min-max when $\snr$ increases. In Fig. \ref{Fig-6-2}, we see a significant outage performance increase when the density increases and the performance gap among the different schemes increases with the density, as we observed in Fig. \ref{Fig-4-2}. In Fig. \ref{Fig-6-2}, we also see that the performance advantage achieved by the optimum scheme over the mid-point scheme is minor when the density $\lambda=0.1$. This means that the mid-point scheme is a near suboptimum selection scheme when the density is very small.

%\begin{figure}
%	\centering
%	\subfloat[$N=8$ dB.]{
%		\label{f_tpvslam2}
%		\includegraphics[width=0.33\textwidth]{Rate_exp_lambda_N8.eps}}
%        	\subfloat[$N=16$ dB.]{
%		\label{f_tpvslam5}
%		\includegraphics[width=0.33\textwidth]{Rate_exp_lambda_N16.eps}}
%         \subfloat[[$N=32$ dB.]{
%		\label{f_tpvslam5}
%		\includegraphics[width=0.33\textwidth]{Rate_exp_lambda_N32.eps}}
%\\
%		\caption{Average Rate achieved by different RIS selection schemes versus $\lambda$ for different values of number of elements $N$, $d=1$, $\alpha=1.037$, $\rho=5$ dB, $\Bar\gamma=0$ dB.}
%	\label{Fig: OutageDiffRS}
%\end{figure}

In Fig. \ref{Fig-7}, we plot the average rate versus the average $\snr$ (i.e., see Fig.~\ref{Fig-7-1}) and the average rate versus density $\lambda$ (i.e., see Fig.~\ref{Fig-7-2}). Analytical values calculated from \eqref{throughput given phi 2} overlap with the simulation results, which verifies \eqref{throughput given phi 2}. Interestingly, we see that the optimum and mid-point selection schemes have very small rate differences for both $N=16 \text{ and } 32$ at $\lambda=0.5$. For example, the rate difference is around 0.07\,[bits/sec/Hz] for $N=16$ at $\snr=5$\,dB. For all three cases $N=8,16,32$ in Fig.~\ref{Fig-7-2}, we have significant rate improvement from $\lambda=0.1$ to $\lambda=0.5$, and then there is a rate floor when $\lambda$ increases further. This is due to the fact that there is a sufficient number of RISs within the neighborhood of TX and RX to support the communication. Thus, it is not worth to densify RISs in a given area beyond a certain limit.

%In Figs. \ref{fig:outage1} and \ref{fig:outage2}, for power-law model, for high-density (e.g., 10), min-min is the best sub-opt and even perform closely to the opt. However, for low density (e.g., 0.1), mid-opt is the best sub-opt.
%
%In Fig. \ref{fig:outage3}, we assume outage means instantaneous $\snr$ in a given fading channel is below a threshold. In this case, we need to derive outage probability using a different method and there is a gap between simulation and analysis.
%
%In Figs. \ref{fig:outage4}, we see ``minimal prod'' selection scheme has a close performance to the optimal scheme.

\subsection{Distributed Network Operation and Limited-Feedback Case}
%\begin{figure}[!htb]
%\centering
%\includegraphics[width=0.7\textwidth]{power-FB-T-dens.eps}
%\caption{Power-law: $\lambda=0.1$, $d = 1$, $N = 16$, $Pt/N0=10$ dB, $\eta=3$}\label{fig:outage4}
%\end{figure}
%
%\begin{figure}[!htb]
%\centering
%\includegraphics[width=0.7\textwidth]{power-FB-T-dens-10.eps}
%\caption{Power-law:$\lambda=10$, $d = 1$, $N = 16$, $Pt/N0=10$ dB, $\eta=3$}\label{fig:outage4}
%\end{figure}

\begin{figure}
	\centering
	\subfloat[Power-Law: $\eta=4$]{
		\label{Fig-8-1}
		\includegraphics[width=0.45\textwidth]{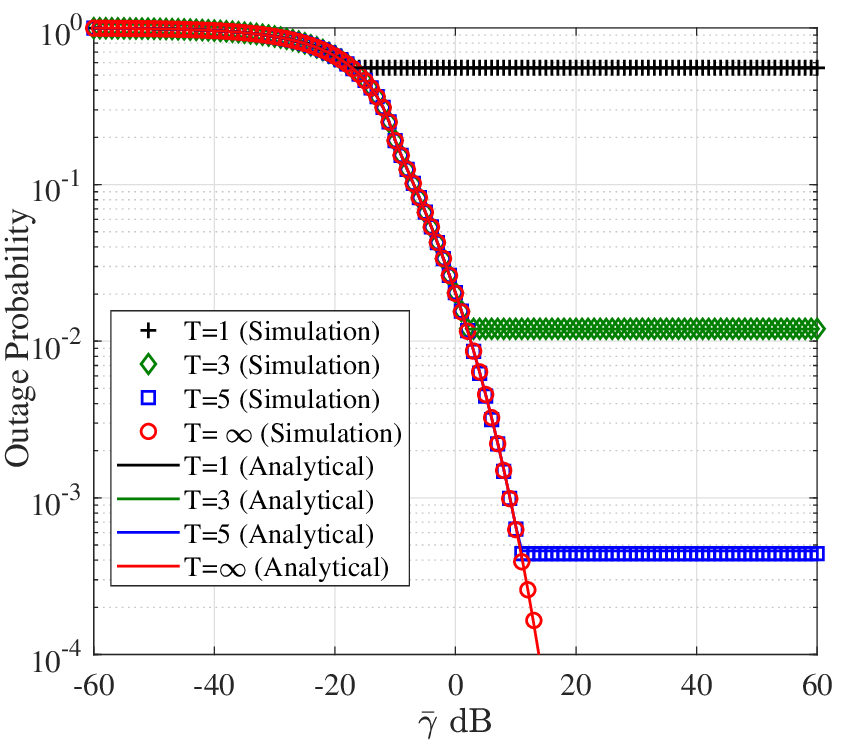}}
        	%\subfloat[Power-Law: $\lambda=10$, $\eta=3$]{
%		\label{f_tpvslam5}
%		\includegraphics[width=0.33\textwidth]{power-FB-T-dens-10.eps}}
\subfloat[Exp-Law: $\alpha=1.037$]{
		\label{Fig-8-2}
		\includegraphics[width=0.45\textwidth]{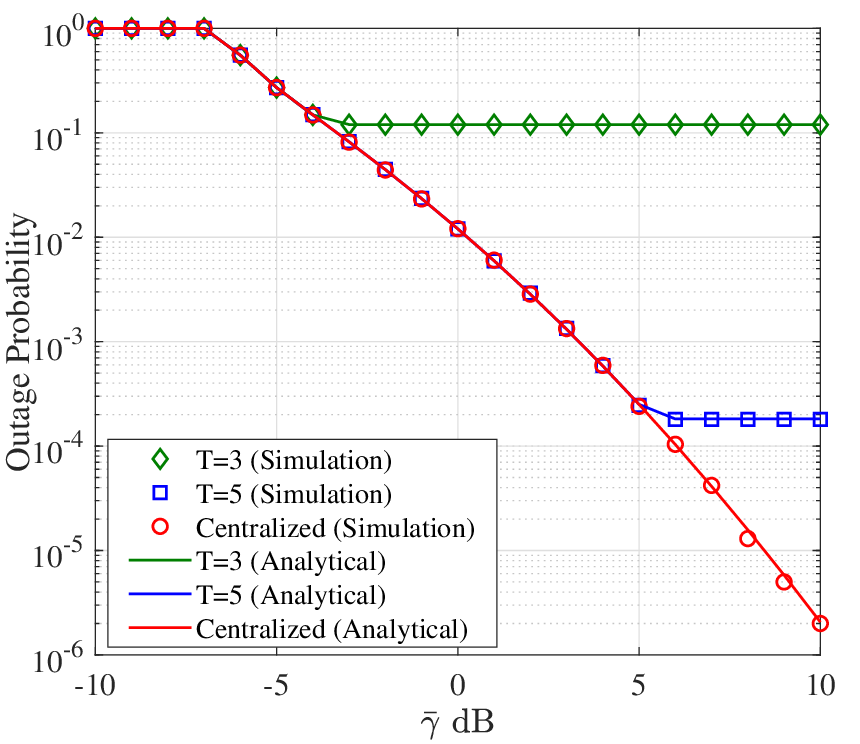}}
\\
		\caption{Outage probability achieved by limited-feedback RIS
selection versus the average $\snr$ $\Bar\gamma$ for different values of the threshold $T$. $d = 1.2$, $N = 16$, $\lambda=0.5$, and $\rho=5$ dB.}
	\label{Fig-8}\vspace{-5mm}
\end{figure}

%In this subsection, we focus on the distributed network performance with limited-feedback RIS selection policies for both power-law and exp-law path-loss models. %We note that $T=\infty $ considered in this subsection corresponds to the non-feedback cases investigated in Section \ref{sec: numerical-nonFB}.

In Fig. \ref{Fig-8}, we plot the outage probability achieved by limited-feedback RIS
selection versus the average $\snr$ $\Bar\gamma$ for the power-law model (Fig. \ref{Fig-8-1}) and for the exp-law model (Fig. \ref{Fig-8-2}). The analytical curves in Figs. \ref{Fig-8-1} and \ref{Fig-8-2} are obtained by \eqref{power-feedback} and \eqref{exp-feedback}, respectively. These curves perfectly match with simulations, thus validating the accuracy of \eqref{power-feedback} and \eqref{exp-feedback}. %the case that feeding back all RIS locations, thus the corresponding performance is equal to the outage probability achieved by the optimum RIS selection policies
We do not consider $T=1$ in Fig. \ref{Fig-8-2} since there is no RIS feeding back when $T<2d$ for the exp-law path-loss. In both figures, we see that outage probability curves first overlap with the centralized case and then keep constant as $\Bar\gamma$ increases. This is because when $\Bar\gamma$ is small, the values of target $\snr$ satisfy the condition $\rho > \frac{\Bar\gamma N(16+(N-1)\pi^2)}{\Tpow^\eta 16}$ for the power-law model and the condition $\rho > \frac{\Bar\gamma N(16+(N-1)\pi^2)}{\exp\paren{\alpha \Texp} 16}$ for the exp-law model. In such conditions, the outage performance is the same as the centralized case and does not depend on $T$ since at least one RIS feeds its location information back to the source node. %and there is no loss of optimality due to a selective feedback RIS selection policy. %is independent of whether is employed or not based on \eqref{power-feedback} and \eqref{exp-feedback}.
When $\Bar\gamma$ continuously increases, these conditions are not satisfied and the outage probability does only depend on the average number of RISs feeding back, without any dependence on the average $\snr$ and fading behavior. This is because the achieved outage probability depends on whether or not there is at least one RIS feeding its location information back to the source. We also see when $T$ is larger, the outage performance is better in the flat region of outage curves. This is because when $T$ is larger, we have a higher probability of at least one RIS feeding back, thus achieving better outage performance. %Based on \eqref{power-feedback} and \eqref{exp-feedback},

%\begin{figure}[!htb]
%\centering
%\includegraphics[width=0.7\textwidth]{rate-RIS-power-SNR-FB.eps}
%\caption{Power-law: $\lambda=0.1$, $d = 1$, $N = 16$, $\eta=3$}\label{fig:outage4}
%\end{figure}

\begin{figure}
	\centering
	\subfloat[Power-Law: $\eta=4$]{
		\label{Fig-9-1}
		\includegraphics[width=0.45\textwidth]{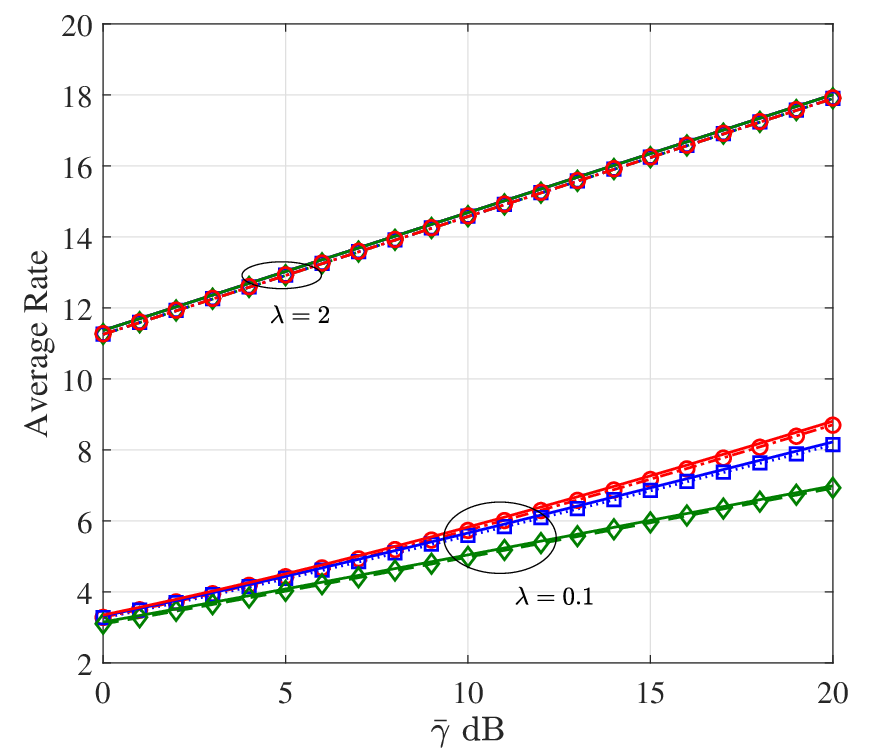}}
        	\subfloat[Exp-Law: $\alpha=1.037$]{
		\label{Fig-9-2}
		\includegraphics[width=0.45\textwidth]{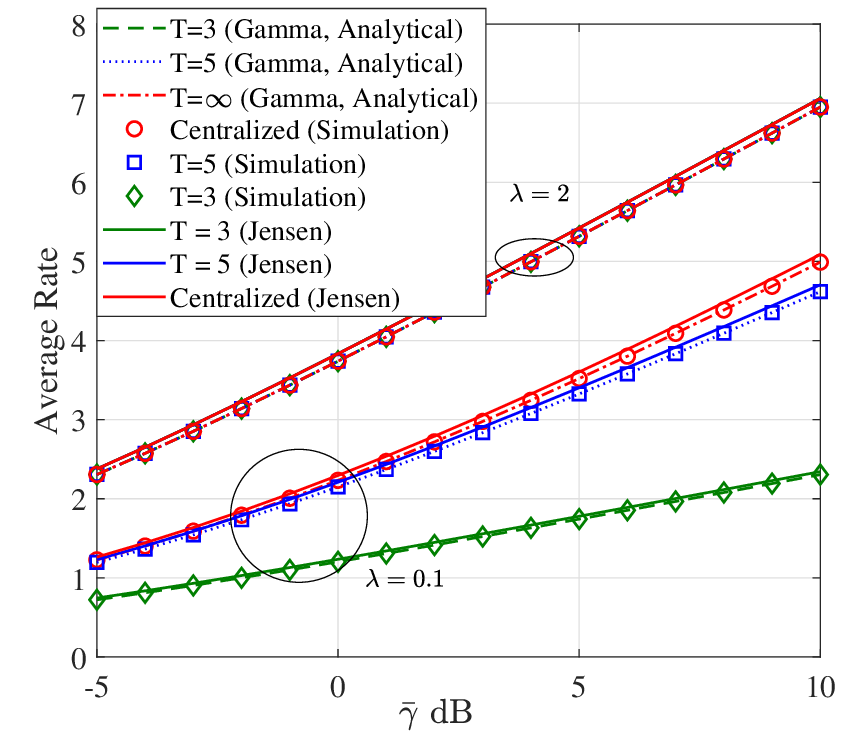}}
%\subfloat[$\lambda=0.1$]{
%		\label{f_tpvslam5}
%		\includegraphics[width=0.33\textwidth]{rate-RIS-power-SNR-FB.eps}}
\\
		\caption{Average rate achieved by limited-feedback RIS
selection for different values of the threshold $T$. $d = 1.2$ and $N = 16$.}
	\label{Fig-9}\vspace{-5mm}
\end{figure}

We plot the average rate for limited-feedback RIS selection versus $\Bar\gamma$ for the power-law model in Fig. \ref{Fig-9-1} and for the exp-law model in Fig. \ref{Fig-9-2}. The perfect agreement between analytical curves obtained by \eqref{throughput given phi,FB} and \eqref{throughput given phi,FB,exp} and simulations verify the accuracy of \eqref{throughput given phi,FB} and \eqref{throughput given phi,FB,exp}. For both Figs. \ref{Fig-9-1} and \ref{Fig-9-2}, we observe that for $\lambda=0.1$, the performance loss from centralized case (which assumes the availability of location information from all RISs) to the case of $T=5$ is much less than that from the case of $T=5$ to that of $T=3$. Based on \eqref{muT}, $T=5$ leads to the average feedback load $\XiPow\approx1.54$.
%This is due to the fact that the probability $\PR{\NPowFB\geq1}=1-\exp\paren{-\XiPow}$ experiences significant reduction from $T=5$ to $T=3$ while this reduction becomes relatively small for values of $T$ higher than $5$.
This observation numerically demonstrates that small average feedback load around 1.54 is enough to experience negligible optimization loss, compared to the non-feedback case. This suggests that we can achieve a significant reduction in the feedback load in distributed network operation with limited-feedback, {\em whilst} not sacrificing from the data rate performance.

%\begin{figure}[!htb]
%\centering
%\includegraphics[width=0.7\textwidth]{exp-FB-T-6-10-inf.eps}
%\caption{Exponential model: $\lambda=0.1$, $d = 1$, $N = 16$, $Pt/N0=10$ dB, $\alpha=1.037$}\label{fig:outage4}
%\end{figure}
%
%\begin{figure}[!htb]
%\centering
%\includegraphics[width=0.7\textwidth]{rate-RIS-expPath-SNR-FB.eps}
%\caption{Power-law: $\lambda=0.1$, $d = 1$, $N = 16$, $\alpha=1.037$}\label{fig:outage4}
%\end{figure}

\vspace{-5mm}

\section{Conclusion}\label{sec:con}
In this paper, we have considered an RIS-aided wireless network where a single RIS is chosen from multiple PPP-distributed RISs to establish a communication link between the TX and RX. We have analyzed product-scaling and sum-scaling path-loss models in detail, which covers the important cases of RIS-aided end-to-end path-loss models. For each path-loss law, we have proposed an optimum location-based RIS selection policy which aims to maximize the network $\snr$ and derived the distance distribution of the optimum RIS node location. Based on these distributions, we have evaluated the outage probability and the average rate of the optimum RIS selection policies to assess the network performance for the product-scaling and sum-scaling path-loss models. To make the network operation distributed, we have further assumed the extra feedback capability at RISs and proposed limited-feedback RIS selection policies. We have derived the outage probabilities and average rates achieved by limited-feedback RIS selection policies for both path-loss models by deriving the distribution of the number of RISs feeding back. Our numerical results show the performance advantage of the proposed optimum and limited feedback RIS selection policies and reveal the performance gap for sub-optimum policies for product-scaling and sum-scaling path-loss models. Furthermore, the impact of system parameters, e.g., the number of reflecting elements and RIS node density, on the network performance have been quantified thoroughly by means of a comprehensive numerical analysis utilizing the derived analytical results.

\Fr{Interesting future work includes considering the mixture product and sum scaling path-loss models for all RISs in a PPP and analyzing the performance metrics of the optimum selection policies under the mixture path-loss models. This work is an open problem since there are no completely consistent conditions to determine the regime where the product-scaling law holds and the regime where the sum-scaling law holds. Existing work, e.g., \cite{9206044,9433568}, derive the path-loss under different setups and assumptions. Moreover, some conditions for determining these regimes and path-loss functions in \cite{9206044,9433568} do not have straightforward relationships with the TX-RIS and RIS-RX distances. This consequently makes it very challenging to analytically derive the performance metrics of optimum RIS selection policies where potential RISs have mixture path-loss models using the available path-loss models in the current literature. This problem may become tractable when easy-to-use conditions and path-loss expressions for the product-scaling and sum-scaling laws are available. }

%\newpage
%===================================
%\appendix
\numberwithin{equation}{section}
\begin{appendices}

%\section{Proof of Lemma~\ref{lemma:cdf-uniform}}\label{Appendix: cdf-uniform}
\vspace{-5mm}
\section{Proof of Theorem~\ref{Theorem: Optimal dis distribution}}\label{Appendix: optimal dis distribution Proof}
We first present an important lemma that lays the foundations for proving Theorem~\ref{Theorem: Optimal dis distribution}. This lemma will also be used in Appendix \ref{Appendix: feedback-power} for proving Theorem~\ref{theorem:feedback-power} as well.
\begin{lemma}\label{lemma:cdf-uniform}
We denote $\mathcal{B}_{\rm right}\paren{\vec{0}, \tau}$ as the right half disc having non-negative first coordinates with radius $\tau$ centered at the origin $\vec{0}$. We assume that $\vec{U}_{\rm r}$ is a uniformly distributed random node over $\mathcal{B}_{\rm right}\paren{\vec{0}, \tau}$. Let also $\Upsilon = \RISselectPow\paren{\vec{U}_{\rm r}}$. The expression for the CDF of RV $\Upsilon$, $F_{\Upsilon}(\gamma)$, is given by \eqref{Eqn: Distance Distribution}.
\begin{figure*}
{\small
\begin{equation}\label{Eqn: Distance Distribution}
F_{\Upsilon}(\gamma)
=\left\{
\begin{array}{ll}
\frac{2}{\pi \tau^2}\left(\frac{1}{d^2}\left(d^4 E(\frac{\gamma^2}{d^4})+(\gamma^2-d^4)K(\frac{\gamma^2}{d^4})\right)\right)& \mbox{ if }\gamma \leq d^2 \\
\frac{2}{\pi \tau^2}\left(\gamma E(\frac{d^4}{\gamma^2})
\right)  & \mbox{ if } d^2 < \gamma \leq \tau^2-d^2 \\
\frac{2}{\pi \tau^2}\left(-\frac{d^2}{2}\sqrt{1 - \frac{(d^4 - \gamma^2 + \tau^4)^2}{4d^4\tau^4}}
+ \gamma E(\frac{d^4}{\gamma^2})- \frac{\gamma E(\arccos (\frac{d^4+\tau^4-\gamma^2}{2d^2\tau^2}),\frac{d^4}{\gamma^2})}{2}+\theta_2\tau^2\right);
& \mbox{ if } \tau^2-d^2 < \gamma \leq d^2+\tau^2 \\
1
&\mbox{ if } \gamma > d^2+\tau^2
\end{array}\right.
\end{equation}}
\hrule
\end{figure*}
\end{lemma}
\begin{IEEEproof}
By using the law of cosines, $\Upsilon$ can be written as
\begin{eqnarray}
\Upsilon &=& \sqrt{\left({\norm{\vec{U}_{\rm r}}}^2 + 2d\norm{\vec{U}_{\rm r}} \cos\Theta + d^2\right)\left({\norm{\vec{U}_{\rm r}}}^2 - 2d\norm{\vec{U}_{\rm r}} \cos\Theta + d^2\right)}\nonumber \\
&=&\sqrt{d^4+{\norm{\vec{U}_{\rm r}}}^4-2d^2{\norm{\vec{U}_{\rm r}}}^2\cos2\Theta},
\end{eqnarray}
where $\Theta$ is the angle between the non-negative $x$-axis and the line segment connecting $\vec{0}$ and $\vec{U}_{\rm r}$. $\Theta$ is uniformly distributed over $\sqparen{-\frac{\pi}{2}, \frac{\pi}{2}}$, and is independent of $\norm{\vec{U}_{\rm r}}$ because $\vec{U}_{\rm r}$ is uniformly distributed. Thus, we derive the conditional CDF of $\Upsilon$ given $\brparen{\Theta = \theta}$ as
\begin{eqnarray}\label{cond-CDF}
F_{\Upsilon | \Theta}\paren{\gamma | \theta} = \PR{\Upsilon^2 \leq \gamma^2 | \Theta = \theta}
=\PR{d^4+{\norm{\vec{U}_{\rm r}}}^4-2d^2{\norm{\vec{U}_{\rm r}}}^2\cos2\theta\leq \gamma^2 | \Theta = \theta}.
%&=& \PR{\norm{\vec{U}_{\rm r}} \leq \sqrt{\gamma^2 - d^2\sin^2\theta} - d\cos\theta} \nonumber
\end{eqnarray}

To solve \eqref{cond-CDF}, we study the monotonicity of the function $f\left(\norm{\vec{U}_{\rm r}}\right) = d^4+{\norm{\vec{U}_{\rm r}}}^4-2d^2{\norm{\vec{U}_{\rm r}}}^2\cos2\theta$. We obtain the first derivative of $f\left(\norm{\vec{U}_{\rm r}}\right)$ with respect to $\norm{\vec{U}_{\rm r}}$ as
%\begin{eqnarray}\label{cond-cdf-deri}
$f^{'}\left(\norm{\vec{U}_{\rm r}}\right) = 4{\norm{\vec{U}_{\rm r}}}^3-4d^2\norm{\vec{U}_{\rm r}}\cos2\theta$.
%\end{eqnarray}
Based on the expression of $f^{'}\left(\norm{\vec{U}_{\rm r}}\right)$, we find that $f^{'}\left(\norm{\vec{U}_{\rm r}}\right)>0$ holds when the case 1) $\cos2\theta\leq0$ and $\norm{\vec{U}_{\rm r}}>0$ or case 2) $\cos2\theta>0$ and $\norm{\vec{U}_{\rm r}}>\sqrt{d^2\cos2\theta}$ is satisfied. That is to say, for
$\theta \in \sqparen{\frac{\pi}{4}, \frac{\pi}{2}}\cup \theta \in \sqparen{-\frac{\pi}{2}, -\frac{\pi}{4}}$, $f\left(\norm{\vec{U}_{\rm r}}\right)$ is an increasing function with $\norm{\vec{U}_{\rm r}}$. For $\theta \in \sqparen{-\frac{\pi}{4}, \frac{\pi}{4}}$, $f\left(\norm{\vec{U}_{\rm r}}\right)$ is an increasing function when  $\norm{\vec{U}_{\rm r}}>\sqrt{d^2\cos2\theta}$ and $f\left(\norm{\vec{U}_{\rm r}}\right)$ is a decreasing function when $\norm{\vec{U}_{\rm r}}<\sqrt{d^2\cos2\theta}$.

We first solve \eqref{cond-CDF} for $\theta \in \sqparen{\frac{\pi}{4}, \frac{\pi}{2}}\cup \sqparen{-\frac{\pi}{2}, -\frac{\pi}{4}}$. Under these conditions, only one positive root $U_1$ for $f\left(\norm{\vec{U}_{\rm r}}\right)-\gamma^2=0$ exists, which is given by
\begin{eqnarray}\label{root-1}
U_1 &=& \sqrt{-d^2+2d^2\cos^{2}\theta+\frac{\sqrt{-d^4+2\gamma^2+d^4\cos4\theta}}{\sqrt{2}}}.
\end{eqnarray}

Since $0\leq\norm{\vec{U}_{\rm r}}\leq\tau$ and $f\left(\norm{\vec{U}_{\rm r}}\right)$ is an increasing function with $\norm{\vec{U}_{\rm r}}$, thus we obtain $d^4\leq f\left(\norm{\vec{U}_{\rm r}}\right) \leq d^4+\tau^4-2d^2\tau^2\cos2\theta$. Given $U_1$ and the range of $f\left(\norm{\vec{U}_{\rm r}}\right)$, we solve \eqref{cond-CDF} for $\theta \in \sqparen{\frac{\pi}{4}, \frac{\pi}{2}}\cup \sqparen{-\frac{\pi}{2}, -\frac{\pi}{4}}$ as
\begin{equation}\label{Eqn: Conditional Distance CDF}
F_{\Upsilon | \Theta}\left(\gamma | \theta, \theta \in \sqparen{\frac{\pi}{4}, \frac{\pi}{2}}\cup  \sqparen{-\frac{\pi}{2}, -\frac{\pi}{4}}\right)
=\left\{
\begin{array}{ll}
0 \qquad\qquad\mbox{ if } \gamma < d^2\\
\PR{\norm{\vec{U}_{\rm r}}\leq U_1}=\frac{{U_1}^2}{\tau^2}\\
 \quad\mbox{ if } d^2 \leq \gamma \leq \sqrt{d^4+\tau^4-2d^2\tau^2\cos2\theta} \\
1 \qquad\quad\mbox{ if } \gamma > \sqrt{d^4+\tau^4-2d^2\tau^2\cos2\theta}
\end{array}\right.
\end{equation}
where $\PR{\norm{\vec{U}_{\rm r}}\leq U_1}=\frac{{U_1}^2}{\tau^2}$ is based on the CDF of $\norm{\vec{U}_{\rm r}}$, i.e., $F_{\norm{\vec{U}_{\rm r}}}(u) = \frac{u^2}{\tau^2}$.

We next solve \eqref{cond-CDF} for $\theta \in \sqparen{-\frac{\pi}{4}, \frac{\pi}{4}}$. Under these conditions, two positive roots $U_1$ and $U_2$ for $f\left(\norm{\vec{U}_{\rm r}}\right)-\gamma^2=0$ exist, where $U_1$ is given by \eqref{root-1} and $U_2$ is given by\\
%\begin{eqnarray}\label{root-2}
$U_2 = \sqrt{-d^2+2d^2\cos^{2}\theta-\frac{\sqrt{-d^4+2\gamma^2+d^4\cos4\theta}}{\sqrt{2}}}$.
%\end{eqnarray}
Since $f\left(\norm{\vec{U}_{\rm r}}\right)$ is an increasing function when  $\left(\norm{\vec{U}_{\rm r}}\right)>\sqrt{d^2\cos2\theta}$ and $f\left(\norm{\vec{U}_{\rm r}}\right)$ is a decreasing function when $\left(\norm{\vec{U}_{\rm r}}\right)<\sqrt{d^2\cos2\theta}$, $f\left(\norm{\vec{U}_{\rm r}}\right)$ attains the minimal value $f\left(\sqrt{d^2\cos2\theta}\right)=d^4-d^4\cos^{2}2\theta$ at $\norm{\vec{U}_{\rm r}}=\sqrt{d^2\cos2\theta}$. Thus, the range of $f\left(\norm{\vec{U}_{\rm r}}\right)$ is $d^4-d^4\cos^{2}2\theta \leq f\left(\norm{\vec{U}_{\rm r}}\right)\leq d^4+\tau^4-2d^2\tau^2\cos2\theta$. Given $U_1$, $U_2$, and the range of $f\left(\norm{\vec{U}_{\rm r}}\right)$, we solve \eqref{cond-CDF} when $\theta \in \sqparen{-\frac{\pi}{4}, \frac{\pi}{4}}$ as
\begin{equation}\label{Eqn: Conditional Distance CDF-1}
F_{\Upsilon | \Theta}\left(\gamma | \theta,\theta \in \sqparen{-\frac{\pi}{4}, \frac{\pi}{4}}\right)
=\left\{
\begin{array}{ll}
0 \qquad\qquad\mbox{ if } \gamma < \sqrt{d^4-d^4\cos^{2}2\theta}\\
\PR{U_2\leq\norm{\vec{U}_{\rm r}}\leq U_1}=\frac{{U_1}^2-{U_2}^2}{\tau^2}\\
 \quad\mbox{ if } \sqrt{d^4-d^4\cos^{2}2\theta} \leq \gamma \leq
 d^2\\
 \PR{\norm{\vec{U}_{\rm r}}\leq U_1}=\frac{{U_1}^2}{\tau^2}\\
 \quad\mbox{ if } d^2\leq \gamma \leq
  \sqrt{d^4+\tau^4-2d^2\tau^2\cos2\theta} \\
1 \qquad\quad\mbox{ if } \gamma > \sqrt{d^4+\tau^4-2d^2\tau^2\cos2\theta}
\end{array}\right.
\end{equation}

We will obtain $F_{\Upsilon}(\gamma)$ by averaging $F_{\Upsilon | \Theta}(\gamma | \theta)$ over $\Theta$. For $\theta \in \sqparen{\frac{\pi}{4}, \frac{\pi}{2}}\cup \sqparen{-\frac{\pi}{2}, -\frac{\pi}{4}}$, we have
\begin{equation}\label{Eqn: Conditional Distance CDF-int1}
F_{\Upsilon}\left(\gamma |  \theta \in \sqparen{\frac{\pi}{4}, \frac{\pi}{2}}\cup  \sqparen{-\frac{\pi}{2}, -\frac{\pi}{4}}\right)
=\left\{
\begin{array}{ll}
0 \qquad\qquad\mbox{ if } \gamma < d^2\\
\frac{2}{\pi}\paren{\int_{\frac{\pi}{4}}^{\frac{\pi}{2}} \frac{{U_1}^2}{\tau^2} d\theta}
\quad\mbox{ if } d^2 \leq \gamma \leq \sqrt{d^4+\tau^4}\\
\frac{2}{\pi \tau^2}\paren{\int_{\theta_2}^{\frac{\pi}{2}} {U_1}^2 d\theta+\int_{\frac{\pi}{4}}^{\theta_2} \tau^2 d\theta}
\quad\mbox{ if } \sqrt{d^4+\tau^4} \leq \gamma \leq d^2+\tau^2\\
\frac{2}{\pi\tau^2}\paren{\int_{\frac{\pi}{2}}^{\frac{\pi}{4}}\tau^2  d\theta}
\quad\mbox{ if }\gamma \geq d^2+\tau^2
\end{array}\right.
\end{equation}

For $\theta \in \sqparen{-\frac{\pi}{4}, \frac{\pi}{4}}$, we have
\begin{equation}\label{Eqn: Conditional Distance CDF-int2}
F_{\Upsilon}\left(\gamma |  \theta \in \sqparen{-\frac{\pi}{4}, \frac{\pi}{4}}\right)
=\left\{
\begin{array}{ll}
\frac{2}{\pi}\paren{\int_{\theta_1}^{\frac{\pi}{4}} 0 d\theta+\int_{0}^{\theta_1} \frac{{U_1}^2-{U_2}^2}{\tau^2} d\theta}  \qquad\qquad\mbox{ if }
\gamma < d^2\\
\frac{2}{\pi}\paren{\int_{0}^{\frac{\pi}{4}} \frac{{U_1}^2}{\tau^2} d\theta}
\quad\mbox{ if } d^2 \leq \gamma \leq \sqrt{\tau^2-d^2}\\
\frac{2}{\pi}\paren{\int_{\theta_2}^{\frac{\pi}{4}} \frac{{U_1}^2}{\tau^2} d\theta+\int_{0}^{\theta_2} 1 d\theta}
\quad\mbox{ if } \tau^2-d^2 \leq \gamma \leq \sqrt{d^4+\tau^4}\\
\frac{2}{\pi}\paren{\int_{0}^{\frac{\pi}{4}}1  d\theta}
\quad\mbox{ if }\gamma \geq\sqrt{d^4+\tau^4}
\end{array}\right.
\end{equation}

Combining \eqref{Eqn: Conditional Distance CDF-int1} and \eqref{Eqn: Conditional Distance CDF-int2}, for $\theta \in \sqparen{-\frac{\pi}{2}, \frac{\pi}{2}}$, we have
\begin{equation}\label{Eqn: Conditional Distance CDF-int3}
F_{\Upsilon}\left(\gamma |  \theta \in \sqparen{-\frac{\pi}{2}, \frac{\pi}{2}}\right)
=\left\{
\begin{array}{ll}
\frac{2}{\pi}\paren{\int_{0}^{\theta_1} \frac{{U_1}^2-{U_2}^2}{\tau^2} d\theta}  \qquad\qquad\mbox{ if }
\gamma < d^2\\
\frac{2}{\pi}\paren{\int_{0}^{\frac{\pi}{4}} \frac{{U_1}^2}{\tau^2} d\theta+\int_{\frac{\pi}{4}}^{\frac{\pi}{2}} \frac{{U_1}^2}{\tau^2} d\theta}
\quad\mbox{ if } d^2 \leq \gamma \leq \tau^2-d^2\\
\frac{2}{\pi}\paren{\int_{\theta_2}^{\frac{\pi}{4}} \frac{{U_1}^2}{\tau^2} d\theta+\int_{0}^{\theta_2} 1 d\theta
+\int_{\frac{\pi}{4}}^{\frac{\pi}{2}} \frac{{U_1}^2}{\tau^2} d\theta}
\quad\mbox{ if } \tau^2-d^2 \leq \gamma \leq \sqrt{d^4+\tau^4}\\
\frac{2}{\pi}\paren{\int_{0}^{\frac{\pi}{4}} 1 d\theta+\int_{\theta_2}^{\frac{\pi}{2}} \frac{{U_1}^2}{\tau^2} d\theta
+\int_{\frac{\pi}{4}}^{\theta_2} 1 d\theta}
\quad\mbox{ if } \sqrt{d^4+\tau^4} \leq \gamma \leq d^2+\tau^2\\
\frac{2}{\pi}\paren{\int_{0}^{\frac{\pi}{4}}1  d\theta+
\int_{\frac{\pi}{4}}^{\frac{\pi}{2}}1  d\theta
}
\quad\mbox{ if }\gamma \geq d^2+\tau^2
\end{array}\right.
\end{equation}

Applying \cite[eq.~(2.576)]{Gradshteyn07_book} given by
%\begin{align}\label{int-1}
$\int \sqrt{a+b\cos x}dx = \frac{2}{b}\paren{(a-b)F(c_1,\frac{1}{c_2})+2bE(c_1,\frac{1}{c_2})}$
%\end{align}
and applying
%\begin{align}\label{int-2}
$\int \sqrt{a+b\cos x}dx = 2\sqrt{a+b}E(\frac{x}{2},c_1)$
%\end{align}
to \eqref{Eqn: Conditional Distance CDF-int3}, where $c_1=\arcsin\sqrt{\frac{b(1-\cos x)}{a+b}}$ and $c_2= \sqrt{\frac{2b}{a+b}}$, we obtain $F_{\Upsilon}(\gamma)$ as in \eqref{Eqn: Distance Distribution} in Lemma \ref{lemma:cdf-uniform}. The proof of Lemma 1 ends here. Finally, using
the relation
%\begin{eqnarray}\label{express-1}
$F_{\Upsilon_{\rm opt}}\paren{\gamma} = 1 - \paren{\lim_{\tau \ra \infty} \exp\paren{{-\frac{\lambda \pi \tau^2}{2}F_{\Upsilon}\paren{\gamma}}}}^2$
%\end{eqnarray}
given by \cite{atapattu2020locationbased} and the result of $F_{\Upsilon}(\gamma)$ given in Lemma~\ref{lemma:cdf-uniform}, we arrive at \eqref{cdf-product}, which concludes the proof.
\end{IEEEproof}

\vspace{-5mm}
\section{Proof of Theorem~\ref{theorem: minSum dis distribution}}\label{Appendix: minSum dis distribution Proof}
We assume $\vec{U}_{\rm r}$ is a uniformly distributed random RIS location. We also assume that $\Lambda = \RISselectExp\paren{\vec{U}_{\rm r}}$, which is the sum of the distance between the source node and $\vec{U}_{\rm r}$ and the distance between $\vec{U}_{\rm r}$ and the destination node. By using the law of cosines, we write $\Lambda$ as
\begin{eqnarray}
\Lambda &=& \sqrt{{\norm{\vec{U}_{\rm r}}}^2 + 2d\norm{\vec{U}_{\rm r}} \cos\Theta + d^2}+\sqrt{{\norm{\vec{U}_{\rm r}}}^2 - 2d\norm{\vec{U}_{\rm r}} \cos\Theta + d^2}
\end{eqnarray}
The conditional CDF of $\Lambda$ given $\brparen{\Theta = \theta}$ can be expressed as
\begin{eqnarray}\label{cond-cdf-minSum}
F_{\Lambda | \Theta}\paren{\gamma | \theta} =\PR{\Lambda^2 \leq \gamma^2 | \Theta = \theta} %\nonumber \\
=\PR{w\paren{\norm{\vec{U}_{\rm r}}}\leq \gamma^2 | \Theta = \theta},
%&=& \PR{\norm{\vec{U}_{\rm r}} \leq \sqrt{\gamma^2 - d^2\sin^2\theta} - d\cos\theta} \nonumber
\end{eqnarray}
where
\begin{eqnarray}\label{w-u}
w\paren{\norm{\vec{U}_{\rm r}}}=2d^2+2\norm{\vec{U}_{\rm r}}^2\!+\!\sqrt{\paren{{\norm{\vec{U}_{\rm r}}}^2 \!+\! 2d\norm{\vec{U}_{\rm r}} \cos\theta + d^2}\paren{{\norm{\vec{U}_{\rm r}}}^2 - 2d\norm{\vec{U}_{\rm r}} \cos\theta + d^2}}.
%&=& \PR{\norm{\vec{U}_{\rm r}} \leq \sqrt{\gamma^2 - d^2\sin^2\theta} - d\cos\theta} \nonumber
\end{eqnarray}

We note that only one positive root $W_1$ for $w\left(\norm{\vec{U}_{\rm r}}\right)-\gamma^2=0$ exist, where $W_1$ is given by
\begin{eqnarray}\label{root-1-mSum}
W_1 &=& \frac{\sqrt{-4d^2\gamma^2+\gamma^4}}{2\sqrt{\gamma^2-4d^2\cos^{2}\theta}}.
\end{eqnarray}

Based on $w\paren{\norm{\vec{U}_{\rm r}}}$ is the increasing function with respect to $\norm{\vec{U}_{\rm r}}$ and $0\leq\norm{\vec{U}_{\rm r}}\leq\tau$, we obtain the range of $w\left(\norm{\vec{U}_{\rm r}}\right)$ is $4d^2 \leq w\left(\norm{\vec{U}_{\rm r}}\right)\leq w\paren{\tau}$. Given $W_1$ and the range of $w\left(\norm{\vec{U}_{\rm r}}\right)$, we solve \eqref{cond-cdf-minSum} as
\begin{equation}\label{Eqn: Conditional Distance CDF-mSum}
F_{\Lambda | \Theta}\left(\gamma | \theta\right)
=\left\{
\begin{array}{ll}
0 \qquad\qquad\mbox{ if } \gamma < 2d\\
\PR{\norm{\vec{U}_{\rm r}}\leq W_1}=\frac{{W_1}^2}{\tau^2}
 \quad\mbox{ if } 2d \leq \gamma \leq \sqrt{w\paren{\tau}} \\
1 \qquad\quad\mbox{ if } \gamma > \sqrt{w\paren{\tau}}
\end{array}\right.
\end{equation}

We will obtain $F_{\Lambda}(\gamma)$ by averaging $F_{\Omega | \Theta}(\gamma | \theta)$ over $\Theta$. We need to consider four cases separately, as follows:
\begin{equation}\label{Eqn: Distance Distribution-mSum}
F_{\Lambda}(\gamma)
=\left\{
\begin{array}{ll}
0& \mbox{ if }\gamma \leq 2d \\
\frac{2}{\pi\tau^2} \int_{0}^{\frac{\pi}{2}} {W_1}^2 d\theta
& \mbox{ if } 2d < \gamma \leq 2\tau \\
\frac{2}{\pi\tau^2} \int_{\phi_1}^{\frac{\pi}{2}} {W_1}^2 d\theta+\frac{2}{\pi} \int_{0}^{\phi_1} 1 d\theta
& \mbox{ if } 2\tau < \gamma \leq 2\sqrt{d^2+\tau^2} \\
1& \mbox{ if } \gamma > 2\sqrt{d^2+\tau^2}
\end{array},\right.
\end{equation}
where
%\begin{eqnarray}\label{phi1}
$\phi_1 = \frac{1}{2}\arccos\paren{\frac{d^4+\tau^4-\frac{{\paren{\gamma^2-2d^2-2\tau^2}}^2}{4}}{2d^2\tau^2}}$
%\end{eqnarray}
and $\phi_1$ is obtained by solving $\gamma^2=w\paren{\tau}$. Applying \cite[eq.~(2.562.2)]{gradshteyn2007}, we obtain $F_{\Lambda}(\gamma)$ as in \eqref{Eqn: Distance Distribution-mSum1}.
\begin{figure*}
{\small
\begin{equation}\label{Eqn: Distance Distribution-mSum1}
F_{\Lambda}(\gamma)
=\left\{
\begin{array}{ll}
0& \mbox{ if }\gamma \leq 2d \\
\frac{2}{\pi\tau^2} \paren{\frac{1}{8}(-1)^{\Floor{\frac{\pi-2\Arg{\gamma}+\Arg{-4d^2+\gamma^2}}{2\pi}}}\pi\gamma\sqrt{-4d^2+\gamma^2}}& \mbox{ if }2d < \gamma \leq 2\tau \\
\frac{2}{\pi\tau^2} \paren{-\frac{\arctang\paren{\cot\paren{\theta}\sqrt{\frac{-4d^2+\gamma^2}{\gamma^2}}}\sqrt{\gamma^2\paren{-4d^2+\gamma^2}}}{4}+\phi_1\tau^2}
& \mbox{ if } 2\tau < \gamma \leq 2\sqrt{d^2+\tau^2}  \\
1& \mbox{ if } \gamma > 2\sqrt{d^2+\tau^2}
\end{array}\right.
\end{equation}}
\hrule
\end{figure*}
%and \eqref{relation2}
Using $F_{\Lambda_{\rm opt}}\paren{\gamma} = 1 - \paren{\lim_{\tau \ra \infty} \exp\paren{{-\frac{\lambda \pi \tau^2}{2}F_{\Lambda}\paren{\gamma}}}}^2$ \cite{atapattu2020locationbased}, we arrive at
\eqref{cdf-d-minSum}. This completes the proof.

\vspace{-5mm}
\section{Proof of Theorem~\ref{theorem:feedback-power}}\label{Appendix: feedback-power}
We denote $\mathcal{B}\paren{\vec{0}, \tau}$ as the disc centered at the origin $\vec{0}$ with radius $\tau$. We assume $\XiPow(\tau)$ as the average number of RISs located in $\mathcal{B}\paren{\vec{0}, \tau}$ that feedback their channel quality indicators. We also assume that $\vec{U}$ is a uniformly distributed random node over $\mathcal{B}\paren{\vec{0}, \tau}$. Thus, we have
% Thus, we have $\XiPow(\tau)=\EW_\Phi\sqparen{\sum_{\vec{x}\in\phi\bigcap\mathcal{B}\paren{\vec{0}, \tau}}\I{\RISselectPow\paren{\vec{x}}\leq \Tpow}}$. We assume the number of RISs is $n$ in particular a realization $\phi$. Since all the RISs are independently and uniformly distributed over $\mathcal{B}\paren{\vec{0}, \tau}$, we have $\EW_\Phi\sqparen{\sum_{\vec{x}\in\phi\bigcap\mathcal{B}\paren{\vec{0}, \tau}}\I{\RISselectPow\paren{\vec{x}}\leq \Tpow}}=n\PR{\RISselectPow\paren{\vec{U}}\leq \Tpow}$, where $\vec{U}$ is a uniformly distributed random node over $\mathcal{B}\paren{\vec{0}, \tau}$. Since $\EW_\Phi\sqparen{n}=\lambda\pi\tau^2$, we can write $\XiPow(\tau)$ as
\begin{eqnarray}\label{aver-feedback-tau}
\XiPow(\tau) = \lambda\pi\tau^2 \PR{\RISselectPow\paren{\vec{U}}\leq \Tpow}.
\end{eqnarray}

We next obtain $\PR{\RISselectPow\paren{\vec{U}}\leq \Tpow}$. We recall that $\vec{U}_{\rm r}$ is defined in Lemma \ref{lemma:cdf-uniform} and $\vec{U}_{\rm r}$ is a uniformly distributed random node over right half disc $\mathcal{B}_{\rm right}\paren{\vec{0}, \tau}$. Similarly, we define $\mathcal{B}_{\rm left}\paren{\vec{0}, \tau}$ as the left half disc that centered at the origin $\vec{0}$ with radius $\tau$ having negative first coordinates. We let $\vec{U}_{\rm l}$ is a uniformly distributed random node over left half disc $\mathcal{B}_{\rm left}\paren{\vec{0}, \tau}$.

Since the distribution of $\vec{U}_{\rm r}$ over $\mathcal{B}_{\rm right}$ is same as the distribution of $\vec{U}_{\rm l}$ over $\mathcal{B}_{\rm left}$, $\vec{U}$, $\vec{U}_{\rm r}$, and $\vec{U}_{\rm l}$ are identically distributed RVs. As such, we have
\begin{eqnarray}\label{relation-feedback}
\PR{\RISselectPow\paren{\vec{U}}\leq \Tpow} = \PR{\RISselectPow\paren{\vec{U}_{\rm r}}\leq \Tpow} = \PR{\RISselectPow\paren{\vec{U}_{\rm l}}\leq \Tpow}.
\end{eqnarray}

We note that the CDF of $\Upsilon$, $F_{\Upsilon}(\gamma)$, is given by \eqref{Eqn: Distance Distribution}, where $\Upsilon = \RISselectPow\paren{\vec{U}_{\rm r}}$. Thus, the expression for $\PR{\RISselectPow\paren{\vec{U}}\leq \Tpow}$ can be obtained by replacing $\gamma$ with $\Tpow$ in \eqref{Eqn: Distance Distribution}. Finally, applying the expression for $\PR{\RISselectPow\paren{\vec{U}}\leq \Tpow}$ to \eqref{aver-feedback-tau} and taking the limit $\XiPow=\lim_{\tau\rightarrow}\XiPow(\tau)$, we arrive at \eqref{muT}. The Poisson distribution property for the number of RISs feeding back can be established by using characteristic functions as in \cite{atapattu2020locationbased}, which completes the proof. %This completes the proof.

\section{Simulations with Mixture Path-loss Models}\label{Appendix: simu}

In this appendix, we conduct simulations with mixture product-scaling and sum-scaling path-loss models and compare these simulations with our analysis with a single path-loss models for all potential RISs, to show the reasonableness of the used assumption. We recall that \cite{9206044} shows the path-loss is product-scaling if the TX and RX are both in the far-field of RIS in the beamforming case and shows the path-loss is sum-scaling if the TX and RX are both or only one of them in the near-field of RIS, in the broadcasting case. Based on the conditions where product-scaling and sum-scaling path-loss models hold in \cite{9206044}, in our simulations, we calculate the path-loss of each RIS using sum-scaling law if the TX and RX are both or only one of them in the near-field of RIS or using the product-scaling law if the TX and RX are both in the far-field of RIS. Fortunately, by conducting new simulations, we show that i) the results derived by assuming all RISs have the same sum-scaling path-loss model can be good approximations of the results that assumes mixture path-loss models for large RISs, and ii) the results assuming all RISs have the same product-scaling path-loss model can be good approximations of the results assuming mixture path-loss models for small RISs.
Please see the results below:

 \begin{figure}[H]
	\centering
\subfloat[$\lambda=0.1$, $N=16$]{
		\label{Fig-path-loss-2-e}
		\includegraphics[width=0.45\textwidth]{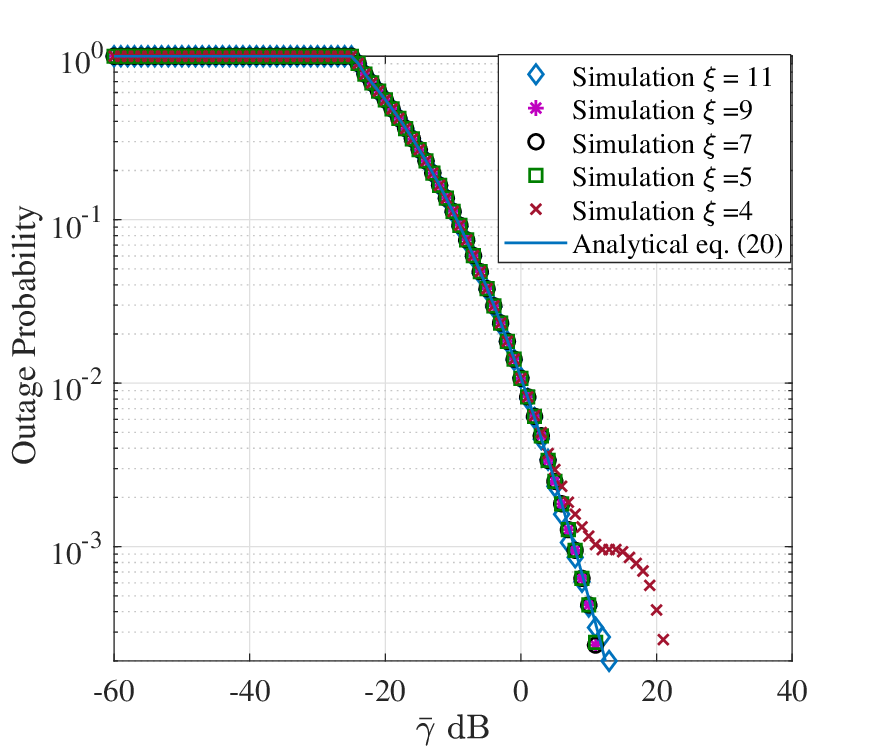}}
	\subfloat[$\lambda=0.1$, $N=128$]{
		\label{Fig-path-loss-1-e}
		\includegraphics[width=0.45\textwidth]{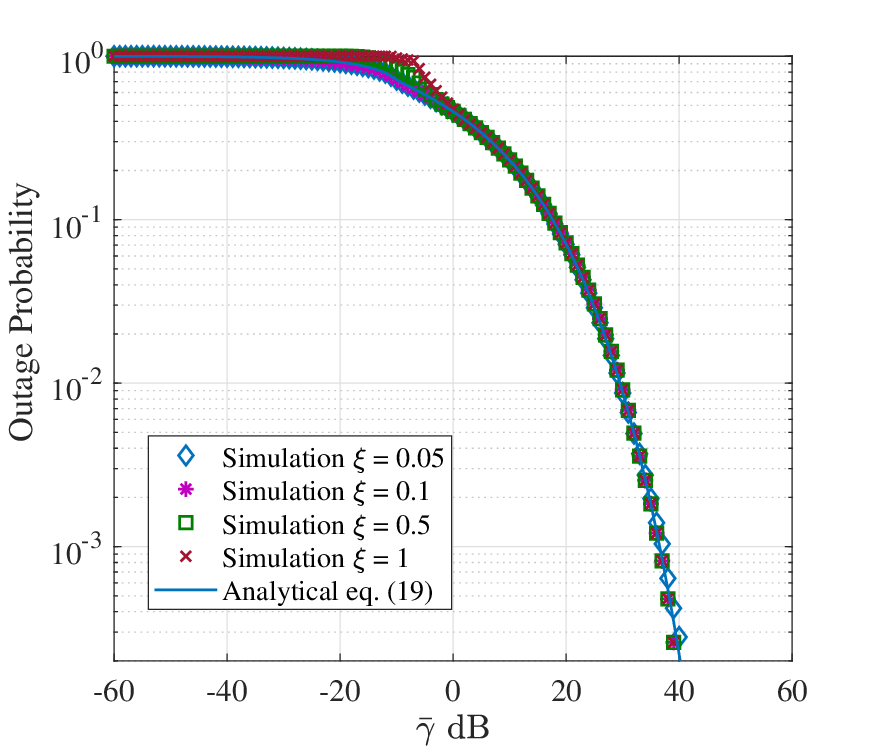}}

\vspace{-3mm}
		\caption{In simulations, the path-loss of RIS-aided links follow the power-law path-loss model if the RIS located at $\vec{X}_i$ satisfy $\norm{\xs -\vec{X}_i}>\xi$ and $\norm{\vec{X}_i -\xd}>\xi$; otherwise, RIS-aided links follow the exp-law path-loss model. $\xi=\frac{8L_{x}L_{y}}{\lambda_s}$ is the boundary of the far-field and near-field defined in \cite{9206044} and other parameters are the same as those in the manuscript.}%is used to generate this figure.
	\label{Fig-path-loss-e}
\end{figure}

In Fig. \ref{Fig-path-loss-2-e}, we observe that the simulations with the boundary values of $\xi=5,7,9,11$ have a good agreement with (20). This means the analytical outage probability with the sum-scaling law is a good approximation of the simulated outage probability with product-scaling and sum-scaling path-loss models co-existing for potential RISs, when the boundary value $\xi$ is relatively large, e.g., $\xi\geq5$. This is because the RISs in the regime of product-scaling path-loss model (i.e., the TX and RX are both in the far-field of the RIS) are less likely to be the optimum one compared with the RISs in the regime of sum-scaling path-loss model (i.e., the TX and RX are both or only one of them in the near-field of RIS). This likelihood decreases when the boundary value $\xi$ of the far-field and near-field increases.
We note that $\xi\geq5$ is the normal range values of $\xi$ of electrically-large RISs. For example, \cite{9206044} considers one large RIS prototype with the size of $1\,\m\times1.2\,\m$ and another large RIS prototype with the size of $0.34\,\m\times0.5\,\m$, and carrier frequency is 10.5 GHz, which leads to $\xi=71.4\,\m$ and $\xi=11.9\,\m$. %They are in the range where our analysis with the single sum-scaling path-loss model is accurate.
In Fig. \ref{Fig-path-loss-1-e}, we observe that the simulations with the boundary values of $\xi=0.05,0.1$ have a good agreement with (19). This means the analytical outage probability with the product-scaling law is a good approximation of the simulated outage probability with product-scaling and sum-scaling path-loss models co-existing for potential RISs, when the boundary value $\xi$ is relatively small, e.g., $\xi\leq0.1$. This is because most of RISs are in the regime of product-scaling path-loss model, i.e., the TX and RX are both in the far-field of most of RISs, when $\xi$ is small. We note that $\xi\leq0.1$ is the normal range values of $\xi$ of electrically-small RISs. For example, \cite{9206044} considers a small RIS prototype with the size of $0.384\,\m\times0.096\,\m$, which leads to $\xi=0.1\,\m$ when the carrier frequency is 425 MHz.

\end{appendices}
\vspace{-5mm}
%\bibliographystyle{IEEEtran}
%\bibliography{references,IEEEabrv}

\end{document}